\documentclass[a4paper,11pt]{article}
\usepackage{jheppub} 
\usepackage{lineno}

\usepackage[utf8]{inputenc}
\usepackage{amsmath}
\usepackage{amssymb}
\usepackage{amsfonts}
\usepackage{amsthm}
\usepackage{xcolor}
\usepackage{hyperref}
\usepackage{rotating}
\usepackage{array}
\usepackage{appendix}
\usepackage{bbold}
\usepackage{textcomp}
\usepackage{float}
\usepackage{comment}
\usepackage{diagbox}
\newcolumntype{C}{>{$}c<{$}}
\usepackage{multirow}
\allowdisplaybreaks

\usepackage{caption}
\usepackage{subcaption}

\usepackage{tikz}
\usetikzlibrary{arrows,chains,matrix,positioning,scopes}

\theoremstyle{definition}

\def\fz{f^{\mathcal{Z}}}
\def\fp{f^{\langle\phi^2\rangle}}
\def\Ifp{I_f^{\langle\phi^2\rangle}}
\def\sc{\textit{sc}}

\def\llb{\langle\langle}
\def\rrb{\rangle\rangle}

\def\Measure{\mathcal{W}}

\title{Heavy-Heavy-Light Asymptotics from Thermal Correlators}
\author[a]{Ilija Buri\'c,}
\author[b]{Francesco Mangialardi,}
\author[c,d]{Francesco Russo,}
\author[b,e]{Volker Schomerus}
\author[d]{and Alessandro Vichi}

\affiliation[a]{School of Mathematics and Hamilton Mathematics Institute, Trinity College, Dublin 2, Ireland}
\affiliation[b]{Deutsches Elektronen Synchroton DESY, Notkestr. 85, 22607 Hamburg, Germany}
\affiliation[c]{CPHT, CNRS, \'Ecole polytechnique, Institut Polytechnique de Paris, 91120 Palaiseau, France}
\affiliation[d]{Department of Physics, University of Pisa and INFN, Largo Pontecorvo 3, I-56127 Pisa, Italy}
\affiliation[e]{II. Institut f\"{u}r Theoretische Physik, Universit\"{a}t Hamburg, Luruper Chaussee 149,D-22761 Hamburg, Germany}

\abstract{We revisit the calculation of spectral densities and heavy-heavy-light (HHL) operator product expansion (OPE) coefficients in three-dimensional conformal field theories using thermal one-point functions on $S^1 \times S^2$. A central element of our analysis is a new inversion formula for one-point functions which is derived via Casimir differential equations. We develop systematic expansions of the spectral density and HHL OPE coefficients in the regime of large $\Delta_H$. We validate our analytic tools by comparing the results with the partial wave expansions of thermal one-point functions in free field theories. The algorithms developed for these expansions make full use of Casimir recursion relations, thereby extending their applicability into the heavy exchange regime. In the end, we observe excellent agreement with our analytic predictions and an improvement of up to three orders of magnitude compared to all previous leading order estimates of the CFT data even for moderate values of $\Delta_H$.}

\emailAdd{burici@tcd.ie}\emailAdd{francesco.mangialardi@desy.de}
\emailAdd{francesco.russo@polytechnique.edu}
\emailAdd{volker.schomerus@desy.de}
\emailAdd{alessandro.vichi@unipi.it}

\begin{document}

\maketitle

\section{Introduction and summary}
\label{S:Introduction}

Finite temperature correlation functions are interesting observables in conformal field theories (CFTs) for a variety of experimental and theoretical reasons. Quantum critical points that these theories describe are always experimentally realised at non-zero temperature. In the context of holography, thermal correlators are related to field propagation on black hole backgrounds, \cite{Witten:1998zw}. From a broader perspective, finite temperature CFTs may be regarded as some of the simplest examples of conformal theories on backgrounds other than $\mathbb{R}^d$. It is expected that a consistent CFT may be put on any manifold - however, different manifolds may bring to the fore different properties of the theory. The simplest and most studied background other than the flat space has been $S^1 \times \mathbb{R}^{d-1}$. Two-point functions on this geometry contain a wealth of CFT data and satisfy a consistency condition in the form of KMS invariance, \cite{El-Showk:2011yvt}. The KMS condition has been used to constrain two-point functions and extract thermal one-point coefficients in various models, \cite{Iliesiu:2018fao,Iliesiu:2018zlz,Marchetto:2023xap,Barrat:2024aoa,Barrat:2024fwq,Alday:2020eua,Dodelson:2022yvn,Karlsson:2022osn,Huang:2022vet,Dodelson:2023vrw,Esper:2023jeq,Ceplak:2024bja,Buric:2025anb,Barrat:2025nvu}.
\smallskip

In addition to $S^1 \times \mathbb{R}^{d-1}$, other spaces that have been explored include various products of higher-dimensional tori, spheres and flat spaces \cite{Shaghoulian:2015kta,Belin:2016yll,Horowitz:2017ifu,Belin:2018jtf,Luo:2022tqy,Marchetto:2023fcw,Allameh:2024qqp}, and ‘genus two' manifolds, \cite{Benjamin:2023qsc}. In this work, we will focus on the geometry $S^1 \times S^{d-1}$. On the one hand, the kinematics of correlators on this background is more complicated than that of $S^1 \times \mathbb{R}^{d-1}$, making them more difficult to study. However, there are several advantages that the above geometry brings: already one-point functions are not kinematically fixed and therefore present interesting observables; conformal block decomposition of correlators involves the {\it flat space} CFT data rather than infinite averages thereof; angular {\it chemical} potentials are naturally introduced in addition to the temperature and allow one to access CFT data resolved in spin. In theories with holographic duals, two-point functions on $S^1\times S^{d-1}$ also exhibit a new type of singularities, termed the {\it bulk-cone} singularities, \cite{Hubeny:2006yu,Dodelson:2020lal,Dodelson:2023nnr}, that are not present on $S^1\times \mathbb{R}^{d-1}$. Despite the considerable interest in the spherical geometry, there have been very few works on the latter, \cite{Shaghoulian:2016gol,Gobeil:2018fzy,Alkalaev:2024jxh,David:2024pir}.
\smallskip

Thermal correlation functions are interesting for a further reason - their high temperature behaviour controls the CFT data at large scaling dimensions.
Understanding the asymptotic behaviour of CFT data has been the central focus of numerous works, beginning with Cardy's formula for the density of states in 2D CFTs, \cite{Cardy:1986ie}, which follows from the modular invariance of the genus-one partition function. This result was later generalised in \cite{Cardy:2017qhl}, where a formula for the average squared OPE coefficients $\overline{\lambda_{HHH}^2}$ of three heavy Virasoro primaries was derived using the modular invariance of the genus-two partition function. Extensions to cases in which at least one of the operators is light (i.e., heavy-light-light (HLL) and heavy-heavy-light (HHL) configurations) in 2D CFTs were obtained in \cite{Kraus:2016nwo,Collier:2019weq}. By contrast, much less is known in higher dimensions. The constraining power of modular invariance is much reduced in higher dimensions, making it significantly more challenging to derive universal results. Some progress has been made using techniques such as the Lorentzian inversion formula, large spin perturbation theory, and the lightcone bootstrap, \cite{Komargodski:2012ek,Fitzpatrick:2012yx,Caron-Huot:2017vep}, which allow one to access the HLL asymptotics of certain families of OPE coefficients, namely those associated to multi-trace operators of finite twist. More general statements make use of Tauberian theorems applied to four-point functions in flat space and related techniques, \cite{Pappadopulo:2012jk,Qiao:2017xif,Pal:2022vqc,Mukhametzhanov:2018zja} (see also \cite{Karlsson:2019qfi} for related work). In the above context, thermal one-point functions in dimensions greater than two were first considered in \cite{Gobeil:2018fzy} and subsequently in \cite{Buric:2024kxo}.

A novel approach to the above questions, which formalises the intuition that high temperature behaviour of a theory should control the CFT data at large scaling dimensions by the notion of the {\it thermal effective field theory} and applies in any $d$, has recently been proposed, \cite{Benjamin:2023qsc,Benjamin:2024kdg} (see also \cite{Bhattacharyya:2007vs,Banerjee:2012iz,Jensen:2012jh,Kang:2022orq}). The thermal EFT gives a high temperature expansion of the CFT partition function on $S^1 \times S^{d-1}$, which in turn provides the leading asymptotics of the large $\Delta$ density of primaries, as well as its subleading corrections. Exploring the partition function on a genus-two manifold also led to an asymptotic formula for the density of heavy-heavy-heavy (HHH) OPE coefficients. Observables other than the partition function are expected to similarly constrain OPE coefficients when scaling dimensions of some of the operators become large. This reasoning was applied in several related contexts to derive asymptotic behaviour of CFT data, including one-point functions in the presence of a defect, \cite{Diatlyk:2024qpr,Kravchuk:2024qoh}. 
\smallskip

In the present work, we will study the asymptotics of HHL OPE coefficients, with identical heavy operators, which are controlled by one-point functions on $S^1\times S^{d-1}$. These coefficients have appeared in the context of the eigenstate thermalisation hypothesis (ETH) when specified to CFTs, \cite{Lashkari:2016vgj,Delacretaz:2020nit}. Our analysis of thermal one-point functions ties together and proves predictions from these works and extends them to {\it non-dominant tensor structures}, as we shall describe below.

\medskip

\subsection*{Summary of results} Some of the tools we advance in this work have been put forward in previous related work \cite{Buric:2024kxo} involving some of the authors. In particular, in \cite{Buric:2024kxo}, we initiated a systematic study of thermal conformal block expansions for one-point functions of a scalar field $\phi$ with weight $\Delta_\phi$ on $S^1 \times S^2$, 
\begin{equation}\label{block-decomposition_intro}
\text{tr}_\mathcal{H}\left(\phi(x) q^D y^M\right) 
= 
    r^{-\Delta_\phi} \sum_{\mathcal{O},a} \lambda_{\phi\mathcal{OO}}^a\, 
    g_{\Delta,\ell}^{\Delta_\phi,a}(q,y,s)\ .
\end{equation}
Here $D$ and $M$ denote the generators of dilations and rotations around the 
$x^1$-axis, respectively. Such one-point functions depend on three cross ratios - the temperature $T = 1/\beta$, an angular chemical potential $\mu\in[-\pi,\pi]$ and the angle $\theta$ of $x$ with the $x^1$-axis. Throughout most of this work 
we shall use 
\begin{equation}
q = e^{-\beta} \ , \quad y = e^{i\mu} = e^{i \beta \Omega}\ , \quad s = \sin^2 \theta\  \quad\quad \mathrm{and} \quad u = y + \frac{1}{y} -2\ .
\end{equation} 
The dependence on the length $r$ and the second angle on the sphere is constrained
by Ward identities. The blocks $g(q,u,s)$ carry their quantum numbers. These
include the conformal weight $\Delta=\Delta_\mathcal{O}$ and the spin $\ell = 
\ell_\mathcal{O}$ of the primaries that are exchanged along the thermal circle. 
In addition, there appears the label $a=1, \dots, \ell,$ that refers to the choice 
of a tensor structure for the $\phi\mathcal{O}\mathcal{O}$ three-point function. 
\medskip 

\paragraph{One-point blocks.} 
The full characterisation of the blocks $g = g(q,u,s)$ involves two key
ingredients: the set of Casimir differential equations they satisfy, as well as
asymptotic conditions at low temperature. The relevant Casimir equations are
reviewed in eqs.\ (\ref{Casimir-eqn}-\ref{D2-p-variable}). Concerning the low 
temperature behaviour we formulate a new prescription of the \textit{asymptotic conditions} in terms of Jacobi polynomials $P^{\alpha,\beta}_n$, 
\begin{equation} 
\lim_{q \rightarrow 0} q^{-\Delta} g^{\Delta_\phi,a}_{\Delta,\ell}(q,u,s) 
    =c_{\ell ,a}\, u^a 
    P_{\ell -a}^{\left(2a+\frac12,-\frac12\right)}\left(1+\frac{u}{2}\right)\, 
    P_a^{\left(0,-\frac12\right)}(1-2s)\,, \qquad a=0,\dots,\ell \,,
\end{equation} 
where $\Delta_\phi$ denotes the weight of the external scalar field, $(\Delta,\ell)$ are the weight and spin of the exchanged field and $a$ enumerates a basis of tensor structures, as described above. The constant prefactors $c_{\ell,a}$ are given in eq.\ \eqref{eq:cla}. One of the features that this prescription ensures is \textit{orthonormality} of the blocks with respect to the \textit{inner product} we spell out in eqs.~\eqref{inner-product} and \eqref{measure-blocks}. The resulting \textit{inversion formula} 
\begin{equation}\label{inversion-formula-introduction}
   \rho(\Delta,\ell) \overline{\lambda_{\phi\mathcal{O}\mathcal{O}}^a}(\Delta,\ell)
   = 
   \frac{r^{\Delta_\phi}}{2\pi i} \int \Measure\  
   g^{3-\Delta_\phi,a}_{3-\Delta,\ell}(\beta,\mu,\theta)\,  \text{tr}_\mathcal{H}\left(\phi(x) q^D y^M\right)\,,
\end{equation}
extracts averages of OPE coefficients $\lambda^a_{\phi\mathcal{OO}}$ 
from thermal one-point functions of scalar fields on $S^1 \times S^2$. The  definition of the average is given in eq.\ \eqref{averaged-OPE-coefficients}. Formulas for the spectral density $\rho$ and the integration measure $\Measure$
can be found in eqs. \eqref{inverse-character-transform} and \eqref{measure-blocks}, respectively. 
\smallskip

In addition to orthogonality relations, we will study two types of expansion of thermal one-point blocks. The first one, already developed in \cite{Buric:2024kxo}, is the low temperature expansion. Compared to \cite{Buric:2024kxo}, we shall derive and use {\it recursion relations} to speed up the evaluation of conformal blocks as power series in variables $q,u,s$. These results allow one to efficiently decompose thermal partition and one-point functions and to read off low-lying (averaged) OPE coefficients $\lambda^a_{\phi\mathcal{OO}}$. The second expansion of blocks that we study is that in powers of $1/\Delta$. In particular, we provide a simple formula \eqref{zeroth-order-solution} for the leading term of this expansion that dominates the heavy exchange.
\smallskip

\paragraph{Spectral densities.}
While the main focus of this work is on the OPE coefficients, we also obtain new 
results for the spectral density in gapped conformal field theories and for the free scalar field. Recall that the adjective ‘gapped' refers to theories that 
are gapped upon compactification on the thermal circle. For the leading behaviour
of the spectral density in gapped theories we find
\begin{equation}\label{rho-gapped-intro}
   \rho^{\text{gapped}}(\Delta, \ell) =\, \frac{8\pi^{\frac23}f^{\frac53}(2\ell+1)}{\sqrt{3}} 
   e^{-8\pi c_1} \Delta^{-\frac{11}{3}}\,e^{3\pi^{\frac13}f^{\frac13} \Delta^\frac23} 
   \left( 1 - \frac{3\pi^\frac13 f^\frac13}{\Delta^\frac13} + O\left(\Delta^{-\frac23}\right)\right)\,,
\end{equation}
at large $\Delta$ and finite spin $\ell$. The second term in this expansion is 
new compared to the formula for the spectral density in \cite{Benjamin:2023qsc}.
We note that it depends on the same EFT data as the leading term, namely the free 
energy $f$ and the parameter $c_1$. We can also compute higher order corrections
systematically, see e.g.\ formula \eqref{rho-gapped} for the NNL contribution, 
but these involve additional EFT parameters. 

For the bosonic free field theory, the spectral density possesses a very similar asymptotic expansion, see eq.\ \eqref{eq::rho_refined_leading}. But in this case we can push the high temperature expansion of the thermal partition function to very high orders using some techniques from \cite{Zagier}, see Section \ref{sec:PFathighT}. When combined with the methods developed in Section \ref{SS:Asymptotic density of primaries}, this allows us to push the asymptotic 
expansion of the spectral density for the free field to very high orders. Indeed we shall present results for the first seven orders. Using explicit low temperature expansions, we can compute the spectral density up to $\Delta \sim 200$ for any spin $\ell$. At $\Delta = 210$, the spectral density for scalar fields that we find agrees with our asymptotic formula
to very high precision ($0.06 \% $ if we use the asymptotic expansion to order seven). The comparison also shows that the NL correction we displayed in eq.\ \eqref{rho-gapped-intro} (or rather the free field theory 
analogue in eq.~\eqref{eq::rho_refined_leading}) is quite relevant. Indeed, 
at $\Delta = 210$ the leading term from \cite{Benjamin:2023qsc} deviates by 
$ 40 \% $ form the numerical value. Once we add the NL term, the discrepancy 
drops to $5.5 \%$. 
\smallskip

\paragraph{Averaged OPE coefficients.}
With the inversion formula and the large-$\Delta$ expansion in hand, we can derive an asymptotic expansion for the averaged OPE coefficients $\overline{\lambda^a_{\phi\mathcal{OO}}}$, which reads
\begin{equation}\label{OPE_coeff_asymptotics_gapped_intro}
\overline{\lambda_{\phi\mathcal{O}\mathcal{O}}^a}^{\, \text{gapped}}(\Delta,\ell) 
=  b_{0,2a,a}\left(\frac{\Delta}{8\pi f}\right)^{\frac{\Delta_\phi}
{3}}\mathcal{N}_{a,\ell}\,
\Delta^{-2a}\left(1+O(\Delta^{-\frac13})\right)\,,
\end{equation}
at least for $a\leq4$. Once again, $f$ is the free energy density of the CFT, $b_{0,2a,a}$ a dynamical coefficient entering in the high-temperature behaviour 
of the one-point function of $\phi$ and $\mathcal{N}_{a,\ell}$ a normalisation constant, see eqs.~\eqref{1pt-coeff-s-expansion} and \eqref{coefficient_N_a_l}. 
We also compute subleading corrections to eq.\  \eqref{OPE_coeff_asymptotics_gapped_intro}, which are parametrised by higher analogues of the coefficient $b_{0,2a,a}$. An interesting observation that follows directly from eq.~\eqref{OPE_coeff_asymptotics_gapped_intro} is that the OPE coefficients exhibit a hierarchy at large $\Delta$, which is controlled by the tensor structure label $a$. This property of OPE coefficients was numerically observed in \cite{Buric:2024kxo} and its understanding was one of the motivations for the present work. The special case $a=0$ of the formula \eqref{OPE_coeff_asymptotics_gapped_intro} has often appeared in discussions of ETH, \cite{Lashkari:2016vgj,Gobeil:2018fzy,Delacretaz:2020nit}.

Our analytic formulas for the HHL OPE coefficients (more precisely, slight generalisations thereof, see eq.~\eqref{OPE-coefficients-dominant-asymptotics-free}) are finally checked against for free fields, similarly to the checks we
described above for the spectral density. In Section \ref{S:Generalised free theory at finite temperature}, we analyse the one-point function of $\phi^2$ for the free scalar theory on $S^1\times S^2$. On the one hand, we obtain low-temperature expansions of this observable and read off OPE coefficients up to $\Delta=30$. On the other hand, we develop all-order series expansions at high temperature and derive asymptotic CFT data with several subleading orders. The comparison of exact CFT data and asymptotic formulas is shown in Figures \ref{fig:ope_spin2} - \ref{fig:ope_spin6SubdominantWAsymp}, showing excellent agreement, both for the dominant tensor structure $a=0$ and subdominant ones. 
These results once again demonstrate the importance of subleading terms - without the latter the asymptotic formulas only become accurate at very large values of $\Delta$.
\medskip

The paper is organised as follows. In Section \ref{S:Thermal one-point conformal blocks}, we review results of \cite{Buric:2024kxo} regarding kinematics of one-point functions on $S^1\times S^2$ and introduce the basis of three-point tensor structures that is used in the definition of conformal blocks. Orthogonality relations for blocks and the inversion formula for one-point functions are derived. In Section \ref{S:Expansions of Thermal One-point Blocks} we develop small $q$ and large $\Delta$ expansions of one-point blocks. Section \ref{S:Asymptotic CFT data} is dedicated to the derivation of asymptotic CFT data at large $\Delta$. Section \ref{S:Generalised free theory at finite temperature} tests our general results on the example of the free scalar theory. We end in Section \ref{S:Conclusions} with a summary and a discussion of some future directions.

\section{Thermal One-point Conformal Blocks}
\label{S:Thermal one-point conformal blocks}

This section is devoted to thermal conformal blocks for one-point functions of primary fields on $S^1_\beta \times S^2_R$. After reviewing how these appear in the decomposition of one-point functions along with some of their basic properties from \cite{Buric:2024kxo} in the first subsection, we shall discuss the inner product on the space of blocks in the second. The inner product is derived by analysing Casimir differential equations obeyed by the blocks. In the third subsection, we discuss the precise relation between thermal blocks and three-point tensor structures and define a special orthonormal basis for the former. This basis exhibits various nice properties and will be used in the subsequent analysis. The final subsection gives the inversion formula that extracts OPE coefficients from a thermal one-point function.

\subsection{Review: One-point functions on \texorpdfstring{$S^1 \times S^2$}{S1 x S2}}

This introductory section gives the background concerning one-point functions on the geometry $S^1 \times S^2$ and their associated conformal blocks. It serves to establish our conventions and notation. We will mostly follow \cite{Buric:2024kxo}, to which the reader is referred for more details.\footnote{There are slight differences in conventions and notation between \cite{Buric:2024kxo} and the present work.}
\smallskip

The Euclidean conformal group $G=SO(4,1)$ in three dimensions has rank two, i.e. its Lie algebra $\mathfrak{g} = \text{Lie}(G)$ possesses two Cartan elements which we shall denote by $D$ and $M$. Here, $D$ is the generator of dilations and $M$ is the generator of rotations around one of the coordinate axes in 3-dimensional Euclidean space. Given a conformal field theory and primary fields $\phi_1,\dots,\phi_n$, we consider {\it thermal correlation functions}, by which we mean traces of the form
\begin{equation}\label{def:Gn} 
    G_n(x_i,q,y) = \text{tr}_\mathcal{H} \left(\phi_1(x_1)\dots \phi_n(x_n) q^D y^M\right)\ .
\end{equation}
Here, $\mathcal{H}$ is the Hilbert space of the theory, $x_i$ are points in $\mathbb{R}^3$ and $q,y$ are complex variables referred to as chemical potentials. We will sometimes parametrise the latter in terms of other sets of variables $(\beta,\mu)$ or $(\beta,\Omega)$, related to the above by
\begin{equation}\label{q-y-mu-beta-Omega}
    q = e^{-\beta}, \qquad y = e^{i \mu} = e^{i \beta\Omega}\ .
\end{equation}
Here, $\beta$ is related to the temperature via $\beta = T^{-1}$. By using the plane-cylinder 
map, one can think of correlators \eqref{def:Gn} as $n$-point functions on $S^1 \times S^{2}$, 
see \cite{Buric:2024kxo}. However, the definition \eqref{def:Gn} is all that is required for 
the present work. We will mostly study zero- and one-point functions. The zero-point function 
is the partition function
\begin{equation}
  {\mathcal Z(q,y)} = \text{tr}_\mathcal{H} \left( q^D y^M\right) \ .
\end{equation}
It is often convenient to divide the higher correlation functions by the partition function and work with the normalised correlators which we denote by 
 \begin{equation}\label{thermal-correlators}
    \langle \phi_1(x_1)\dots \phi_n(x_n)\rangle_{q,y} = 
    \frac{1}{\mathcal Z(q,y)}\text{tr}_\mathcal{H}\left(\phi_1(x_1)\dots \phi_n(x_n) q^D y^M\right) \ .
\end{equation}
Let us now review the definition and properties of conformal blocks for the case of interest, 
namely one-point functions $(n=1)$ of scalar external operators $\phi_1 = \phi$ of weight 
$\Delta_\phi$. Thermal conformal blocks are defined by inserting a projector $P_\mathcal{O}$ 
to the space of $\mathfrak{so}(4,1)$ descendants of the primary field $\mathcal{O}$ inside 
the trace
\begin{equation}\label{thermal-block-definition}
    G_1(x,q,y)\big|_\mathcal{O} = \text{tr} \left(P_\mathcal{O}\phi(x) q^D y^M \right) 
    \equiv \text{tr}_{\mathcal{H}_\mathcal{O}} \left(\phi(x) q^D y^M\right)\ .
\end{equation}
By convention, the generator $M$ is that of rotations around the $x^1$-axis. To be somewhat more precise, we actually define the thermal blocks by dividing the functions \eqref{thermal-block-definition} by a simple factor that depends only on the scaling weight $\Delta_\phi$ of the external field
\begin{equation} \label{eq:Ggrelation}
    G_1(x,q,y)\big|_\mathcal{O} =  
    r^{-\Delta_\phi} g_{\Delta,\ell}^{\Delta_\phi,a}(q,y,s)\ .
\end{equation}
Here, $(r,\theta,\varphi)$ are the spherical polar coordinates on $\mathbb{R}^3$
\begin{equation}
    x^1 = r \cos{\theta}\,, \qquad x^2 = r \sin{\theta}\cos\varphi\,, \qquad  x^3=
    r \sin{\theta}\sin\varphi\,, 
\end{equation}
and $s=\sin^2\theta$. The lower indices $(\Delta,\ell)$ of the block $g$ are quantum numbers of the primary field $\mathcal{O}$ that we chose when we inserted the projection operator $P_\mathcal{O}$. The first upper index is the scaling dimension of the external field $\phi$. Recall that we assumed $\phi$ to be a scalar so that the spin label $\ell_\phi =0$ is trivial and does not appear in our notation. Finally, the last upper index $a$ labels $\langle\phi\mathcal{OO}\rangle$ tensor structures, see \cite{Buric:2024kxo} for details. We will also elaborate on the tensor structure dependence of blocks below. Finally, conformal Ward identities along with the $r$-dependent prefactor we extracted ensure that the remaining function $g$ does not depend on $r$ or $\varphi$. Hence, the only dependence of $g$ on the insertion point $x$ is through the $\theta$ or $s = \sin^2 \theta$, as our notation suggests.

Putting everything together, the conformal block decomposition of thermal one-point functions takes the form
\begin{equation}\label{block-decomposition}
    \mathcal{Z}(q,y) \langle\phi(x)\rangle_{q,y} = 
    r^{-\Delta_\phi} \sum_{\mathcal{O},a} \lambda_{\phi\mathcal{OO}}^a\, 
    g_{\Delta,\ell}^{\Delta_\phi,a}(q,y,s)\,,
\end{equation}
where $\lambda_{\phi\mathcal{OO}}^a$ are the OPE coefficients. In \cite{Buric:2024kxo}, we obtained low-temperature expansions for thermal conformal blocks, 
that in the case under consideration take the form\footnote{We will give a refined version of the ansatz in \eqref{eq:gqpower}.}
\begin{equation}\label{block-series-expansion}
    g^{\Delta_\phi,a}_{\Delta,\ell}(q,y,s) = q^{\Delta} \sum_{n_i} \tilde A_{n_1 n_2 n_3} q^{n_1} y^{n_2} s^{n_3}\,,
\end{equation}
where $n_i$ are integers with $n_1\geq0$ and
\begin{equation}
    -n_1-\ell\leq n_2 \leq n_1 + \ell\,, \qquad 0\leq n_3 \leq n_1 + \ell\ .
\end{equation}
The coefficients $\tilde A_{n_1 n_2 n_3}$ are determined by solving a Casimir-type equation order by order in $q$. At the leading order, $n_1=0$, the number of linearly independent solutions of the form \eqref{block-series-expansion} is equal to the number of three-point tensor structures $\langle\phi\mathcal{OO}\rangle$. Each of them is then uniquely extended to higher orders in $q$. While the method is completely systematic, due to the large number of coefficients $\tilde A_{n_1 n_2 n_3}$, it proves time-consuming to go to high orders in $q$. This is the one of the problems that we address in the present work by developing an efficient solution method through recursion relations in Section \ref{SS:Recursion relations for thermal blocks}.

\subsection{Thermal blocks: Casimir equations and inner product}
\label{SS:Orthogonality of one-point conformal blocks}

In order to perform the abstract sum over conformal multiplets that appeared in our description of one-point blocks in the previous subsection we follow the standard strategy and characterise blocks through the Casimir differential equations that they satisfy. In the case of scalar external fields that we are interested in here, the second order Casimir equation reads, see \cite{Buric:2024kxo} for a detailed derivation, 
\begin{equation}\label{Casimir-eqn}
    \mathcal{C} _{\Delta_\phi} \, g^{\Delta_\phi,a}_{\Delta,\ell}(q,y,s) = 
    -2\Big(\Delta (\Delta-3) + \ell
    (\ell+1)\Big) g^{\Delta_\phi,a}_{\Delta,\ell}(q,y,s)\,,
\end{equation}
where 
\begin{align}
    &\mathcal{C} _{\Delta_\phi} = -2q^2 \partial_q^2-2y^2\partial_y^2 +\frac{4q^2}{1-q}\partial_q +
    \frac{4y^2}{1-y}\partial_y-2\frac{q+y}{q-y}(q\partial_q-y\partial_y)-2\frac{qy+1}{qy-1}
    (q\partial_q+y\partial_y)\nonumber\\[2pt]
    &+ \frac{q}{(q-1)^2}\mathcal{D}^{(1)}_{\Delta_\phi}(s) + \frac{y}{(y-1)^2}
    \mathcal{D}^{(2)}_{\Delta_\phi}(s) + \left( \frac{qy}{(q-y)^2} + 
    \frac{qy}{(qy-1)^2}\right) \mathcal{D}^{(3)}_{\Delta_\phi}(s)\,, \label{Laplace-Casimir-op}
\end{align}
and the differential operators $\mathcal{D}^{(i)}_{\Delta_\phi}$ contain all derivatives with respect to the 
variable $s$ as well as the dependence on the conformal weight $\Delta_\phi$ of the external scalar. Explicitly, 
they are given by 
\begin{align}\label{D1-p-variable}
    \mathcal{D}^{(1)}_{\Delta_\phi} & = 8s^2(1-s)\partial_s^2 +4\big(1+2\Delta_\phi 
    -2s(\Delta_\phi+1)\big) s\partial_s - 2\Delta_\phi \big(1 + (s-1)\Delta_\phi \big) 
    \, ,\\[2pt]
    \mathcal{D}^{(2)}_{\Delta_\phi} & = 8s(s-1)\partial_s^2 + 4(3s-2)
    \partial_s\,,\quad \mathcal{D}^{(3)}_{\Delta_\phi} = - 
    \frac{\mathcal{D}^{(1)}_{\Delta_\phi} + \mathcal{D}^{(2)}_{\Delta_\phi}}{2} + \Delta_\phi 
    (\Delta_\phi-3)\ .\label{D2-p-variable}
\end{align}
Here we have kept a reference to the scaling weight of the external scalar field $\phi$ in the subscript of these operators. Note that the expressions we displayed describe the action of the quadratic Casimir element on the functions $g$ that are related to the projected correlation function through a $\Delta_\phi$-dependant prefactor, see eq.\ \eqref{block-decomposition}. Let us agree to denote by $\mathcal{H}_{\Delta_\phi}$ the space of functions $g$ on which the 
Casimir operator $\mathcal{C}_{\Delta_\phi}$ acts. 
\smallskip 

The space of thermal conformal blocks carries an inner product which takes the following form 
\begin{equation}\label{inner-product}
    \langle g_1, g_2 \rangle = \int\limits_{\gamma-i\infty}^{\gamma+i\infty} d\beta \int\limits_{-\pi}^{\pi} d\mu  
    \int\limits_0^\pi d\theta \ \Measure (\beta,\mu,\theta)\, g_1(\beta,\mu,\theta)\, \tilde g_2(\beta,\mu,\theta)\ .
\end{equation}
Here, $\gamma$ is an arbitrary positive real number. We have written this inner product in terms of the functions $g \in \mathcal{H}_{\Delta_\phi}$ on which our Casimir operator acts. Note that in the integrand on the right hand side, the function $g_2$ is considered as an element of $\mathcal{H}_{\tilde \Delta_\phi}$ where $\tilde \Delta_\phi = 3-\Delta_\phi$ is the weight of the shadow $\tilde \phi$. This is why we wrote $\tilde g_2$ instead of $g_2$.\footnote{In other words, $g\mapsto\tilde g$ is the map defined by $g^{\Delta_\phi,a}_{\Delta,\ell}\mapsto g^{3-\Delta_\phi,a}_{\Delta,\ell}$ and extended to arbitrary functions $g(q,u,s)$ by linearity.} Up to an overall normalisation, the measure $\Measure$ is uniquely fixed by demanding that the Casimir operator is symmetric with respect to the inner product, 
\begin{equation} 
\langle \mathcal{C}_{\Delta_\phi} g_1,g_2 \rangle = \langle g_1 , \mathcal{C} _{\Delta_\phi} g_2 \rangle\ . 
\end{equation} 
In order to evaluate this condition we need to take into account that the $\sim$-transform 
acts on the Casimir operator as 
\begin{equation} 
\widetilde{(\mathcal{C}_{\Delta_\phi}g)} = \mathcal{C} _{3-\Delta_\phi} \tilde g\ . 
\end{equation} 
It is then easy to check that the measure $\Measure = \Measure(\beta,\mu,\theta) d\beta d\mu d\theta$ must take the following factorised form 
\begin{equation}\label{measure-blocks}
    \Measure(\beta,\mu,\theta) = \left(\frac{8}{\sqrt{\pi}} \sinh\frac{\beta}{2} \sin\frac{\mu}{2} 
    \sinh\frac{\beta+i\mu}{2} \sinh\frac{\beta-i\mu}{2} \right)^2 \frac{\sin\theta}{2}\ 
    .
\end{equation}
A special case that deserves attention appears if $\phi = {\bf 1}$ is the identity field. With this choice, the one-point function $G_1$ coincides with the partition function, is independent of the variable $\theta$ and possesses an expansion into characters $\chi_{\Delta,\ell}(q,y)$ of the 3-dimensional conformal group. The canonical measure $\omega = \omega(\beta,\mu) d\beta d\mu$ coincides with the $\Measure$ up to the $\theta$-dependent factor, i.e. 
\begin{equation}\label{Haar-measure}
    \omega(\beta,\mu) = \left(\frac{8}{\sqrt{\pi}} \sinh\frac{\beta}{2} \sin\frac{\mu}{2} 
    \sinh\frac{\beta+i\mu}{2} \sinh\frac{\beta-i\mu}{2} \right)^2\ .
\end{equation}
Also, the integration contours in $\beta$ and $\mu$ are the same as in our definition for the inner product \eqref{inner-product} for the more general one-point blocks. While there are no closed-form formulas for the one-point blocks, the characters are easy to write down explicitly. For a unitary representation $V_{\Delta,\ell}$ with weight $\Delta$ and spin $\ell$ of the Lorentzian conformal group $SO(3,2)$, the character reads
\begin{equation}\label{characters}
    \chi_{\Delta,\ell}(q,y) = \text{tr}_{V_{\Delta,\ell}}\left(q^D y^M\right) = 
    \frac{q^\Delta y^{-\ell} \left(1 - y^{2\ell+1}\right)}{(1-q)(1-y)(1-qy)(1-qy^{-1})}\ .
\end{equation}
We will slightly abuse the notation and write characters as $\chi_{\Delta,\ell}(q,y)$ or $\chi_{\Delta,\ell}(\beta,\mu)$, depending on which variables are being used. The two sets of variables are related by \eqref{q-y-mu-beta-Omega}. Given these explicit formulas for characters it is not difficult to verify the following orthonormality relations 
\begin{equation}\label{orthogonality-characters}
   \int\limits_{\gamma-i\infty}^{\gamma+i\infty} d\beta \int\limits_{-\pi}^{\pi} d\mu\ \omega(\beta,\mu)\ 
   \chi_{\Delta_1,\ell_1}(\beta,\mu) \chi_{3-\Delta_2,\ell_2}(\beta,\mu) = 
   2\pi i \delta(\Delta_1-\Delta_2)  \delta_{\ell_1,\ell_2}\ . 
\end{equation}
Thermal one-point blocks $g$ should satisfy a similar set of orthogonality relations with respect to the inner product \eqref{inner-product}, i.e. 
\begin{equation}\label{orthogonality_blocks_integrated}
    \int \Measure\ g^{\Delta_\phi,a}_{\Delta_1,\ell_1}(\beta,\mu,\theta)\, g^{3-\Delta_\phi,b}_{3-\Delta_2,\ell_2}(\beta,\mu,\theta) = 2\pi i\, \delta(\Delta_1-\Delta_2)\delta_{\ell_1,\ell_2} \delta_{a,b}\ .
\end{equation}
For \eqref{orthogonality_blocks_integrated} to be true, one needs to work in the appropriate basis of three-point tensor structures defining the labels $a,b$. This basis is specified in the next subsection.

\subsection{Thermal blocks: Boundary conditions and orthogonality}
\label{SS:Blocks and tensor structures}

In the previous subsection we characterised the thermal blocks through the Casimir differential equation \eqref{Casimir-eqn} they satisfy. But this does not fully fix the blocks. Indeed, the equation must be supplemented by some boundary 
conditions at zero temperature, i.e. we need to fix the behaviour of the blocks in the limit $q \to 0$. The behaviour of the block near $q = 0$ is given by a function of the other two variables $y,s$ which depends on the spin of the exchanged field $\mathcal{O}$, as well as the choice of the tensor structure $\langle\phi\mathcal{OO}\rangle$. In order to make this more precise, let us briefly return to our original construction of a block through projection of the trace over all states of the theory 
\begin{small}
\begin{align}
    & \text{tr}_{\mathcal{O}} \left(\phi(x) q^D y^M \right) = 
    \sum_{m=-\ell}^\ell \langle m|\phi(x) q^D y^M |m\rangle 
    + \text{desc} = q^\Delta \sum_{m=-\ell}^\ell y^m \langle m|\phi(x) |m\rangle + \text{desc}\nonumber\\[2mm]
    & = q^\Delta \sum_{m=-\ell}^\ell y^m 
    \sum_{m'=-\ell}^\ell \lim_{x_2\to\infty} 
    \left(\langle\mathcal{O}^m(x_2)\mathcal{O}^{m'}(0)\rangle \langle \mathcal{O}_{m'}
    (x_2)\phi(x)\mathcal{O}_m(0)\rangle \right) + \text{desc}\ . \label{limit-3pt-fn}
\end{align}
\end{small}

Here, we have separated the contribution of states with the lowest scaling weight in the multiplet of the primary $\mathcal{O}$ from the rest of the sum. The former have conformal weight  $\Delta$ and are labelled by the eigenvalue $m = -\ell,\dots,\ell$ of the rotation generator $M$. The remainder of the sum is over descendants. In going to the second line we made use of the state-field correspondence for the bra and the ket states in order to express the leading term in terms of three-point functions or primary multiplets.  
\smallskip

Let $\{\mathbb{T}_3^a\}$ denote the set of independent $\langle\phi \mathcal{OO}\rangle$ tensor structures. We recall that the number of independent parity even tensor structures is $\ell+1$, so we may take $a=0,1,\dots, \ell$. Introduce
\begin{equation}\label{tensor structures definition}
    \tau^a_m (x) = r^{\Delta_\phi}\sum_{m'=-\ell}^\ell \lim_{x_2\to\infty} 
    \left( \langle\mathcal{O}^m(x_2)\mathcal{O}^{m'}(0)\rangle(\mathbb{T}_3^a)_{m'm}(x_2,x,0)\right)\ . 
\end{equation}
With this definition, our expansion of the thermal one-point block takes the form  
\begin{equation}\label{blocks-definition-exp}
     g^{\Delta_\phi,a}_{\Delta,\ell}(q,y,s) = 
      q^\Delta \sum_m y^m \tau^a_{m}(x) + \text{desc} \ .
\end{equation}
From eqs.~\eqref{tensor structures definition} and \eqref{blocks-definition-exp}, it is clear that each of the tensor structures gives rise to a different conformal block. The blocks expand as $q^\Delta$ multiplied by a powers series in $q$, and the choice of the tensor structure fixes the leading term in the series. The standard way to keep track of the possible tensor structures, and hence thermal blocks, passes through the embedding space formalism. Using the notation of \cite{Costa:2011mg}, we can construct all the $\ell+1$ independent parity even tensor structures from the 
building blocks $H_{ij}$ and $V_{i,jk}$ as 
\begin{equation} \label{eq:TS}
   \mathbb{T}_3^\alpha \sim  (H_{12})^\alpha(V_{1,23}V_{2,31})^{\ell-\alpha},\qquad 
   \alpha=0,\dots,\ell\ .
\end{equation}
Note that we have changed the label for tensor structures from $a$ to $\alpha$ in order to emphasise that the basis of structures \eqref{eq:TS} is not the one we shall adopt below. But given the standard basis of tensor structures one can work out the associated polynomials $$h^\alpha_0(y,s) =  \sum_m y^m \tau^\alpha_m(x)\,,$$ that appear as the coefficient of the leading term $q^\Delta$ in the expansion \eqref{blocks-definition-exp}. While we do not have closed formulas, it is not difficult to find these on case-by-case basis. For $\ell=2$, for example, the three polynomials that are 
associated with the tensor structures \eqref{eq:TS} are given by 
\begin{align} 
h^0_0 &=  \frac{3 s^2 (y-1)^4+12 s y (y-1)^2+8 y^2}{y^2} \,,\\
h^1_0 &= \frac{3 s \left(-2 y^4+y^3+2 y^2+y-2\right)-2 y (y (3 y+4)+3)}{y^2} \,, \\
h^2_0 &= \frac{1+y+y^2+y^3+y^4}{y^2} \, .
\end{align}
A code that calculates these polynomials is described in Appendix F of \cite{Buric:2024kxo}, and it is available online. 
\medskip 

For our purposes, however, it is more convenient to use another basis of three-point structures in which the leading terms of conformal blocks exhibit a number of very nice properties. The basis we choose can be inferred from the Laplace-Casimir operator \eqref{Laplace-Casimir-op} after expanding the $q$-dependence of our blocks. We start with a power series which slightly refines the result in 
eq.~\eqref{block-series-expansion},
\begin{align}\label{eq:gqpower}
    g^{\Delta_\phi,a}_{\Delta ,\ell }(q,u,s) & = 
    q^{\Delta }\sum_{n_1=0}^{\infty} q^{n_1}\, f_{n_1}(u,s) \,, \\
    f_{n_1}(u,s)=\sum_{n_2,n_3} A_{(n_1,n_2,n_3)} u^{n_2} s^{n_3}\,, \qquad & \text{with} \qquad  
    0\leq n_2\leq n_1+ \ell  \,, \ \ 0\leq n_3 \leq n_2\,,        
\end{align}
where $u = y + y^{-1} - 2$. By inserting this general ansatz into the Casimir equation one finds the following differential equation for the leading term $f_0 = f_0(u,s)$
\begin{equation}\label{diff_eq_block_leading}
\left(4s(1-s) \, \partial_s^2 + (4-6s) \, \partial_s + u^2 (u+4) \, 
\partial_u^2 + 2u(u+3) \, \partial_u \right)f_0(u, s)
= \ell  (\ell +1) u\, f_0(u, s)\ ,
\end{equation}
which, note, does not depend on the quantum numbers $\Delta_\phi$ and $\Delta$. There are $\ell+1$ linearly independent solutions of the form \eqref{eq:gqpower}, corresponding to even three-point tensor structures. Among many possible choices for the basis of solutions, we pick the one that takes the factorised form
\begin{equation}\label{tensor_structures_basis}
    f_0^{\ell ,a}(u,s)=c_{\ell ,a}\, u^a 
    P_{\ell -a}^{\left(2a+\frac12,-\frac12\right)}\left(1+\frac{u}{2}\right)\, 
    P_a^{\left(0,-\frac12\right)}(1-2s)\,, \qquad a=0,\dots,\ell \,,
\end{equation}
where $P_n^{(\alpha,\beta)}$ are Jacobi polynomials and $c_{\ell ,a}$ is a normalisation constant that will be specified in a moment. The fact that such separable solutions exist can be seen from the form of the differential equation \eqref{diff_eq_block_leading} and one can easily check that eq.\ \eqref{tensor_structures_basis} indeed solves the equation. 

Apart from this nice factorisation property of our leading polynomials, one can also show that the resulting blocks possess the orthonormality property \eqref{orthogonality_blocks_integrated}.\footnote{Let 
us stress that the leading term $f_0^a$ fixes the associated block uniquely. Indeed, all subleading 
contributions $f_n$ can be computed recursively with the help of the Casimir differential equation.} 
In other words, picking this basis makes the conformal blocks \eqref{eq:gqpower} fully
orthogonal with respect to the inner product \eqref{orthogonality_blocks_integrated}, meaning that 
they are orthogonal not only in the quantum numbers $\Delta ,\ell $, but also 
in the tensor structure label $a$. Indeed, denoting
the arguments of Jacobi polynomials by $v=1+u/2$ and $t=1-2s$, we find
\begin{align}
    &\langle g_{\Delta ,\ell }^{\Delta_\phi,a},g_{3-\Delta' ,\ell' }^{\Delta_\phi,b}\rangle  = 
    \int \Measure\,  q^{\Delta -\Delta' +3} \left(f_0^{\ell ,a}(u,s) + O(q)\right) 
    \left(f_0^{\ell' ,b}(u,s) + O(q)\right) \\[2mm]
    & = \frac{(-2)^{a+b}}{4\sqrt{2}\pi} c_{\ell ,a} c_{\ell' ,b} \int\limits_{\gamma-i\infty}^{\gamma+i\infty}
    d\beta\, e^{-\beta(\Delta -\Delta' )} \int\limits_{-1}^1 \frac{dv}{\sqrt{1-v^2}}(1-v)^{a+b+1}
    \int\limits_{-1}^1\, \frac{dt}{\sqrt{1+t}}\nonumber\\[2mm]
    & \hspace{4cm }\Bigg(P_{\ell -a}^{\left(2a+\frac12,-\frac12\right)}\left(v\right)\, 
    P_{\ell' -b}^{\left(2b+\frac12,-\frac12\right)}\left(v\right)\, P_a^{\left(0,-\frac12\right)}(t)\, 
    P_b^{\left(0,-\frac12\right)}(t) + O(q)\Bigg)\nonumber\ .
\end{align}
By pushing the $\beta$-contour to infinity, i.e. taking $\gamma\to\infty$, we can drop the subleading terms in $q$. The remaining integral factorises into three one-dimensional integrals. Using orthogonality relations satisfied by Jacobi polynomials, we obtain 
\begin{align}
    &\langle g_{\Delta,\ell }^{\Delta_\phi,a},g_{3-\Delta',\ell' }^{\Delta_\phi,b}\rangle 
    = i\frac{2^{4a+2}\Gamma(\ell -a+\frac12)\Gamma(\ell +a+\frac32)}{(4a+1)(2\ell +1)(\ell +a)!(\ell -a)!} 
    c_{\ell ,a}^2 \delta\left(\Delta  - \Delta' \right)\delta_{ab} \delta_{\ell \ell' }\ .
\end{align}
Therefore, requiring the normalisation condition \eqref{orthogonality_blocks_integrated} fixes the constant prefactor to be 
\begin{equation} \label{eq:cla}
    c_{\ell ,a}= \,\sqrt{\frac{\pi(2\ell +1)(4a+1)(\ell +a)!
    (\ell -a)!}{2^{4a+1}\,\Gamma(\ell -a+\frac12)\Gamma(\ell +a+
    \frac32)}}\ .
\end{equation}
Before closing this section on boundary conditions and tensor structures, let us make a couple of 
additional remarks concerning the basis \eqref{tensor_structures_basis} that will be useful when 
discussing asymptotics of OPE coefficients later on. For any value of the spin $\ell $, there is 
only one function $f^{\ell ,a}_0$ that is independent of the position space coordinate $s$, namely 
the one with $a=0$. Switching from the $u$ coordinate back to $y$, we see that this function is 
nothing else but the character of $SO(3)$,
\begin{equation}\label{dominant-q-expansion}
    f_0^{\ell ,a=0}(y) = \frac{y^{-{\ell }}
    \left(1-y^{2\ell +1}\right)}{(1-y)} =\chi^{SO(3)}_{\ell }(y) \ .
\end{equation}
Therefore, the $q$-expansion of the associated thermal conformal blocks reads
\begin{equation}
        g^{\Delta_\phi,a=0}_{\Delta ,\ell }(q,y,s) = 
        q^{\Delta } \left(\chi^{SO(3)}_{\ell }(y) + O(q)\right)\ .
\end{equation}
In the following, we will refer to this particular type of block as the \textit{dominant} one, while 
calling all the other ones, corresponding to $a=1,\dots,\ell $, \textit{subdominant}. The 
reason for such terminology will become clear in Sections \ref{S:Expansions of Thermal One-point Blocks} and \ref{S:Asymptotic CFT data}. An important property of the subdominant blocks, which is a direct consequence of the orthogonality of Jacobi
polynomials, is that they are annihilated by integration against the measure \eqref{measure-blocks} in the $s$ (or equivalently $\theta$) space, i.e.\ 
\begin{equation}\label{zero-integral-subdominant}
    \int_0^\pi d\theta\,\Measure(\beta,\mu,\theta)\,f_0^{\ell ,a}(u,\theta)=0\,, 
    \quad \text{for} \quad a=1,\dots,\ell \ .
\end{equation}
It is not difficult to verify this claim using the explicit formulas \eqref{tensor_structures_basis}, \eqref{measure-blocks} for the functions $f^{\ell ,a}_0$ and the measure.

\subsection{Inversion formulas for thermal one-point functions}

Orthogonality relations allow for conformal block decompositions of one-point functions and, as a particular case, the character decomposition of thermal partition functions. Let us first discuss the simpler case of character decompositions, before turning to thermal one-point functions with non-trivial insertions of scalar fields $\phi$. 

To formulate the inversion formula for partition functions we begin with the character decomposition of the latter,
\begin{equation}\label{function-character-decomposition}
    \mathcal{Z}(\beta,\mu) = \int_0^\infty d\Delta \sum_{\ell=0}^\infty \rho(\Delta,\ell)\chi_{\Delta,\ell}(\beta,\mu)\ .
\end{equation}
Here, $\rho(\Delta,\ell)$ denotes the density of primary states of the theory. Using the orthogonality relation \eqref{orthogonality-characters}, we can write the projection of $\mathcal{Z}$ to the representation $(\Delta,\ell)$ as 
\begin{equation}\label{inverse-character-transform}
    \rho(\Delta,\ell) = \frac{1}{2\pi i} \int\limits_{\gamma-i\infty}^{\gamma+i\infty} 
    d\beta \int\limits_{-\pi}^{\pi}d\mu\ \omega(\beta,\mu)\, \chi_{3-\Delta,\ell}(\beta,\mu)\, 
    \mathcal{Z}(\beta,\mu)\ .
\end{equation}
We shall make use of this inversion formula later on. It extracts the character density $\rho$ given a function $\mathcal{Z}$ with the decomposition \eqref{function-character-decomposition}.
\smallskip 

Before we write down the inversion formula for thermal one-point functions, let us rewrite the equation \eqref{block-decomposition} a little bit. In the original formulation, we sum over primaries rather than their weight and spin. In the case the spectrum of primaries has degeneracies, the same block can appear with multiplicity $n(\Delta,\ell)$. We pass from the sum over $\mathcal{O}$ to a sum over $(\Delta,\ell)$ by introducing the following averaged OPE coefficients 
 \begin{equation}\label{averaged-OPE-coefficients}
    \overline{\lambda_{\phi \mathcal{O}\mathcal{O}}^a} = 
    \left( \sum_{i=1}^{n(\Delta,\ell)} 
    \lambda_{\phi \mathcal{O}_i \mathcal{O}_i}^a \right) / n(\Delta,\ell)\ .
\end{equation}
With this notations set up, we can now rewrite the decomposition \eqref{block-decomposition} in the following form 
\begin{equation}\label{1pt-function-decomposition}
   \mathcal{Z}(\beta,\mu) \langle\phi(x)\rangle_{\beta,\mu} = 
   r^{-\Delta_\phi}\int_0^\infty d\Delta \sum_{\ell=0}^\infty \,  \rho(\Delta,\ell)\sum_{a=0}^\ell \overline{\lambda_{\phi\mathcal{O}\mathcal{O}}^a}
   (\Delta,\ell)\,g_{\Delta,\ell}^{\Delta_\phi,a}(\beta,\mu,\theta)\ .
\end{equation}
Now that we have collected all the terms that are multiplied by the same conformal block, 
it is easy to invert the block decomposition \eqref{block-decomposition} or rather its 
reformulation in eq.\ \eqref{1pt-function-decomposition}. Using the orthogonality 
relation \eqref{orthogonality_blocks_integrated} we obtain
\begin{equation}\label{inversion-formula}
   \rho(\Delta,\ell) \overline{\lambda_{\phi\mathcal{O}\mathcal{O}}^a}(\Delta,\ell)
   = 
   \frac{r^{\Delta_\phi}}{2\pi i} \int \Measure\  
   g^{3-\Delta_\phi,a}_{3-\Delta,\ell}(\beta,\mu,\theta)\, \mathcal{Z}(\beta,\mu) 
   \langle\phi(x)\rangle_{\beta,\mu}\ .
\end{equation}
This simple formula will be very useful later on for determining the asymptotic behaviour of the averaged OPE coefficients in the regime where $\Delta$ becomes large. 

\section{Expansions of Thermal One-point Blocks}
\label{S:Expansions of Thermal One-point Blocks}

In the previous section we have fully defined our basis of thermal one-point blocks through the differential 
equation (\ref{Casimir-eqn}-\ref{D2-p-variable}), along with the asymptotic behaviour \eqref{tensor_structures_basis}. 
Now we would like to study important features of these functions. The first subsection addresses the systematic computation of corrections to the leading term given in eq.\ \eqref{tensor_structures_basis}. Such expansions are needed to decompose low-temperature expansions of thermal one-point functions and read off low lying CFT data - we shall use them in Section \ref{S:Generalised free theory at finite temperature} in the example of the free scalar theory. We shall turn the Casimir differential equations into recursion relations that will speed up the necessary construction of blocks compared to \cite{Buric:2024kxo}. In the second subsection we shall then investigate how our blocks behave in the limit of large $\Delta$, developing a systematic $1/\Delta$ expansion. We shall apply this analysis later to deduce properties of HHL OPE coefficients.

\subsection{Recursion relations for thermal blocks}
\label{SS:Recursion relations for thermal blocks}
 
The computation of blocks used in \cite{Buric:2024kxo} turns out to be rather slow when going to high powers of 
the $q$ expansion and high values of $\ell$. To address this problem, we shall now implement recursion 
relations analogous to what was done for ordinary four-point blocks in \cite{Hogervorst:2013sma,Costa:2016xah}. 
This indeed speeds up the computations drastically and it allows us to push the exact decomposition 
of thermal one-point functions to relatively high values of $\Delta$, high enough to reach the asymptotic 
regime that we can treat with analytic methods. 

For illustration, let us consider the case of an internal scalar, $\ell=0$. The first step in implementing 
recursion relations is to bring the right hand side of the Casimir equation \eqref{Casimir-eqn} to the left. 
We then insert the expansion \eqref{eq:gqpower} of the conformal blocks to obtain 
\begin{equation}\label{eq:actionCasimir_scalar}
    \Big(\mathcal{C}_{\Delta_\phi}  
    +2(\Delta (\Delta-3) + \ell
    (\ell+1))\Big)  \left(q^\Delta \sum_{n_1,n_2,n_3}A_{(n_1,n_2,n_3)} f^{(n_1,n_2,n_3)}\right) = 0 \ .
\end{equation}
Here we have introduced the shorthand $f^{(n_i)} = f^{(n_1,n_2,n_3)} \equiv q^{n_1} u^{n_2} s^{n_3}$ for the 
monomials in the three conformal invariants. In order to rewrite this differential equations as a recursion 
relation we must first get rid of the denominators present in the Casimir operator. This can be done by 
multiplying the entire expression \eqref{eq:actionCasimir_scalar} by the factor 
$$  \mathcal{P}(q,u) = (1-q^2)\left(q u-(1-q)^2\right)\ . $$ 
The first term of this product is the denominator of the coefficient of the operator
$\mathcal{D}^{(1)}_{\Delta_\phi}$ in the expression \eqref{Laplace-Casimir-op} for the Casimir operator. 
Similarly, the second factor in $\mathcal{P}$ arises from the coefficient of the operator
$\mathcal{D}^{(3)}_{\Delta_\phi}$ after the change of variables from $y$ to $u$. The coefficient of 
the operator $\mathcal{D}^{(2)}_{\Delta_\phi}$ is simply $1/u$ and hence there is no 
need to remove it from the differential operator. This is why we did not include this factor into the 
function $\mathcal{P}$. Once all the coefficients of the derivatives in the differential operator are 
written as sums of monomials, we can use the following obvious properties of the functions $f^{(n_1,n_2,n_3)}$,   
\begin{align}
    &q f^{(n_i)} = f^{(n_1+1,n_2,n_3)} \,, & u f^{(n_i)} = f^{n_1,n_2+1,n_3} \,,\quad & s f^{(n_i)} = f^{(n_1,n_2,n_3+1)} \,,\nonumber\\[2pt]
    &\partial_q f^{(n_i)} = n_1 f^{(n_1-1,n_2,n_3)} \, ,  &\!\!\!\partial_u f^{(n_i)} = n_2 f^{(n_1,n_2-1,n_3)} \,, \quad & \partial_s f^{(n_i)} = n_3 f^{(n_1,n_2,n_3-1)}. \nonumber
\end{align}
to rewrite the left hand side of eq. \eqref{eq:actionCasimir_scalar} in the form 
\begin{align}
    & q^{-\Delta} \mathcal{P}(q,u) \Big(\mathcal{C} _{\Delta_\phi}  
    +2(\Delta (\Delta-3) + \ell
    (\ell+1))\Big)  \left( q^\Delta \sum_{n_1,n_2,n_3} A_{(n_1,n_2,n_3)} f^{(n_1,n_2,n_3)}\right)=  \nonumber \\[2mm] 
    &\quad =\sum_{n_1,n_2,n_3}A_{(n_1,n_2,n_3)} \Big( 8 n_3^2 f^{(n_1,n_2-1,n_3-1)}-4(n_3-n_2)(1+2n_3+2n_2)f^{(n_1,n_2-1,n_3)}\nonumber\\[2mm]
    & \hspace{2cm} +2(n_2(n_2+1)-\ell(\ell+1)+n_1(n_1+2\Delta-3))f^{(n_1,n_2,n_3)}+\dots \Big)\ .
\end{align}
\begin{figure}[h!]
    \centering
    \begin{minipage}[b]{0.32\textwidth}
        \includegraphics[width=\textwidth]{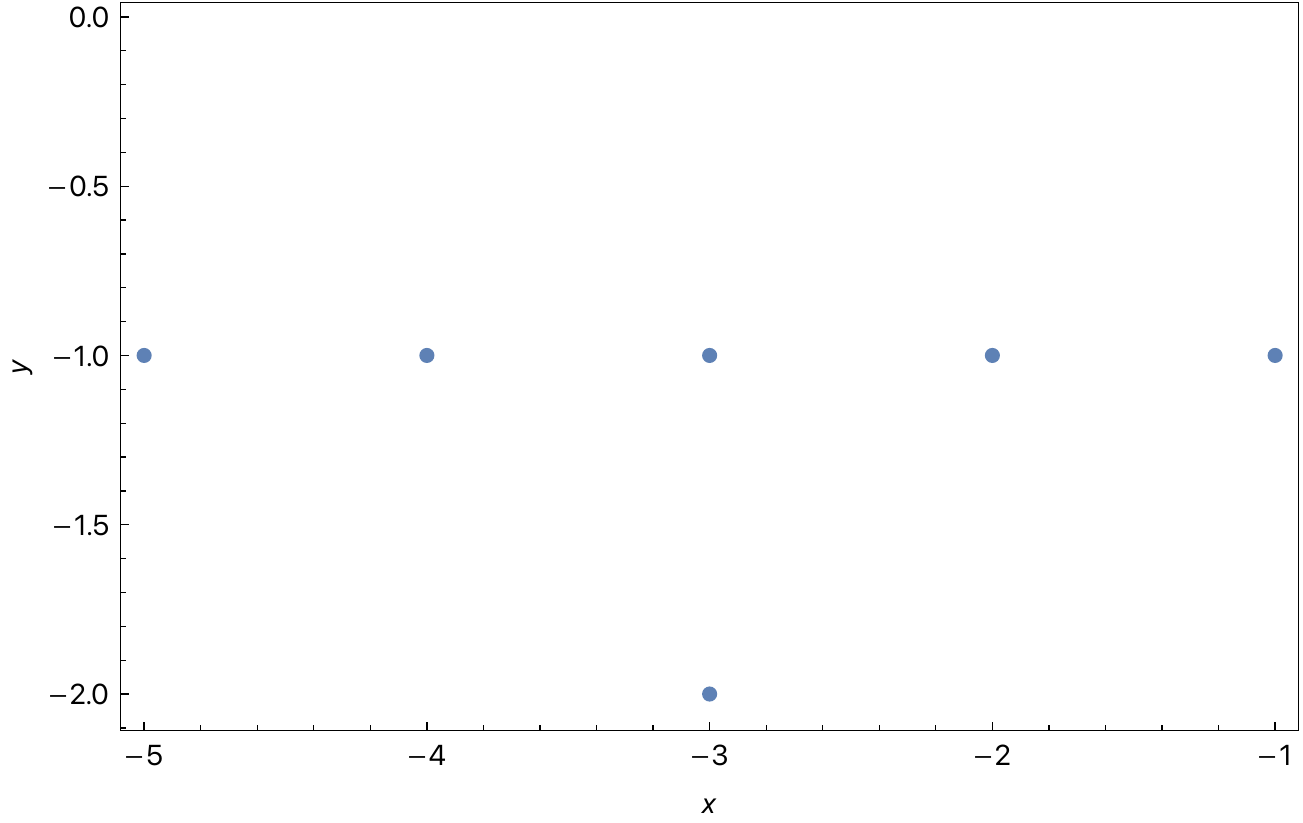}
        \caption*{$n_3 -1$}
    \end{minipage}
    \hfill
    \begin{minipage}[b]{0.32\textwidth}
        \includegraphics[width=\textwidth]{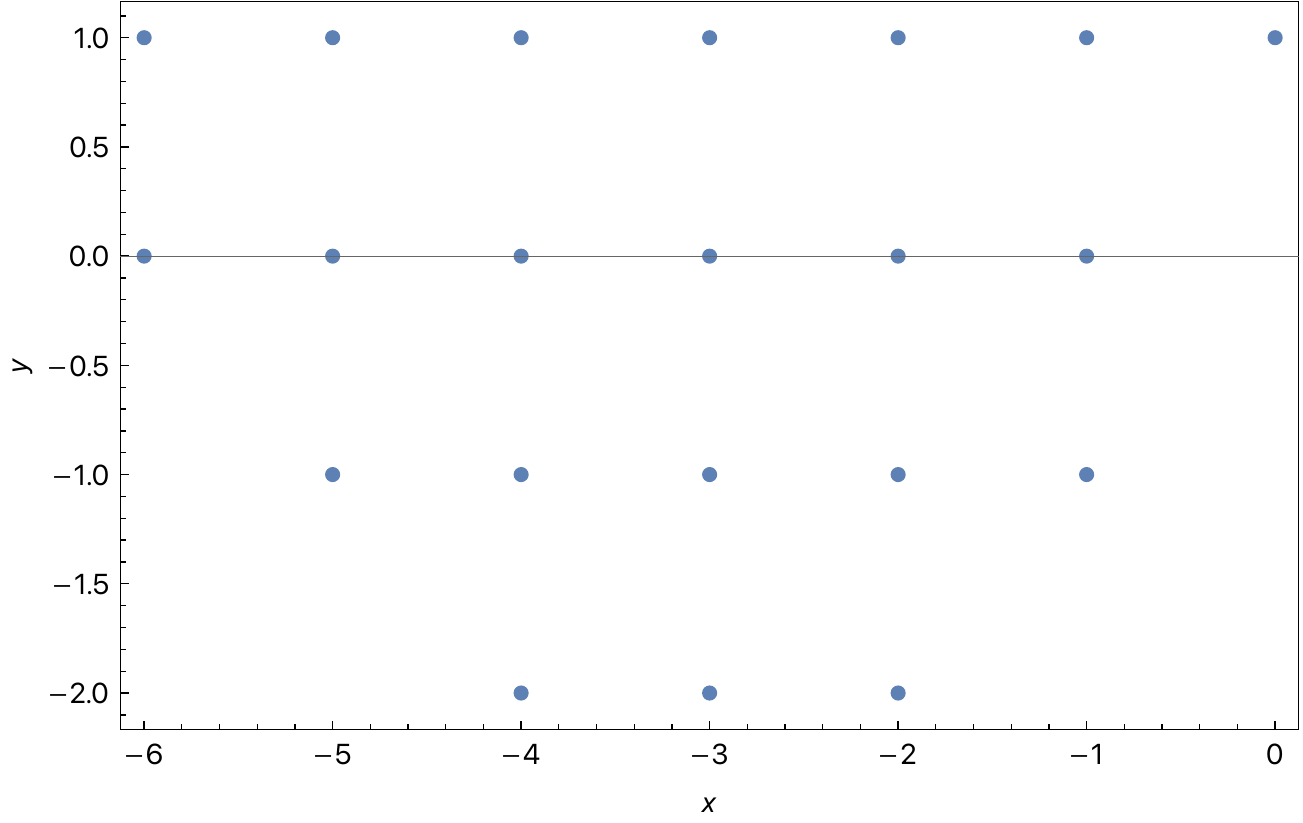}
        \caption*{$n_3  $}
    \end{minipage}
    \hfill
    \begin{minipage}[b]{0.32\textwidth}
        \includegraphics[width=\textwidth]{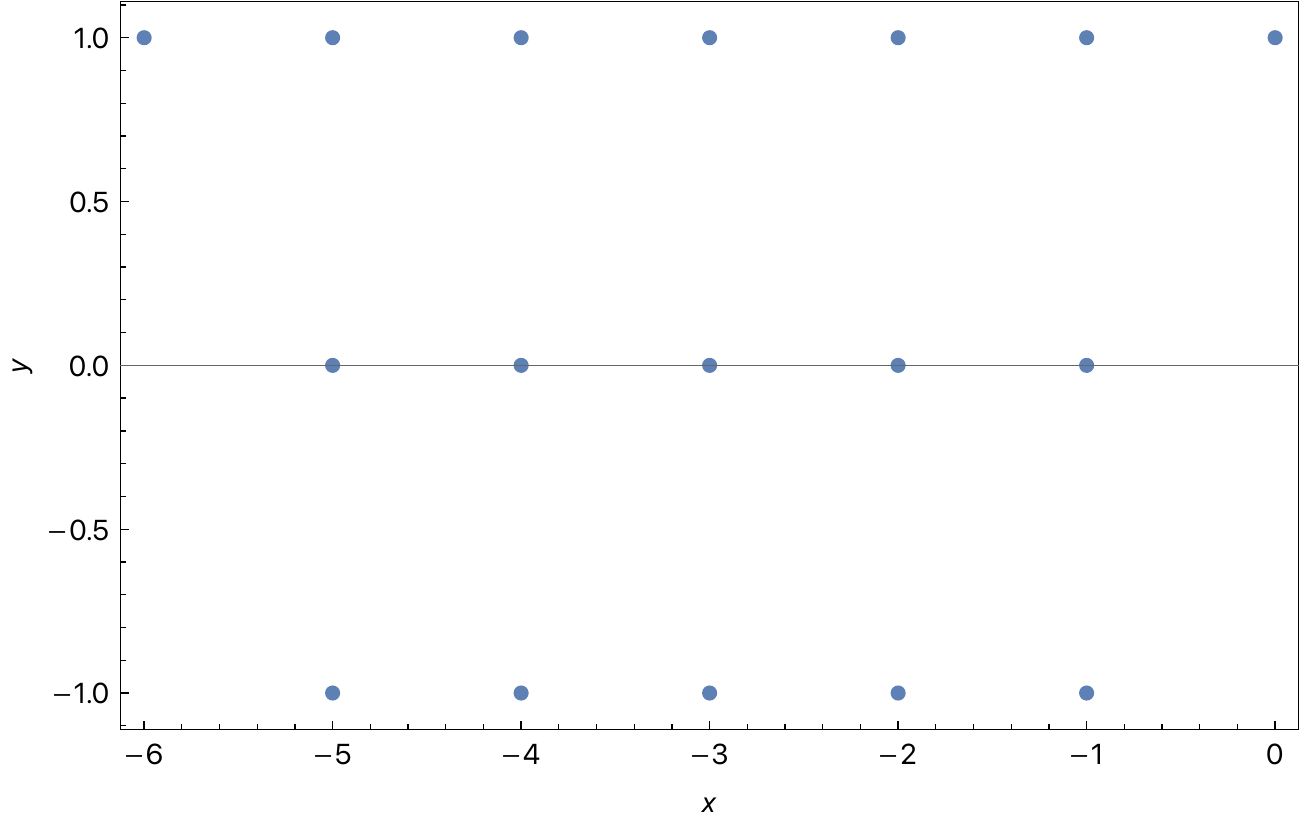}
        \caption*{$n_3 + 1$}
    \end{minipage}
    \caption{Contributions to the recursion relations of the type $A_{(n_1+y,n_2+x, \cdot)}$.}
    \label{fig:recursion_plots}
\end{figure}

We have only displayed a few of the summands that appear. In total there are $44$ terms involving 
shifts of the indices $n_i$ that can go to values as high as 6. By definition, in order for 
$A_{(n_1,n_2,n_3)}$ to be the coefficients of a thermal block, the sum in the previous equation 
has to vanish which implies that 
 \begin{align} \label{eq:recursion} 
   & \hspace{-1cm} A_{(n_1,n_2,n_3)} =  -2\big(18-\ell(\ell+1) +n_2(n_2+1)- 6\Delta+n_1(n_1+2\Delta-9)\big)
   A_{(n_1-6,n_2,n_3)} \nonumber \\[2mm]
    & +4(-n_2+n_3-1)(2n_2+2n_3+3)A_{(n_1-6,n_2+1,n_3)}-8(n_3+1)^2A_{(n_1-6,n_2+1,n_3+1)} \nonumber \\[2mm]
    & + (2n_3-2+\Delta)^2 A_{(n_1-5,n_2-1,n_3-1)} + \dots \,,
\end{align}
see Figure \ref{fig:recursion_plots} for the $A_{(n_1,n_2,n_3)}$ contributing to \eqref{eq:recursion}.
Once again, there are 43 terms on the right hand side of which we merely displayed 4. We have organised the
equations such that all the terms on the right hand side involve coefficients $A$ in which the three indices 
are shifted down by some strictly positive net amount. In the first line, the overall shift in $n_1,n_2,n_3$ 
is by six units, the second line contains terms in which we shift by $6-1 = 5$ and $6-1-1=4$ units, 
respectively, etc.
\smallskip 

\begin{figure}[h!]
    \centering
    \includegraphics[width=0.7\textwidth]{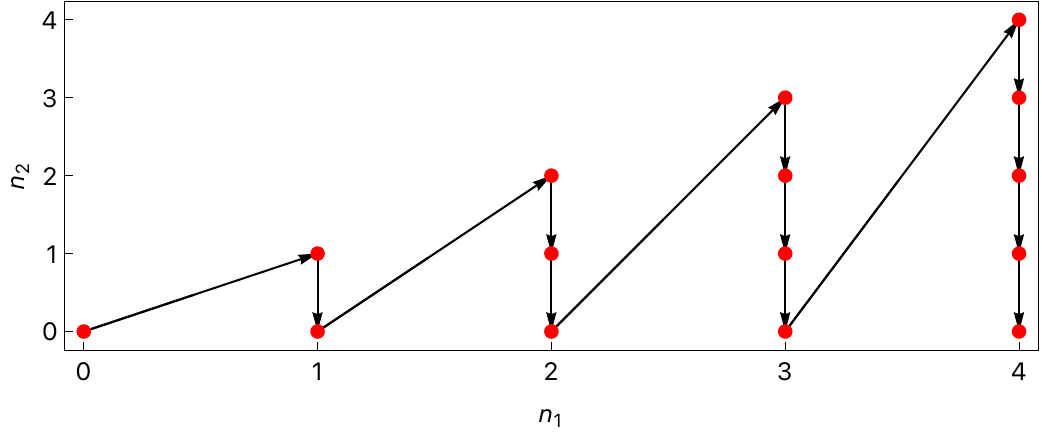}
    \caption{The order in which we proceed to solve the recursion relations. Each red dots stands for 
    a linear recursion in which we determine the $n_3$ dependence at fixed $n_1$ and $n_2$. In the 
    recursion we start with the largest value $n_3 = n_2$ and go down until we reach $n_3=0$.}
    \label{fig:recursion_pattern}
\end{figure}

We need to supplement the previous recursion relation \eqref{eq:recursion} with boundary conditions. 
In particular, we set coefficients $A_{(n_1,n_2,n_3)} = 0$ unless the indices $n_1,n_2,n_3$ satisfy 
the following inequalities 
\begin{equation}
    n_1\geq 0 \,, \qquad 0\leq n_2\leq n_1+ \ell  \,, \qquad 0\leq n_3 \leq n_2\ .
\end{equation}
In addition, we also fix the finite set of non-vanishing coefficients $A_{(0,n_2,n_3)}$ for which 
the first index is $n_1=0$. These coefficients can be read off directly from the leading term on 
the $q$-expansion of the thermal block that we fixed in eq.\ \eqref{tensor_structures_basis}. From 
these boundary conditions we can then reconstruct all the coefficients $A_{(n_1,n_2,n_3)}$ through 
the recursion relation \eqref{eq:recursion}. This requires some prescription for the order in which 
the coefficients are computed. We proceed in the lexicographic order, i.e. given $n_1$ and $n_2$ we 
first lower the index $n_3$ in steps of one from $n_2$ down to zero. Once that 
has been reached, we lower $n_2$ by one unit, keeping $n_1$ fixed and setting $n_3$ back to the maximal value. Then we increase $n_3$ again, one step at a time and so on, see Figure \ref{fig:recursion_pattern} as a graphical example. Starting from $n_1=1$, we can iteratively solve the Casimir equations and obtain the $q$-expansion of the thermal block. The biggest improvement compared to \cite{Buric:2024kxo} lies in the fact that we do not solve a large system of linear equations, instead obtaining individual coefficients one by one. A code which computes the blocks is available at \href{https://gitlab.com/russofrancesco1995/thermal_blocks}{gitlab.com/russofrancesco1995/thermalblocks}.

\subsubsection*{Example} 
As an example, we now compute the first orders of the block $g^{\Delta_\phi}_{\Delta,0}
(q,u,s)$ for which with both internal and external fields are scalar. The boundary condition from 
which we start is only $A_{(0,0,0)}=1$, since there are no tensor structures in this case. Some of the next coefficients to compute from the recursion relations are
\begin{align}
    A_{(1,1,1)} & = \frac{\Delta_\phi^2}{4\Delta} A_{(0,0,0)} = \frac{\Delta_\phi^2}{4\Delta} \, , \\
    A_{(1,1,0)} & = \frac{2\Delta - \Delta_\phi}{2\Delta} A_{(0,0,0)} = \frac{2\Delta - \Delta_\phi^2}{2\Delta} \, , \\
    A_{(1,0,0)} & = \frac{(\Delta_\phi^2 - 3\Delta_\phi + 6\Delta)A_{(0,0,0)} -6 A_{(1,1,0)} - 4 A_{(1,1,1)}}{2(\Delta-1)} = \frac{\Delta_\phi^2 - 3\Delta_\phi + 6\Delta}{2\Delta} \, , \\
    A_{(2,2,2)} & = \frac{(\Delta_\phi+2)^2}{8(\Delta+1)} A_{(1,1,1)} = \frac{\Delta_\phi^2 (\Delta_\phi+2)^2}{32 \Delta (\Delta+1)} \, , \\
    \dots \ .\nonumber
\end{align}
For internal spinning fields, the procedure is analogous, but with starting conditions which are governed 
by eq.~\eqref{tensor_structures_basis}. We conclude this subsection with two short remarks.

\subsubsection*{Remarks} 
\paragraph{1)} In Section \ref{S:Generalised free theory at finite temperature}, we compute blocks 
with generic parameters $(\Delta_\phi,\Delta)$. This is slower that the calculation of blocks for fixed 
values of $(\Delta_\phi,\Delta)$. But it turns out that the recursion with specific values does nor always 
produce a unique block. A simple case where this phenomenon occurs is for the blocks $g^{1,a}_{6,3}(q,u,s)$. 
If one tries to use the recursion relations with the parameters specified as indicated, then $A_{1,1,1}$ 
cannot be computed. The problem is eliminated if we first construct the block for generic parameters and 
then take the limit, i.e. we can set 
\begin{equation}
 g^{1,a}_{6,3}(q,u,s) :=  \lim_{\Delta_\phi \rightarrow 1,\Delta \rightarrow 6} 
 g^{\Delta_\phi,a}_{\Delta,3}(q,u,s)\,,
\end{equation}
which is well-defined. It turns out that the representation as limits of blocks with generic parameters is 
always possible. 
\smallskip

\paragraph{2)} We have seen that powers of $u$ and $s$ in the low temperature expansion of blocks satisfy 
$n_2\geq n_3$, independently of $n_1$. In terms of the variable $\Omega$, our coordinate $u$ reads
\begin{equation}
    u = 2\left(\cos(\beta\Omega)-1\right) = -\beta^2 \Omega^2 + \frac{1}{12} \beta^4\Omega^4 + \dots\ .
\end{equation}
Therefore, when rewritten as functions of the variables $(q,\Omega,s)$ the conformal blocks can be expanded 
as
\begin{equation}\label{Omega-s-powers-blocks}
    g^{\Delta_\phi,a}_{\Delta,\ell}\left(q,\Omega,s\right)  = \sum_{i\geq j\geq 0} \,B_{ij}(q) \,\Omega^{2i} s^j\ 
\end{equation}
with even powers $2i$ of $\Omega$ only and the order $j$ of $s$ being bounded by $j \leq i$.  This 
observation will be useful in the next section.

\subsection{Blocks at large internal dimensions}
\label{SubS:Blocks at large internal dimensions}

We turn to the expansion of thermal conformal blocks in the limit where the dimension $\Delta$ of the exchanged field goes to infinity. The method described in this subsection follows the idea proposed in \cite{Gobeil:2018fzy}. Our starting point is again the Casimir equation \eqref{Casimir-eqn}, with the Laplace-Casimir operator $\mathcal{C}_{\Delta_\phi}$ given in eq.\ \eqref{Laplace-Casimir-op}. It is convenient to redefine 
our functions by extracting a prefactor, 
\begin{equation}\label{g-f-large-Delta}
    g^{\Delta_\phi,a}_{\Delta,\ell}(q,y,s) = 
    q^{\Delta} F^{\Delta_\phi,a}_{\Delta,\ell}(q,y,s)\,.
\end{equation}
This simple change ensures that the resulting differential equation for $F$ is linear rather 
than quadratic in $\Delta$. Moreover, the term linear in $\Delta$ is a first order differential operator with derivatives only with respect to $q$. In more detail, the equation takes the form
\begin{equation}\label{large-DeltaO-structure}
    2\Delta \left(q\partial_q +
    \frac{q\left(3 q^2-2 q (u+3)+u+3\right)}{(q-1)^3-(q-1) q u}\right) F + \mathcal{D} F = 0\,,
\end{equation}
where $\mathcal{D}$ is a differential operator that does not dependent on $\Delta$. We have dropped all the indices on $F$ for notational simplicity. The structure of equation \eqref{large-DeltaO-structure} implies that $F$ can be systematically computed order by order in $1/\Delta$
\begin{equation}\label{large-Delta-expansion}
    F(q,y,s) = \sum_{n=0}^\infty F^{(n)}(q,y,s) \Delta^{-n}\ .
\end{equation}
At each order, $F^{(n)}$ is obtained by a simple integration of $F^{(n-1)}$ using eq. \eqref{large-DeltaO-structure}. Finally, the ambiguity that arises upon performing the integration is fixed by the block at $q=0$.
\smallskip

The leading order function $F^{(0)}$ satisfies the equation \eqref{large-DeltaO-structure} with $\mathcal{D}=0$, whose general solution reads
\begin{equation}
    F^{(0)}(q,u,s) = \frac{c_0(u,s)}{(1-q)^3-(1-q) q u}\ .
\end{equation}
Here, $c(u,s)$ is an arbitrary function. At $q=0$, the result reduces to
\begin{equation}
    F^{(0)}(0,u,s) = c_0(u,s)\ .
\end{equation}
This is to be identified with the leading order function in the low temperature expansion. Each of the linearly independent blocks $g^{\Delta_\phi,a}_{\Delta,\ell}$, gives rise to different leading order function $F^{(0),a}$ via the boundary condition at $q=0$. The zero-th order solution is given by 
\begin{equation}\label{zeroth-order-solution}
    \left(F^{\Delta_\phi,a}_{\Delta,\ell}\right)^{(0)}(q,u,s) = \frac{f^{\ell,a}_0(u,s)}{(1-q)^3-(1-q) q u}\,,
\end{equation}
where the function $f^{\ell,a}_0$ was given in \eqref{tensor_structures_basis}. In particular, for the dominant solution with $a=0$, the result \eqref{dominant-q-expansion} implies that the leading order block $g^{(0)}$ coincides with the conformal character,
\begin{equation}\label{dominant-Delta-expansion}
    \left(g^{\Delta_\phi,a=0}_{\Delta,\ell}\right)^{(0)} = q^\Delta \left(F^{\Delta_\phi,a=0}_{\Delta,\ell}\right)^{(0)}=\chi_{\Delta,\ell}\ .
\end{equation}
Let us illustrate the algorithm for obtaining the subleading terms $F^{(n)}$ on the simplest example of scalar exchange blocks, $\ell=0$. In this case the zero-th order solution reads
\begin{equation*}
    F^{(0)}(q,u,s) = \frac{1}{(1-q)^3-(1-q) q u}\ .
\end{equation*}
If we substitute $F^{(0)}$ into the differential equation and solve for $F^{(1)}$, we obtain the general solution
    \begin{align}
    F^{(1)} & = \frac{\Delta_\phi  \left(2 \Delta_\phi (q-1)^2 + \Delta_\phi  q u (q s+s-2)-2 (q-2) q (u+3)-6\right)}{4 \left((q-1)^3-(q-1) q u\right)^2}\\
    & \hskip6cm +  \frac{4 (q-1) \left(q^2 - q (u+2)+1\right)}{4 \left((q-1)^3-(q-1) q u\right)^2} c_1(u,s)\,, \nonumber
\end{align}
which contains an arbitrary function $c_1(u,s)$. By setting $q=0$ and equating the right hand side to $f^{\ell=0,a=0}_0=1$, we can infer $c_1 = \Delta_\phi(\Delta_\phi-3)/2$. Hence, we have found that 
\begin{equation}\label{first-order-solution}
    F^{(1)} = \frac{\Delta_\phi  q \left(-6 q^2+2 \Delta_\phi  (q-1)^2+\Delta_\phi  u (q (s-2)+s)+4 q (u+3)-2 (u+3)\right)}{4 \left((q-1)^3-(q-1) q u\right)^2}\ .
\end{equation}
One proceeds to higher orders in a similar way.
\medskip

We end this subsection by noting a property of the functions $F^{(n)}$ that will play an important role in the following. Given a tensor structure label $a$ and a general polynomial $p_j(s)$ of order $j$ in the variable $s$, the first $a-j$ terms $F^{(0)},\dots,F^{(a-j)}$ in the expansion of a block have vanishing integral over the $\theta$-part of the measure, when paired with the polynomial $p_j$
\begin{equation}\label{block_hierarchy}
    \int_0^\pi d\theta\,\Measure(\beta,\mu,\theta)\,p_j(s) \left(F^{\Delta_\phi,a}_{\Delta,\ell}\right)^{(n)}(q,u,s) = 0\,, \qquad \text{for} \qquad n<a-j\ .
\end{equation}
Setting $p_j$ to be trivial, $p_j(s)=1$, this statement tells us that, upon integration over $\theta$, blocks with higher $a$ label are more suppressed as $\Delta\to\infty$. This property and its extension \eqref{block_hierarchy} with non-trivial polynomials $p_j$ will be responsible for various asymptotic properties of HHL OPE coefficients $\overline{\lambda^a_{\phi\mathcal{OO}}}$. The simplest instance of \eqref{block_hierarchy}, stating that for all subdominant blocks the integral of the leading function $F^{(0)}$ vanishes,
\begin{equation}
    \int_0^\pi d\theta\,\Measure(\beta,\mu,\theta)\, \left(F^{\Delta_\phi,a}_{\Delta,\ell}\right)^{(0)}(q,u,s) = 0\,, \quad \text{for} \quad a>0\,,
\end{equation}
readily follows from properties \eqref{zero-integral-subdominant} and \eqref{zeroth-order-solution}. We give a simple proof of the more general statement \eqref{block_hierarchy} in Appendix \ref{A:Proof-hierarchy}.

\section{Asymptotic Conformal Field Theory Data}
\label{S:Asymptotic CFT data}

In this section, we derive the asymptotic density of averaged OPE coefficients $\overline{\lambda^a_{\phi\mathcal{OO}}}$ in the limit in which the weight $\Delta$ of the intermediate field is sent to infinity, while keeping its spin $\ell$ fixed. The first subsection is dedicated to the derivation of the asymptotic density $\rho(\Delta,\ell)$ of primary states in the same regime of $(\Delta,\ell)$. While the leading order asymptotics of the density is known, we will also obtain systematically subleading corrections to it. The first correction turns out to depend on the same EFT parameters as the leading one, while all higher order terms depend on additional dynamical coefficients. Our derivation will also serve to illustrate the methods employed in the analysis of OPE coefficients to which we turn in the second subsection. In the latter, we derive an asymptotic formula for the coefficients $\overline{\lambda^a_{\phi\mathcal{OO}}}$, at least for $a\leq4$. Again, both the leading behaviour and the subleading corrections are obtained. The leading asymptotics is controlled by the label $a$, with the ratio of OPE coefficients for two tensor structures $a$ and $b$ going to zero if $a > b$. 

\subsection{Asymptotic density of primaries}
\label{SS:Asymptotic density of primaries}

The high temperature behaviour of the partition function of any CFT constrains the density of primaries in the 
limit of large $\Delta$. In this subsection, we derive the relation between the two. The resulting asymptotic 
density is tested against exact results in free theory in the next section. Our starting point is the 
character decomposition of the partition function of some CFT,
\begin{equation}
    \mathcal{Z}(\beta,\mu) = \int_0^\infty d\Delta \sum_{\ell=0}^\infty \rho(\Delta,\ell)\, 
    \chi_{\Delta,\ell}(\beta,\mu)\ .
\end{equation}
Here, $\rho(\Delta,\ell)$ denotes the density of primary states of the theory. Using the inverse character 
transform \eqref{inverse-character-transform}, we may express the density as an integral over
$\mathcal{Z}(\beta,\mu)$
\begin{equation}\label{density-integral-1}
\rho\left(\Delta, \ell\right)=\frac{1}{2 \pi i} \int\limits_{\gamma-i \infty}^{\gamma+i \infty} d \beta
\int\limits_{-\pi}^{\pi} d \mu\, \omega(\beta, \mu)\, \chi_{3-\Delta, \ell}(\beta, \mu)\, \mathcal{Z}(\beta,\mu)\,,
\end{equation}
where the measure $\omega(\beta,\mu)$ is given in eq.~\eqref{Haar-measure}. To describe the high temperature behaviour of the partition function, we will write the latter as a product 
\begin{equation}\label{general-CFT-leading-Z}
    \mathcal{Z}(\beta,\Omega) = A(\beta,\Omega)\, e^{\frac{4\pi f}{\beta^2 (1+\Omega^2)}}\,,
\end{equation}
with the variable $\Omega$ that was introduced in eq.~\eqref{q-y-mu-beta-Omega}. Here we have split the partition function into a leading universal exponential, \cite{Bhattacharyya:2007vs}, and a factor $A$ that accounts for the subleading contributions. More precisely, as we send $\beta$ to zero, the prefactor $A$ behaves as 
\begin{equation}\label{exponent-definition-limit}
 \lim_{\beta \rightarrow 0} \beta^2 \log A(\beta,\Omega) =  \lim_{\beta \rightarrow 0}  
 \left(\beta^2 \log \mathcal{Z}(\beta,\Omega) 
 -  \frac{4\pi f}{1+\Omega^2} \right)  =  0\ .
\end{equation}
The limit is taken with constant $\Omega$. In theories that are gapped upon compactification on
the thermal circle $S^1_\beta$, corrections to eq.~\eqref{exponent-definition-limit} take the 
form \cite{Benjamin:2023qsc},
\begin{equation}\label{gapped-theories}
    \log A^\textrm{gapped}(\beta,\Omega) = 
    -8\pi c_1 - \frac{32\pi c_2\, \Omega^2}{3\left(1+\Omega^2\right)} + \beta^2 \left(4\pi c_3 + O(\Omega^2)\right) + o\left(\beta^2\right)\ .
\end{equation}
Here, $c_1$ and $c_2$ are certain theory dependent constants that can be captured by a {\it thermal EFT}, see \cite{Benjamin:2023qsc} for more details. We have also kept an additional parameter $c_3$ in the next term. From eq.\ \eqref{gapped-theories}, one reads off the expansion of $A^\textrm{gapped}(\beta,\Omega)$ around $\beta=0$. \footnote{For the analysis that follows, the existence of the thermal EFT is not essential - as long as we assume that $\log A(\beta,\Omega)$ expands in non-negative powers of $\beta$ and $\Omega^2$ (or $\beta$ and $\Omega$), the arguments go through in essentially the same form. The only change is the number of dynamical coefficients entering at various orders of $\Delta^{-1/3}$ in the resulting density.}
\smallskip

If the theory is not gapped after compactification to the thermal circle, the expansion 
\eqref{gapped-theories} is no longer valid. An example of this kind that will play a prominent role 
in the present work is that of the free scalar. In this case, the factor $A(\beta,\Omega)$ reads
\begin{align}\label{free-theory-A}
    \log &A^\textrm{free}(\beta,\Omega)=\\[2pt]
    &= -\frac{\log\beta(1+2\Omega^2)}{12(1+\Omega^2)}-\left(\zeta'(-1)+\frac{\log2}{12}-\frac{5+6\gamma+6\log4}{72}\Omega^2 + O(\Omega^4)\right)+O(\beta^2)\,, \nonumber
\end{align}
where $\gamma$ is Euler's constant. We provide a derivation of \eqref{free-theory-A} in the next section.
\smallskip 

Having reviewed the form of the partition function in the high temperature limit, we can now begin to evaluate the spectral density. Substituting eq. \eqref{general-CFT-leading-Z} into the formula \eqref{density-integral-1} for the density of states, we obtain
\begin{align}\label{definition-g}
        \rho\left(\Delta, \ell\right)& = \frac{1}{2\pi i} \int\limits_{\gamma-i \infty}^{\gamma+i \infty}
        d \beta \int\limits_{-\pi}^{\pi} d\mu\, \omega(\beta, \mu)\, \chi_{3-\Delta, \ell}(\beta, \mu) 
        A(\beta,\mu)\, e^{\frac{4\pi f}{\beta^2+\mu^2}}\\[2pt]
        &\hskip2cm=-\frac{1}{4\pi} \int \hspace*{-5mm}\int\limits_{\gamma-i \infty}^{\gamma+i \infty} d \beta_L d \beta_R\, g(\beta_L,\beta_R) e^{\frac{\beta_L+\beta_R}{2}\Delta+\frac{4\pi f}{\beta_L\beta_R}} + \text{non-pert}\ . \nonumber
\end{align}
In the second line, we have used the new coordinates $\beta_L, \beta_R$ which are related to $\beta$ and 
$\mu$ as
\begin{equation}\label{betaLR-variables}
\beta_L=\beta+i \mu, \quad \beta_R=\beta-i \mu \qquad , \qquad \beta =\frac{\beta_L+\beta_R}{2}, 
\quad \mu =\frac{\beta_L-\beta_R}{2i}\ .
\end{equation} 
Notice that the integral over $\beta_L$ and $\beta_R$ can been extended from its initial range to the whole line $(\gamma-i\infty,\gamma+i\infty)$ at the expense of introducing the term ‘non-pert' in the second line of \eqref{definition-g}. As shall be seen below, see \eqref{chage-xL-xR} and \eqref{density-in-terms-of-h}, at large $\Delta$ the ‘non-pert' term is suppressed by the factor $e^{-\Delta^{4/3}}$ compared to the leading one. It gives nonperturbative corrections to the density of states that are not considered in this work. In \eqref{definition-g} we have also introduced the function $g$ that collects contributions from the measure $\omega$, the characters $\chi$ and the partition function $\mathcal{Z}$, 
\begin{equation}
g(\beta_L,\beta_R) = \omega(\beta_L,\beta_R) \cdot e^{-\frac{\beta_L+\beta_R}{2}\Delta}
\mathcal{\chi}_{3-\Delta,\ell} (\beta_L,\beta_R) \cdot A(\beta_L,\beta_R)\ .
\end{equation} 
Let us note that the function $g$ does not depend on the weight of the intermediate field since neither $\omega$ nor $A$ do and the prefactor of $\chi$ extracts all $\Delta$ dependence from
the character, see eq.\ \eqref{characters}. The limiting behaviour of the integral \eqref{definition-g} 
can be studied using the saddle point analysis. Here, we shall follow a slightly different route in 
which we first perform another change of variables
\begin{equation}\label{chage-xL-xR}
    \beta_L=\left(\frac{8\pi f}{\Delta}\right)^{\frac13}\left(1+ ix_L\Delta^{-\frac13}\right),
    \qquad \beta_R=\left(\frac{8\pi f}{\Delta}\right)^{\frac13}\left(1+ix_R\Delta^{-\frac13}\right)\ .
\end{equation}
In the new coordinates $x_{L,R}$, the integral expression for the density of states becomes
\begin{align}\label{density-in-terms-of-h}
     &\rho(\Delta,\ell) = \\
     & = \frac{(8\pi f)^{\frac23}}{4\pi\Delta^\frac43}
     \int\limits_{-\infty}^{+\infty} dx_L dx_R  \, h\left(x_L,x_R;\Delta\right) \,e^{3(f\pi)^\frac13\Delta^\frac23-(f\pi)^\frac13\left(x_L^2+x_Lx_R+x_R^2\right)}\left(1+O\left(\text{erfc}(\Delta^{\frac23})\right)\right)\ . \nonumber
\end{align}
The term involving the error function results from extending the integral over $x_L$ and $x_R$ to the whole real line. It describes the non-perturbative corrections mentioned in eq. \eqref{definition-g}, and, from now on, will be omitted from our formulas. We have introduced a new function $h$ by substituting the arguments $\beta_L$ and $\beta_R$ of $g$ for the new variables $x_L$ and $x_R$ and further absorbing a part of the exponential factor in \eqref{definition-g}. Since the coordinate transformation between the two sets of variables depends on the weight $\Delta$, the resulting function $h$ does as well, even though $g(\beta_L,\beta_R)$ did not. The function $h$ admits an expansion of the form 
\begin{equation}
    h\left(x_L,x_R;\Delta\right) = \sum_{\alpha\in \mathcal{A}} h_\alpha\left(x_L,x_R\right) \Delta^{-\alpha}\,,
\end{equation}
with the set $\mathcal{A}$ of exponents $\alpha$ bounded from below. To determine the precise exponents that appear, we need to analyse all three factors that contribute to the definition of $g$. We shall consider them one by one. When rewritten in terms of the coordinates $x_{L,R}$, the measure \eqref{Haar-measure} takes the form
\begin{equation}\label{omega-xL-xR}
    \omega(x_L,x_R) = 16 \pi^{\frac53} f^{\frac83} \left(x_L - x_R\right)^2
    \Delta^{-\frac{10}{3}} \left(1 + \frac{3i(x_L + x_R)}{\Delta^{\frac13}} + O(\Delta^{-\frac23})\right)\ .
\end{equation}
The conformal characters \eqref{characters}, on the other hand, are given by
\begin{equation}\label{char-xL-xR}
    q^{\Delta-3}\chi_{3-\Delta,\ell}(x_L,x_R) = \frac{(2\ell+1)\Delta}{8\pi f} 
    \left( 1 + \frac{6\pi^\frac13 f^\frac13 - 3i(x_L + x_R)}{2\Delta^{\frac13}} + O(\Delta^{-\frac23})\right)\ .
\end{equation}
In discussing the contributions from the partition function, we need to distinguish between the two 
cases we discussed above. Let us start with the gapped theory. In this case one has 
\begin{align}\label{Zgapped-xL-xR}
    \mathcal{Z}^\textrm{gapped}(x_L,x_R)  = 
    \ & e^{3\pi^\frac13 f^\frac13 \Delta^\frac23 - 8\pi c_1 - \pi^\frac13 
    f^\frac13 (x_L^2 + x_L x_R + x_R^2)} \\[2mm]
    & \left(1 + i \pi^\frac13 f^\frac13 (6i + x_L^3 + x_L^2 x_R + x_L x_R^2 + x_R^3) \Delta^{-\frac13} + O\left(\Delta^{-\frac23}\right) \right)\ . \nonumber
\end{align}
Putting these elements together and exchanging the order of integration and $\Delta$-expansion in 
eq.~\eqref{density-in-terms-of-h}, we arrive at the expansion for the density at large $\Delta$
\begin{align}\label{rho-gapped}
   \rho^{\text{gapped}}(\Delta, \ell) =\, & \frac{8\pi^{\frac23}f^{\frac53}(2\ell+1)}{\sqrt{3}} 
   e^{-8\pi c_1} \Delta^{-\frac{11}{3}}\,e^{3\pi^{\frac13}f^{\frac13} \Delta^\frac23}\\[2mm]
   & \left( 1 - \frac{3\pi^\frac13 f^\frac13}{\Delta^\frac13} + \frac{\pi^\frac23 f^\frac23 (16\pi c_3 +5) - \pi^{-\frac13}f^{-\frac13}(16\pi c_2 + \frac{35}{9})}{\Delta^\frac23} + O\left(\Delta^{-1}\right)\right)\ . \nonumber
\end{align}
This recovers the result of \cite{Benjamin:2023qsc} and extends it beyond the leading order at large $\Delta$. It is a simple matter to extend the expansion to higher orders in $\Delta^{-1/3}$. All these higher order terms, however, depend on additional thermal EFT coefficients and the number of such non-universal coefficients increases with increasing order of our expansion. 
\smallskip

It now remains to discuss the second case, namely the free scalar theory. The analysis is similar to the above, 
except for the factor $A$ is now given by eq.~\eqref{free-theory-A}. We only write the final result for the 
density 
\begin{equation}\label{eq::rho_refined_leading}
   \rho^{\text{free}}(\Delta,\ell) = \frac{2^\frac{17}{6}\pi^{\frac{23}{36}}f^{\frac{59}{36}}e^{-\zeta^\prime(-1)}(2\ell+1)}{\sqrt{3}}\Delta^{-\frac{131}{36}}\,e^{3\pi^{\frac13}f^{\frac13}\Delta^\frac23}\left( 1 - \frac{3\pi^\frac13 f^\frac13}{\Delta^\frac13} + O\left(\Delta^{-\frac23}\log\Delta\right)\right)\,,
\end{equation}
where $f$ is given by eq. \eqref{free-energy-free}. We see that the ratio of the spectral density in free and gapped theories behaves as $\Delta^{1/36}$ and hence goes to infinity as we make $\Delta$ large, though with a rather small exponent. In the free theory, there are no undetermined dynamical coefficients and we can extend the expansion \eqref{eq::rho_refined_leading} to arbitrarily high orders in principle and very high orders in practice. This is explained in Section \ref{S:Generalised free theory at finite temperature} and a number of subleading corrections to \eqref{eq::rho_refined_leading} are collected in Appendix \ref{AAA:Free-spectrum-asymptotics}.

\subsection{Asymptotic OPE coefficients}
\label{S:Asymptotic OPE coefficients}

We now turn to the analysis of the asymptotic OPE coefficients $\lambda_{\phi\mathcal{O}\mathcal{O}}^a$. More precisely, what we consider is the average of OPE coefficients \eqref{averaged-OPE-coefficients}. The analysis will be very similar to the one we performed in the previous subsection when we analysed the spectral density. Starting from the inversion formula \eqref{inversion-formula}, we obtain 
\begin{align}\label{eq::OPE_int_rep}
   &\overline{\lambda_{\phi\mathcal{OO}}^a}(\Delta,\ell)=
   \frac{1}{\rho(\Delta,\ell)}\frac{r^{\Delta_\phi}}{2 \pi i}\\
   & \hskip3cm \int\limits_{\gamma-i \infty}^{\gamma+i \infty} d \beta \int\limits_{-\pi}^{\pi} d \mu\int\limits_0^\pi d\theta\, \Measure(\beta, \mu,\theta)\, g_{3-\Delta,\ell}^{3-\Delta_\phi,a}(\beta,\mu,\theta)\,\mathcal{Z}(\beta,\mu)\langle\phi(x)\rangle_{\beta,\mu}\ . \nonumber
\end{align}
In the first step, we insert the change of variables \eqref{chage-xL-xR} and expand the integrand of eq.~\eqref{eq::OPE_int_rep} in powers of $\Delta$. In order to do so, we need to analyse the asymptotic behaviour of the four factors one by one. The measure and the partition function have already been discussed 
in the previous subsection, see eqs.~\eqref{omega-xL-xR} and \eqref{Zgapped-xL-xR} for the case of gapped 
theories\footnote{To be precise, note that the measure here contains an additional factor of $\sin\theta/2$ with respect to the one used in the previous section.}. Therefore, it remains to analyse the blocks $g^{3-\Delta_\phi,a}_{3-\Delta,\ell}$ and the one-point function $\langle\phi(x)\rangle_{\beta,\mu}$. 
\smallskip

For the blocks, we shall make use of the large $\Delta$ expansion obtained in Section \ref{SubS:Blocks at large internal dimensions}. In variables $\beta,\mu$, the expansion reads
\begin{equation}\label{shadow_block_delta_expansion1}
     q^{\Delta-3} g^{3-\Delta_\phi,a}_{3-\Delta,\ell}(\beta,\mu,s) = \sum_{n=0}^{\infty} (3-\Delta)^{-n} \left(F^{3-\Delta_\phi,a}_{3-\Delta,\ell}\right)^{(n)}(\beta,\mu,s)\,,
\end{equation}
where all the dependence on $\Delta$ is encoded in the powers $(3-\Delta)^{-n}$. Since the change of variables \eqref{chage-xL-xR} involves $\Delta$, individual summands in the expansion~\eqref{shadow_block_delta_expansion1} have non-trivial large $\Delta$ expansions when expressed in coordinates $x_L,x_R$; inspecting these terms for the cases $a\leq4$, $n\leq7$, we observe the structure
\begin{equation}\label{shadow_block_delta_expansion2}
    (3-\Delta)^{-n} \left(F^{3-\Delta_\phi,a}_{3-\Delta,\ell}\right)^{(n)} (x_L,x_R,s)= 
    \Delta^{1 - \frac43 a - \frac23 n+\frac23\min\left(\left\lfloor{\frac{n}{3}}\right\rfloor,a\right)}\sum_{k=0}^{\infty}f_{a,\ell,\Delta_\phi}^{(n,k)}(x_L,x_R,s)\Delta^{-\frac{k}{3}}\ .
\end{equation}
The functions $f_{a,\ell,\Delta_\phi}^{(n,k)}(x_L,x_R,s)$ are straightforward to compute, although rather cumbersome to write down. Here we spell out only the very leading term with $n = 0 = k$, 
\begin{align}\label{f_0_0}
   &f_{a,\ell,\Delta_\phi}^{(0,0)}(x_L,x_R,s)=\\[2pt]
   & =(-1)^a\frac{(f\pi)^{\frac23a-1}}{2^{4(a+1)}}\sqrt{\frac{\pi(4a+1)(2\ell+1)
   \Gamma(2(1+\ell+a))}{\Gamma(\frac32+2a)^2\Gamma(1+2(\ell-a))}}P_a^{(0,-\frac12)}(1-2s)(x_L-x_R)^{2a}\ . 
   \nonumber
\end{align}
We shall now turn to the discussion of the one-point function, focusing on gapped theories to begin with. 
The high-temperature behaviour of one-point functions is controlled by the limit to the geometry $S^1 
\times \mathbb{R}^2$. At vanishing chemical potential, this takes the simple form\footnote{For simplicity, we put $r=1$ in equation \eqref{leading-1pt-general} and all subsequent formulas of this subsection.}
\begin{equation}\label{leading-1pt-general}
    \langle\phi(x)\rangle^\textrm{gapped}_{\beta,0} = \frac{b_\phi}{\beta^{\Delta_\phi}} + 
    o\left(\beta^{-\Delta_\phi}\right)\ .
\end{equation}
Let us stress once again that the case of the free theory does not follow \eqref{leading-1pt-general} 
and it will be considered separately below, see eq.~\eqref{free-one-pt-asymptotic-expansion}. At 
non-zero $\Omega$, we will assume that the one-point function expands as
\begin{align}\label{1pt-expansion-ansatz}
\langle\phi(x)\rangle^\textrm{gapped}_{\beta,\Omega} & = \frac{1}{\beta^{\Delta_\phi}} 
    \left( b_{0,0}(s) + b_{0,2}(s) \Omega^2 + b_{0,4}(s) \Omega^4 + O(\Omega^6)\right)\\[2mm] 
    & \qquad + \frac{1}{\beta^{\Delta_\phi - 1}} \left(b_{1,0}(s) + b_{1,2}(s) \Omega^2 + 
    O(\Omega^4)\right) + O\left(\beta^{-\Delta_\phi+2}\right)\ . \nonumber
\end{align}
This expansion involving only even powers of $\Omega$ is motivated by the fact that the one-point 
function must be expandable into conformal blocks. Since the latter only involve even powers of 
$\Omega$ when expanded in the variables $q,\Omega,s$, see eq.~\eqref{Omega-s-powers-blocks}, the 
same must hold for the one-point function. For the very same reason we also assume that the 
coefficients $b_{i,2j}(s)$ are polynomials in the variable $s$ of order $j$,
\begin{equation}\label{1pt-coeff-s-expansion}
    b_{i,2j}(s) = \sum_{k=0}^j b_{i,2j,k}\, s^k\ .
\end{equation}
and we identify $b_{0,0}(s)=b_{0,0,0}=b_\phi$. In coordinates $x_L,x_R$, the expression 
\eqref{1pt-expansion-ansatz} becomes
\begin{equation} \label{eq:gapped1ptfunction-expansion}
    \langle\phi(x)\rangle^\textrm{gapped}_{\beta,\Omega} = \left(\frac{\Delta}{8\pi f}\right)^{\frac{\Delta_\phi}{3}}
    \left(b_{0,0,0}+ \frac{2\pi^\frac13 f^\frac13 b_{1,0,0} -\frac{i}{2}\Delta_\phi(x_L+x_R)b_{0,0,0}}{\Delta^\frac13} + 
    O\left(\Delta^{-\frac23}\right)\right)\ .
\end{equation}
Formulas \eqref{shadow_block_delta_expansion2} and \eqref{eq:gapped1ptfunction-expansion} are the two ingredients we were missing in order to analyse the asymptotics of the average OPE coefficients \eqref{eq::OPE_int_rep}. 
\smallskip 

Putting everything together, and using for the density of primaries the result \eqref{rho-gapped} found in the previous subsection, one obtains the large-$\Delta$ expansion of OPE coefficients by integrating over $x_L,x_R$ and $s$. Let us note an important difference that occurs in this step compared to the analysis of the density of primaries. For the latter, the integration is performed 
over $x_L,x_R$ only, all integrals are Gaussian-like and typically non-zero. In the case of one-point 
functions, on the other hand, many terms of the $\Delta$-expansion in eq.~\eqref{eq::OPE_int_rep} 
(expressed in variables $(x_L,x_R,s)$) have vanishing integrals over $s$. Vanishing of some of 
these terms can be accounted for by the property \eqref{block_hierarchy} of conformal blocks, 
combined with the form \eqref{1pt-coeff-s-expansion} of the one-point function. The full analysis, 
and in particular the identities of conformal blocks that result in vanishing of various terms upon 
integration, is given in Appendix \ref{A:Details on the computations of OPE asymptotics}. 
In the end, the leading behaviour of OPE coefficients is not simply obtained by piecing together 
the leading contribution of the different terms that enter into eq.~\eqref{eq::OPE_int_rep}; rather, 
in most cases, it is the subleading corrections that govern the overall scaling. The complexity of 
the analysis increases as the tensor structure index $a$ increases. We carried out the computations 
for $a\leq 4$, and observed that all these cases are described by the following formula
\begin{equation}\label{OPE_coeff_asymptotics_gapped}
\overline{\lambda_{\phi\mathcal{O}\mathcal{O}}^a}^{\, \text{gapped}}(\Delta,\ell) =  
b_{0,2a,a}\left(\frac{\Delta}{8\pi f}\right)^{\frac{\Delta_\phi}{3}}\mathcal{N}_{a,\ell}\,
\Delta^{-2a}\left(1+O(\Delta^{-\frac13})\right)\,,
\end{equation}
where $\mathcal{N}_{a,\ell}$ is given by
\begin{equation}\label{coefficient_N_a_l}
    \mathcal{N}_{a,\ell}=\frac{\Gamma(1+2a)}{2^{1+4a}\,\Gamma(\frac32+2a)}
    \sqrt{\frac{\pi(4a+1)\Gamma(2(1+\ell+a))}{(2\ell+1)\Gamma(1+2(\ell-a))}}\,,
\end{equation}
and $b_{0,2a,a}$ is one of the dynamical coefficients \eqref{1pt-coeff-s-expansion} entering in the 
expansion of the one-point function. In particular, to each of the tensor structures $a$ corresponds 
a different coefficient. The method also gives subleading corrections in \eqref{OPE_coeff_asymptotics_gapped}, 
which depend on higher dynamical coefficients $b_{i,2j,k}$. The expressions for these higher order coefficients
are generally quite cumbersome - we will only occasionally display some of these coefficients in formulas below.
\smallskip

The equation \eqref{OPE_coeff_asymptotics_gapped} is one of the main results of the present work. It establishes the asymptotic behaviour of HHL OPE coefficients, with the dependence on tensor structures $a$ fully resolved. Let us note that the power of $\Delta$ only depends on $\Delta_\phi$ and the tensor structure label $a$, and it does so in a very simple way. We see that the OPE coefficients with a higher tensor structure label are asymptotically suppressed, 
\begin{equation}\label{OPE-a-b-hirearchy}
\overline{\lambda_{\phi\mathcal{OO}}^a}/\overline{\lambda_{\phi\mathcal{OO}}^{b}} \to 
0 \quad\text{as}\quad \Delta\to\infty, \quad\text{if} \quad a>b\ .
\end{equation}
This justifies the terminology introduced in Section \ref{SS:Blocks and tensor structures} 
where we referred to the tensor structure with $a=0$ as the dominant one. Specialising
eq.~\eqref{OPE_coeff_asymptotics_gapped} to the dominant tensor structure, we obtain
\begin{align}\label{OPE-coefficients-dominant-asymptotics}
    \overline{\lambda_{\phi\mathcal{O}\mathcal{O}}^0}^{\, \text{gapped}}(\Delta,\ell) & =  \left(\frac{\Delta}{8\pi f}\right)^{\frac{\Delta_\phi}{3}} \Bigg( b_{0,0,0} + \frac{2(\pi f)^\frac{1}{3} \,b_{1,0,0}}{\Delta^\frac{1}{3}} + \\ & \hspace{-2cm} +\frac{1}{6(\pi f)^\frac{1}{3}\Delta^\frac{2}{3}}\Big(\Delta_\phi(15-2\Delta_\phi)\,b_{0,0,0}+24 f \pi \,b_{2,0,0} + 9\,b_{0,2,0} +6\,b_{0,2,1} \Big) + O(\Delta^{-1})\Bigg) \, .\nonumber
\end{align}
Here we have also written explicitly the first two corrections. If a more restrictive expansion with $b_{2i+1,2j,0}=0$ can be used for the one-point function, the first subleading term vanishes. The formula \eqref{OPE-coefficients-dominant-asymptotics} recovers the prediction made in relation with the eigenstate thermalisation hypothesis, \cite{Lashkari:2016vgj}, and extends it to subleading orders. The findings \cite{Lashkari:2016vgj} have been refined in \cite{Delacretaz:2020nit} to take a tensor structure label into account. This 
refinement is consistent with our formula, as we shall detail in a remark at 
the end of the present section.%
\medskip

For the free scalar theory, we may follow the same steps as above, just with the modified form of the partition function and the one-point function. All-order asymptotic expansions for 
the partition function and the one-point function of $\langle\phi^2\rangle$ (where $\phi$ 
denotes the fundamental field) for the free theory are derived in Section 
\ref{S:Generalised free theory at finite temperature}. This leads to the modified formula 
for the OPE coefficients
\begin{equation}\label{OPE-coefficients-dominant-asymptotics-free}
    \overline{\lambda_{\phi^2\mathcal{O}\mathcal{O}}^a}^{\, \text{free}}(\Delta,\ell) = \left(\frac{\Delta}{8\pi f}\right)^{\frac13}\Delta^{-2a}\left(k_{a,\ell}\log{\left(\frac{\Delta}{8 \pi f}\right)}+\mathcal{N}_{a,\ell}I_{f,2a,a}^{\langle\phi^2\rangle} + O\left(\Delta^{-\frac13}\log\Delta\right) \right)\ .
\end{equation}
Here we can notice a sum of two contributions. One contribution follows the same rule as in the gapped case \eqref{OPE_coeff_asymptotics_gapped}, and is parametrised by the combination of the kinematical coefficient $\mathcal{N}_{a,\ell}$ given in \eqref{coefficient_N_a_l} and the dynamical coefficient $I_{f,2a,a}^{\langle\phi^2\rangle}$, that is the prefactor of the term $s^a\Omega^{2a}$ in the series expansion \eqref{I_phi_expansion} (and corresponds to the coefficient $b_{0,2a,a}$ in eq.~\eqref{OPE_coeff_asymptotics_gapped}). The other contribution, that has an additional $\log\Delta$ scaling with respect to the previous one, is instead particular to the free theory. It is a direct consequence of the $\log\beta/\beta$ divergence of the one-point function \eqref{free-one-pt-asymptotic-expansion} of $\langle\phi^2\rangle$, which is absent in the gapped case \eqref{1pt-expansion-ansatz}. The associated coefficient $k_{a,\ell}$ is given by
\begin{equation}
    k_{a,\ell}=(-1)^a\frac{\Gamma\left(\frac12+a\right)^2}{2^{2a}\,3\,\Gamma\left(\frac32+2a\right)}
    \sqrt{\frac{(4a+1)\Gamma(2(1+\ell+a))}{\pi(2\ell+1)\Gamma(1+2(\ell-a))}}\ .
\end{equation}
We shall compare the result \eqref{OPE-coefficients-dominant-asymptotics-free} with exact OPE coefficients in the next section. The hierarchy of OPE coefficients \eqref{OPE-a-b-hirearchy} remains valid in the free field theory - this is why we could safely drop the superscript `gapped' in writing that equation. Our result \eqref{OPE-a-b-hirearchy} on the tensor 
structure dependence of the averaged OPE coefficients explains observations made previously in \cite{Buric:2024kxo} by looking at exact CFT data in the free theory. 

\paragraph{Remark} Let us explain how the above results, in particular the equations \eqref{OPE_coeff_asymptotics_gapped} and \eqref{OPE-a-b-hirearchy}, imply a prediction from \cite{Delacretaz:2020nit}, namely the equation (3.8) of that work. In that work, it is predicted that all HHL OPE coefficients, defined with respect to a certain basis of three-point tensor structures, have the same asymptotic behaviour. Let $h_a(u,s)$ be the corresponding leading terms of thermal conformal blocks. We can write
\begin{equation}
    h_a(u,s) = M_{ab} f^{\ell,b}_0(u,s)\,,
\end{equation}
where $f^{\ell,b}$ are the leading terms \eqref{tensor_structures_basis} used in this work and $M_{ab}=M_{ab}(\ell)$ is the matrix realising the change between the two bases. By comparing the bases, one finds the coefficients $M_{a0}(\ell) = 1$. In particular, the coefficients do not depend on $a$, i.e. each $h_a$ contains ‘the same amount' of the dominant polynomial $f^{\ell,0}_0$. Consequently, the leading behaviour of all HHL OPE coefficients in the basis used in \cite{Delacretaz:2020nit} is the same.

\section{Free Theory at Finite Temperature}
\label{S:Generalised free theory at finite temperature}

In this section, we apply and illustrate the above results on the example of the free scalar field theory. After obtaining series expansions for the partition function and the one-point function of $\phi^2$ at low temperature, we expand them in characters and conformal blocks to read off the exact low-lying CFT data. For the spectrum, we compute the data up to $\Delta=210$ and for OPE coefficients up to $\Delta=30$, in all spin and tensor-structure sectors. This is followed  by the computation of the all-order asymptotic series for $\mathcal{Z}$ and $\langle\phi^2\rangle$ around $\beta=0$. Finally, we apply the asymptotic formulas of Section \ref{S:Asymptotic CFT data} and compare them to exact data, finding excellent agreement. The first subsection is concerened with the partition functions and the spectrum, while the second focuses on the one-point function and OPE coefficients.

\subsection{Partition function and the spectrum}
\label{SS:Partition function}

The simplest three-dimensional conformal field theory is that of a free real scalar field $\phi$. The Hilbert space of this CFT is the Fock space built upon the space $\mathcal{H}_1$ of one-particle states,
\begin{equation}
    \mathcal{H} = S^0\mathcal{H}_1 \oplus S^1\mathcal{H}_1 \oplus S^2\mathcal{H}_1 \oplus\dots\,,
\end{equation}
where $\mathcal{H}_1$ is the carrier space of the state representation of the fundamental field $\phi$. In order to compute the partition function at finite temperature and chemical potential, we enumerate the states in the Fock space. To this end, consider a state $|\Delta,m\rangle$ in the one-particle space $\mathcal{H}_1$ of conformal weight $\Delta$ and $M$-spin $m$. Multi-particle states constructed out of this excitation contribute $(1-q^\Delta y^m)^{-1}$ to the partition function (by summing the geometric series). Since each single-particle mode $|\Delta,m\rangle$ is independent, the full partition function is given by the product
\begin{equation}\label{free-partition-function-1}
    \mathcal{Z}^{\text{free}}(q,y) = \prod_{\Delta,m} \frac{1}{1 - q^\Delta y^m} = \prod_{\ell=0}^\infty \prod_{m=-\ell}^\ell \frac{1}{1-q^{\ell+1/2}y^m}\ .
\end{equation}
To pass to the second expression, we used that each state $|\Delta,m\rangle$ appears with unit multiplicity in the free field representation. In the following, we will make use of an alternative expression for eq.~\eqref{free-partition-function-1} which involves a single product. To this end, we can write
\begin{equation}\label{argument-Z-char}
    \log \mathcal{Z}^{\text{free}}(q,y) = -\sum_{\Delta,m}\log(1 - q^\Delta y^m) = \sum_{\Delta,m} \sum_{n=1}^\infty \frac{(q^\Delta y^m)^n}{n} = \sum_{n=1}^\infty\frac{1}{n} \sum_{\Delta,m} q^{n\Delta} y^{nm}\ .
\end{equation}
We used the Taylor expansion of $\log(1-x)$ and exchanged the order of two summations. Notice that the second sum is nothing else but the character $\chi^\text{free}$ of the one-particle space $\mathcal{H}_1$, evaluated at 
shifted values of potentials $q,y$, i.e. $\chi^{\text{free}}(q^n,y^n)$. 
Using the expression for the free character,
\begin{equation}
    \chi^{\text{free}}(q,y) = \chi_{1/2,0}(q,y) - \chi_{5/2,0}(q,y) = \frac{q^{1/2}(1+q)}{(1-qy)\left(1-qy^{-1}\right)}\,,
\end{equation}
we arrive at the following representation of the free partition function
\begin{equation}\label{free-partition-function-2}
    \log \mathcal{Z}^{\text{free}} = \sum_{n=1}^\infty \frac{1}{n} \frac{q^{n/2}(1+q^n)}{\left(1-q^n y^n\right)\left(1 - q^n y^{-n}\right)} = \sum_{n=1}^\infty \frac{1}{n} \frac{q^{n/2}(1+q^n)}{1+q^{2n} - 2q^n T_n\left(1+\frac{u}{2}\right)}\ .
\end{equation}
Here, $T_n$ denotes the Chebyshev polynomial of the first kind. The partition function decomposes into irreducible conformal characters \eqref{characters},
\begin{equation}\label{free-character-decompsition}
    \mathcal{Z}^{\text{free}}(q,y) = \sum_{\Delta,\ell} m_{\Delta,\ell}\, \chi_{\Delta,\ell}(q,y)\ .
\end{equation}
Here, $m_{\Delta,\ell}$ denotes the multiplicity of operators of dimension $\Delta$ and spin $\ell$. A simple way of determining multiplicities is by expanding both sides of \eqref{free-character-decompsition} in powers of $q$ and matching the two expansions. To a few lowest orders
\begin{equation}
    \mathcal{Z}^{\text{free}}(q,y) = 1 + \chi^{\text{free}}(q,y) + \chi_{1,0} (q,y) + \chi_{\frac32,0}(q,y) + \chi_2^{\text{short}}(q,y) + \chi_{2,0}(q,y) + \dots\ .
\end{equation}
Non-trivial multiplicities start appearing higher in the spectrum. The $\chi^\text{short}_\ell$ denote characters of conserved current representations,
\begin{equation}
    \chi^{\text{short}}_\ell(q,y) = \chi_{\ell+1,\ell}(q,y) - \chi_{\ell+2,\ell-1}(q,y)\ .
\end{equation}
The character decomposition of the partition function can be performed very efficiently to high values of the weight $\Delta$.

\subsubsection{Partition function at high temperature}
\label{sec:PFathighT}

In this subsection, we determine the behaviour of the partition function at high temperatures, which, as we have seen, controls the large-$\Delta$ density of primary states. 
\smallskip

The regime we are interested in is $\beta\to0$ with finite $\Omega$, as defined in eq.~\eqref{q-y-mu-beta-Omega}. If instead of $\Omega$ we would keep the chemical potential $\mu$ fixed, the partition function would exhibit fractal-like behaviour, see \cite{Benjamin:2024kdg}. To obtain the asymptotic expansion around $\beta=0$, we start with the expression \eqref{free-partition-function-2} and switch to variables $(\beta,\Omega)$. Then we can write
\begin{equation}\label{Z-free-as-sum-f(n)}
    \beta^{-1}\log \mathcal{Z}^{\text{free}} = \sum_{n=1}^\infty \fz(n\beta,\Omega)\,,
\end{equation}
where
\begin{equation}\label{f-partition-function}
    \fz(\beta,\Omega) = \frac{1}{\beta} \frac{\cosh\frac{\beta}{2} }{\cosh\beta  - \cos(\beta\Omega)} \equiv \sum_{m=-1}^\infty \fz_{2m-1}(\Omega) \beta^{2m-1} \ .
\end{equation}
In the second equality, we expanded the function $\fz(\beta,\Omega)$ in series around $\beta=0$. Next, we perform the sum over $n$ in eq.\ \eqref{Z-free-as-sum-f(n)}. According to the general theory for sums of this kind, see \cite{Zagier} and our 
brief review in Appendix \ref{AA:Asymptotic behaviour of certain infinite sums},  we have
\begin{align}
    &\beta^{-1}\log \mathcal{Z}^{\text{free}} \sim \beta^{-3} \zeta(3)\fz_{-3}(\Omega) -\frac{\log\beta}{\beta} \fz_{-1}(\Omega) + \frac{I_f^{\mathcal{Z}}(\Omega)}{\beta} - \sum_{m=1}^\infty \beta^{2m-1} \zeta(1-2m)\fz_{2m-1}(\Omega)\nonumber\\[3pt]
    & = \frac{2\zeta(3)}{\beta^3 (1+\Omega^2)} - \frac{\log\beta (1+2\Omega^2)}{12\beta(1+\Omega^2)} + \frac{I_f^{\mathcal{Z}}(\Omega)}{\beta} - \sum_{m=1}^\infty \beta^{2m-1} \zeta(1-2m)\fz_{2m-1}(\Omega)\,,\label{Z-free-asymptotic-expansion}
\end{align}
where
\begin{align}\label{IfZ}
    I_f^{\mathcal{Z}}(\Omega) & = \int_0^\infty d\beta\, \left( \fz(\beta,\Omega) - \frac{\fz_{-3}(\Omega)}{\beta^3} - \frac{\fz_{-1}(\Omega) e^{-\beta}}{\beta}\right)\\
    & = \int_0^\infty d\beta\, \left( \frac{1}{\beta} \frac{\cosh\frac{\beta}{2} }{\cosh\beta  - \cos(\beta\Omega)} - \frac{2}{\beta^3(1+\Omega^2)} - \frac{e^{-\beta}(1+2\Omega^2)}{12\beta(1+\Omega^2)}\right)\ . \nonumber
\end{align}
The $\sim$ sign in the first line of eq.~\eqref{Z-free-asymptotic-expansion} indicates that the right hand side is the asymptotic expansion of the left hand side around $\beta=0$. The integral \eqref{IfZ} may be evaluated order by order in $\Omega$ by expanding the integrand and integrating term by term,
\begin{align}\label{IfZ-expansion}
    &I_f^{\mathcal{Z}}(\Omega) = \sum_{k=0}^\infty I_{f,2k}^{\mathcal{Z}} \Omega^{2k} \\
    & = -\zeta'(-1) - \frac{\log2}{12} + \frac{5+6\gamma+6\log4}{72}\Omega^2 + \frac{147 \zeta(3) -120 \gamma -214-240 \log 2}{1440} \Omega^4 + \dots \nonumber\ .
\end{align}
Here, $\gamma$ is Euler's constant. The leading order result in eq.~\eqref{Z-free-asymptotic-expansion} is in accord with generic expectations for any (local) CFT. It allows us to read off the free energy density of the theory,
\begin{equation}\label{free-energy-free}
    f^{\text{free}} = \frac{\zeta(3)}{2\pi}\,,
\end{equation}
which is well-known. The formula \eqref{Z-free-asymptotic-expansion} captures all orders in the asymptotic expansion of the partition function and extends similar expansions \cite{Melia:2020pzd,Benjamin:2023qsc} to non-zero $\Omega$. This is briefly discussed at the end of the subsection.
\smallskip

\begin{figure}[h!]
    \centering
    \begin{minipage}[b]{0.49\textwidth}
        \includegraphics[width=\textwidth]{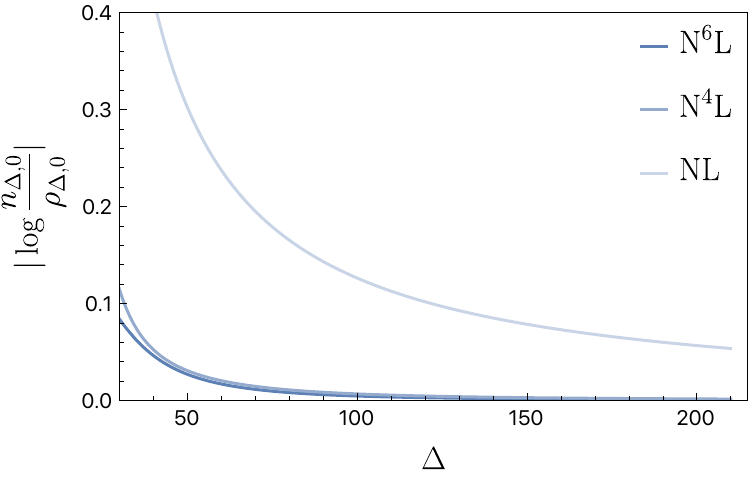}
    \end{minipage}
    \hfill
    \begin{minipage}[b]{0.49\textwidth}
        \includegraphics[width=\textwidth]{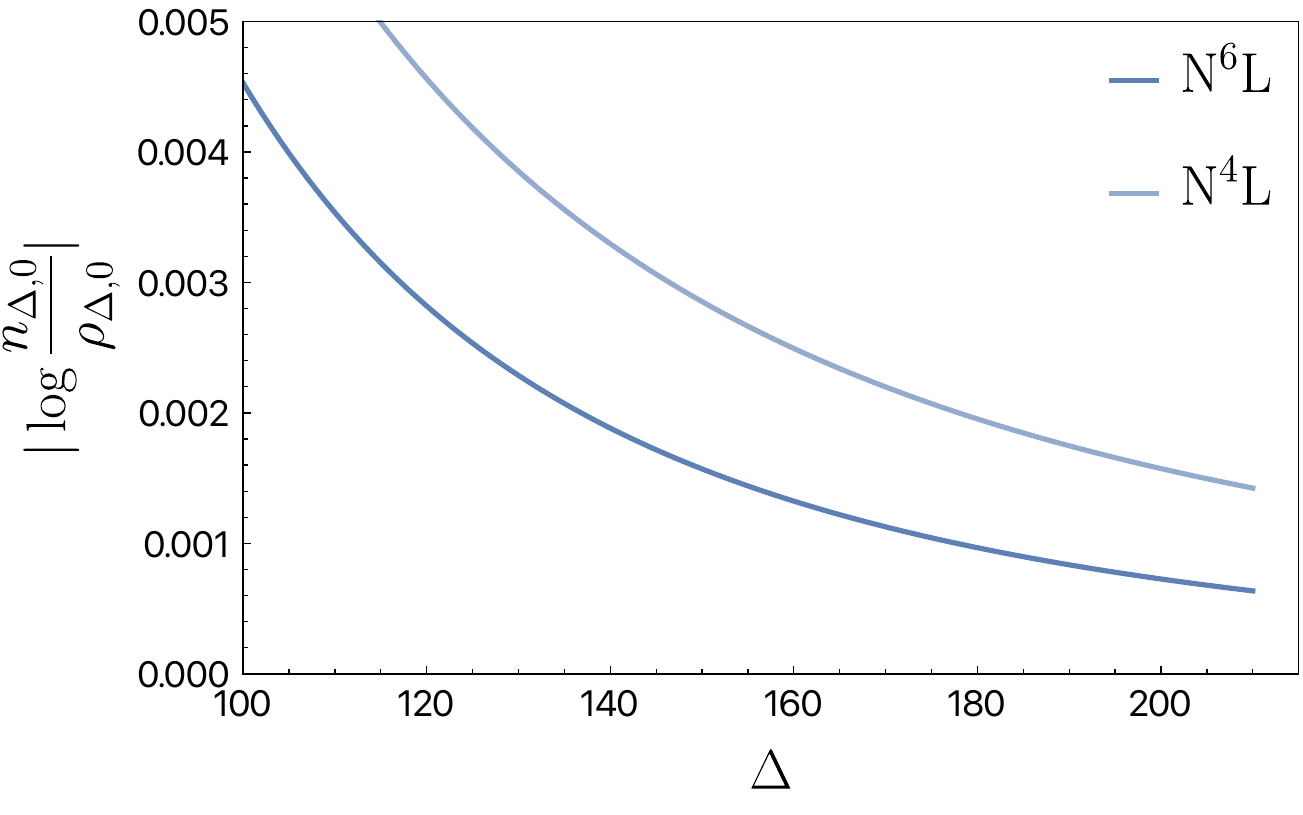}
    \end{minipage}
    \caption{Log of the ratio of exact multiplicities and the asymptotic density of scalars expanded to different orders. The notation $\mathrm{N^n L}$ here stands for $n$ terms in the asymptotic series after the leading order. On the right, a zoomed version of the same plot shows the difference of the asymptotics with five or seven terms in the regime $\Delta>100$.}
    \label{fig:spectrum0_free}
\end{figure}

\begin{figure}[h!]
    \centering
    \begin{minipage}[b]{0.49\textwidth}
        \includegraphics[width=\textwidth]{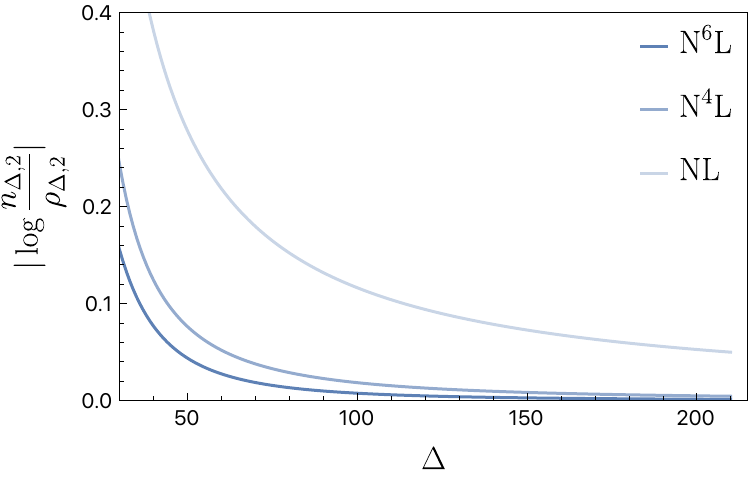}
    \end{minipage}
    \hfill
    \begin{minipage}[b]{0.49\textwidth}
        \includegraphics[width=\textwidth]{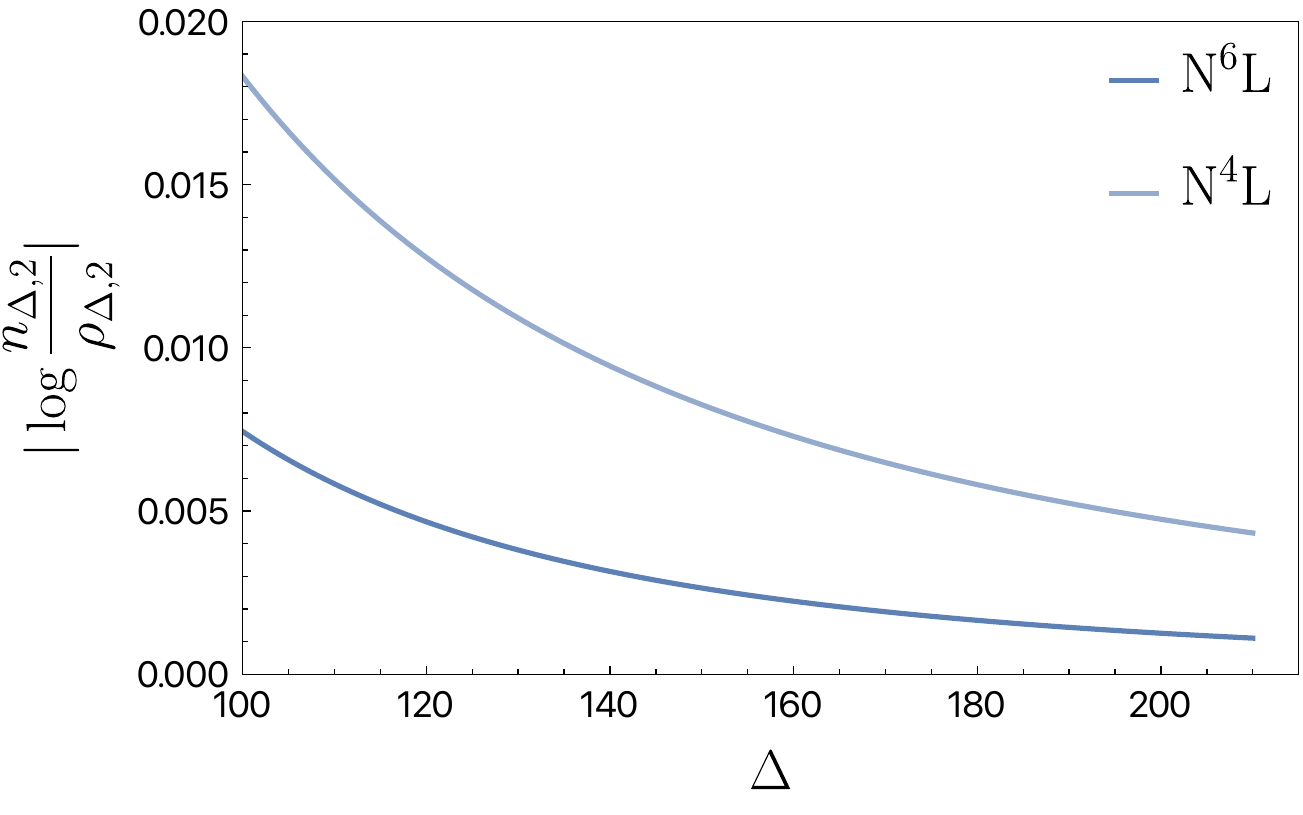}
    \end{minipage}
    \caption{Log of the ratio of exact multiplicities and the asymptotic density of spin two primaries expanded to different orders. The notation $\mathrm{N^n L}$ stands for $n$ terms in the asymptotic series after the leading order. On the right, a zoomed version of the same plot shows the difference of the asymptotics with five or seven terms in the regime $\Delta>100$.}
    \label{fig:spectrum2_free}
\end{figure}

We can now put together all the ingredients discussed above. Having obtained the refined asymptotic formula \eqref{Z-free-asymptotic-expansion} for the partition function, one may apply the theory of the previous section to infer the large-$\Delta$ density of primary states. On the other hand, exact multiplicities are obtained by performing the character decomposition as discussed above. In Figures \ref{fig:spectrum0_free} and \ref{fig:spectrum2_free}, we show the ratio of the exact multiplicities and the asymptotic density of primaries expanded to different orders in the asymptotic series. The coefficients in the expansion are given in Appendix \ref{app:Asymp_freeTheory}. One observes that keeping more terms in the asymptotic expansion significantly improves its accuracy. E.g. for scalar operators, at the highest value $\Delta=210$ that is used for comparison, the errors of the curves are: 5.50\% for the asymptotics with two terms, 0.14\% for the one with five, and 0.06\% for seven. We display the results for operators of spin zero and two. Similar comments apply to other spin sectors.


\paragraph{Remarks} 1) One can also plot the ratio of multiplicities to the leading order asymptotic formula. The rate of convergence of the leading asymptotics to the exact density is very slow - at $\Delta=210$, the error is around 40\%.
\smallskip

2) Setting $\Omega=0$ in \eqref{Z-free-asymptotic-expansion} and using the leading order expansion in eq.~\eqref{IfZ-expansion} recovers the result stated in \cite{Melia:2020pzd},
\begin{equation}\label{free-Z-refined-asymptotics}
    \lim_{\beta\to0}\frac{\mathcal{Z}^{\text{free}}(\beta,\Omega=0)} {(2\beta)^{-\frac{1}{12}} e^{\frac{2\zeta(3)}{\beta^2} - \zeta'(-1)}} = 1 \,,
\end{equation}
that was derived by relating the free partition function to the generating function of plane partitions.
\smallskip

3) At $\Omega=0$, the function $\fz$ becomes
\begin{equation}
    \fz(\beta,0) = \frac{1}{\beta} \frac{\sinh\beta}{4\sinh^3 \frac{\beta}{2}} \ .
\end{equation}
In terminology of \cite{Benjamin:2023qsc}, this is nothing else but ($\beta^{-1}$ times) the generating function of the coefficients $c_{2n}(d=3)$, see equation (C.22) of that work. Therefore, by setting $\Omega=0$, the expansion \eqref{Z-free-asymptotic-expansion} reduces to that stated in equation (C.22) of \cite{Benjamin:2023qsc}. From the $\Omega=0$ result, it is also easy to see that the expansion around $\beta=0$ is asymptotic rather than convergent.

\subsection{One-point function and OPE coefficients}
\label{SS:Two- and one-point functions}

We turn to one-point functions. In order to compute these, we start with the two-point function  $\langle\phi\phi\rangle$ of the fundamental field. In the free theory, this two-point function can be obtained by the method of images. It reads
\begin{equation}\label{two-point-function}
    \langle\phi(x_1)\phi(x_2)\rangle_{q,y} = \sum_{n=-\infty}^\infty \frac{q^{n\Delta_\phi}}{|x_{12}^{(n)}|^{2\Delta_\phi}}\,,
\end{equation}
where $|x_{12}^{(n)}|$ is the distance between the point $x_1$ and the $n$-th thermal image of $x_2$,
\begin{equation*}
    x_{12}^{(n)} = x_1 - q^{nD}y^{n M}\cdot x_2\ .
\end{equation*}
Via the operator product expansion, one can obtain from eq.~\eqref{two-point-function} the thermal one-point functions of all fields appearing in the $\phi\times\phi$ OPE. The simplest one-point function to extract is that of $\phi^2$. To this end, we just need to perform the leading order OPE, which amounts to subtracting the identity contribution $\{n=0\}$ in the sum \eqref{two-point-function} and setting $r_1=r_2=1$, $\theta_1 = \theta_2\equiv\theta$ and $\varphi_1 = \varphi_2 = 0$ in spherical polar coordinates. Thereby we obtain
\begin{align}
    \lambda_{\phi\phi\phi^2} \langle\phi^2(x)\rangle_{q,y} & = \sum_{n\neq0}\frac{q^{n\Delta_\phi}}{\left(1+q^{2n}-2q^n(\cos^2\theta+\sin^2\theta\cos(n\mu))\right)^{\Delta_\phi}} \nonumber \\
    & = \sum_{n=1}^\infty \frac{2}{\left(q^n+q^{-n}-2(\cos^2\theta+\sin^2\theta\, T_n\left(1+\frac{u}{2}\right))\right)^{\Delta_\phi}}\ . \label{phi2-one-pt-function}
\end{align}
In these expressions, $\Delta_\phi = 1/2$ is the scaling dimension of the free field. To access other operators, one performs the OPE expansion to higher orders,  see \cite{Buric:2024kxo} for an explicit expression for the stress tensor one-point function. Yet more operators may be accessed by starting from higher-point functions. How this is done is briefly discussed in Appendix \ref{A:More on the free theory}. In this work, we will focus on the operator $\phi^2$.
\smallskip

\begin{figure}[h!]
    \centering
    \begin{minipage}[b]{0.95\textwidth}
        \includegraphics[width=\textwidth]{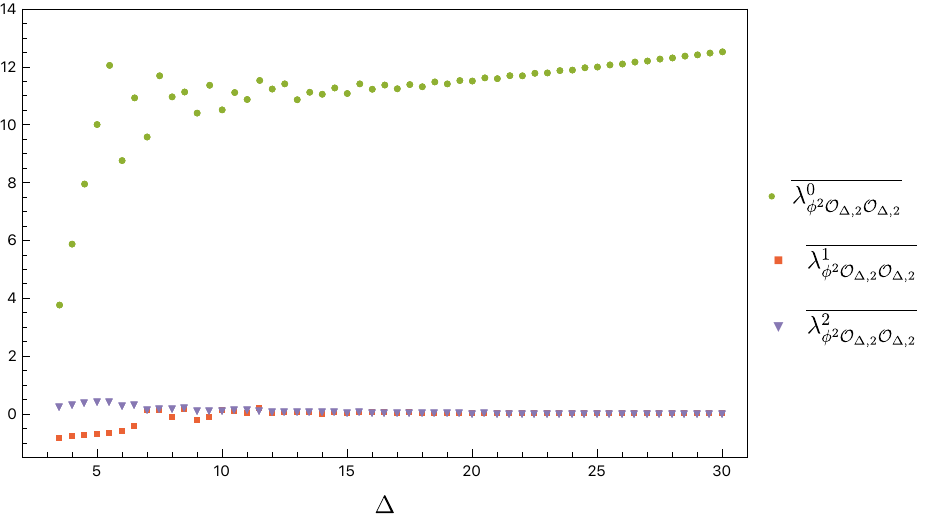}
    \end{minipage}
    \caption{Averaged OPE coefficients for dominant and subdominant tensor structures with internal spin $\ell = 2$.}
    \label{fig:ope_spin2}
\end{figure}

As we have reviewed is Section \ref{S:Thermal one-point conformal blocks}, thermal one-point functions decompose in conformal blocks according to eq.~\eqref{block-decomposition}. Similarly to the character decomposition of the partition function, the one-point functions may be efficiently decomposed by expanding both the correlator and blocks in powers of $q$ and matching the expansions. We carry out this procedure for the one-point function of $\langle\phi^2\rangle$ given in eq.~\eqref{phi2-one-pt-function} using the blocks obtained by recursion relations of Section \ref{SS:Recursion relations for thermal blocks}. We show as an example what happens for $\ell = 2$ in Figure \ref{fig:ope_spin2}, and in the next subsection we compare them with the asymptotics.

\subsubsection{High-temperature behaviour of \texorpdfstring{$\langle\phi^2\rangle$}{<phi2>}}

In this subsection, we derive the asymptotic expansion of $\langle\phi^2\rangle_{q,y}$ around $\beta=0$ and compare the resulting asymptotic OPE coefficients against exact CFT data computed above. At $\Omega=0$, the one-point function \eqref{phi2-one-pt-function} reduces to
\begin{equation}\label{1pt-y=1}
    \lambda_{\phi\phi\phi^2} \langle\phi^2(x)\rangle_q^{\text{free}} = 2 \sum_{n=1}^\infty \frac{q^{n/2}}{1-q^n} \ .
\end{equation}
The sum may be performed exactly. This leads to the following expression 
\begin{equation}\label{leading-divergence-1pt}
    \langle\phi^2(x)\rangle_q^{\text{free}} = 2 \frac{\log(1-q) + \psi_q(1/2)}{\log q}= -2 \left( \frac{\log\beta}{\beta} + \frac{\psi(1/2)}{\beta} + \frac{\beta}{288} + \dots\right)\ .
\end{equation}
Here, $\psi_q(z)$ denotes the $q$-digamma function. In the second expression, we performed the expansion around $\beta=0$. As mentioned before, the leading divergence is more singular than in gapped theories, the latter being $\beta^{-2\Delta_\phi}$.\footnote{For a generalised free field of dimension $\Delta_\phi$ the one-point coefficient is equal to $b_{\phi^2} = \zeta(2\Delta_\phi)$. This expression diverges at $\Delta_\phi = 1/2$, signalling a higher divergence in the free theory.}
\smallskip

Away from $\Omega=0$, we are not aware of an exact expression for the one-point function \eqref{phi2-one-pt-function} in terms of special functions, but the asymptotic expansion around $\beta=0$ can still be obtained. To this end, we note that the one-point function takes the form
\begin{equation}
    \langle\phi^2\rangle_{\beta,\Omega} = \sum_{n=1}^\infty \fp(n\beta,\Omega,s)\,,
\end{equation}
where the function $\fp(\beta,\Omega,s)$ is given by
\begin{equation}\label{f-one-point}
    \fp(\beta,\Omega,s) = \frac{2}{\sqrt{e^{-\beta} + e^\beta - 2 (1 - s + s \cos(\beta\Omega))}} \sim \sum_{m=0}^\infty \fp_{2m-1}(\Omega,s) \beta^{2m-1}\ .
\end{equation}
The second equality defines coefficient functions $\fp_{2m-1}(\Omega,s)$. The $\sim$ sign indicates that the expansion at $\beta=0$ is asymptotic rather than convergent. Using the same method \cite{Zagier} as for the partition function, we obtain the asymptotic expansion of the one-point function around $\beta=0$,
\begin{equation}\label{free-one-pt-asymptotic-expansion}
    \langle\phi^2\rangle_{\beta,\Omega} \sim -\frac{2}{\sqrt{1+\Omega^2 s}}\frac{\log\beta}{\beta} + \frac{\Ifp(\Omega,s)}{\beta} + \sum_{n=1}^\infty \beta^{2n-1}\zeta(1-2n)\fp_{2n-1}(\Omega,s)\,,
\end{equation}
where $\Ifp(\Omega,s)$ is given by
\begin{equation}
   \Ifp(\Omega,s)= \int_0^\infty d\beta\, \left( \fp(\beta,\Omega,s) - \frac{2 e^{-\beta}}{\beta\sqrt{1+\Omega^2 s}}\right)\ .
\end{equation}
One can further expand each term in eq.~\eqref{free-one-pt-asymptotic-expansion} in powers of $\Omega^2$ and $s$. For the most non-trivial of these terms, $\Ifp(\Omega,s)$, the expansion reads
\begin{equation}\label{I_phi_expansion}
    \Ifp(\Omega,s) = 2(\gamma + \log4 ) - \frac14 \Big(6+4\gamma+8\log2 - 7\zeta(3)\Big) s\Omega^2 + \dots\,,
\end{equation}
where $\gamma$ is Euler's constant. With the all-order asymptotic expansion of the one-point function at hand, the results of the previous section allow us to find the asymptotic expansion of averaged OPE coefficients $\overline{\lambda^a_{\phi^2\mathcal{OO}}}$. We have written the leading order result in eq.~\eqref{OPE-coefficients-dominant-asymptotics-free} and further details are given in Appendix \ref{AAA:Free-OPE-asymptotics}. We can now compare these asymptotic expansions against exact OPE coefficients found above. 
\smallskip

\begin{figure}[h!]
    \centering
    \begin{minipage}[b]{0.95\textwidth}
        \includegraphics[width=\textwidth]{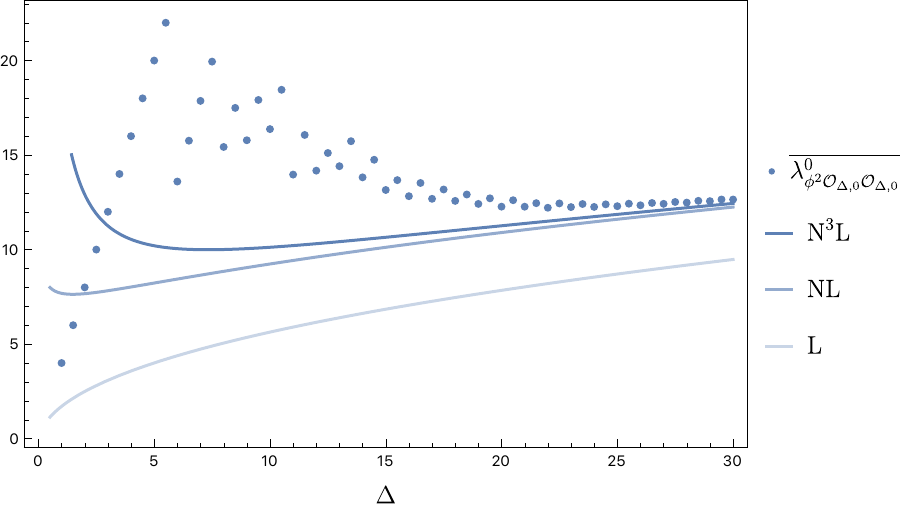}
    \end{minipage}
    \caption{Scalar exchange dominant OPE coefficients against asymptotic formulas with a varying number of terms: leading (L), next-to-leading (NL), etc. The darker the asymptotic curve, the more subleading terms were considered.}
    \label{fig:ope_ScalarAsymp}
\end{figure}
\begin{figure}[h!]
    \centering
    \begin{minipage}[b]{0.95\textwidth}
        \includegraphics[width=\textwidth]{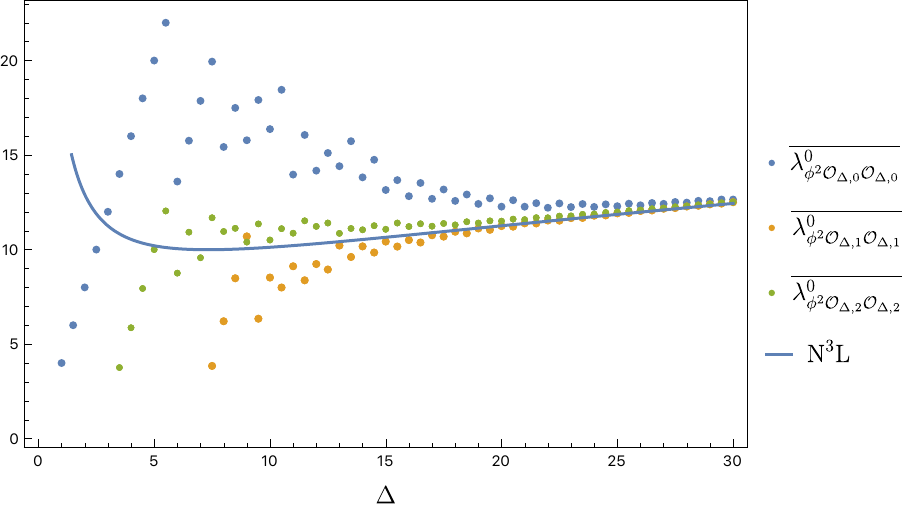}
    \end{minipage}
    \hfill
    \caption{Dominant OPE coefficients in different spin sectors against the asymptotic formula with four terms.}
    \label{fig:ope_dominant}
\end{figure}

To begin with, Figure \ref{fig:ope_spin2} shows the three families $a=0,1,2$ of averaged OPE coefficients for spin two exchanged operators. The distinction between dominant and non-dominant structures is manifest. More quantitatively, in Figure \ref{fig:ope_ScalarAsymp}, we plot the scalar exchange dominant ($a=0$) averaged OPE coefficients against different orders of the asymptotic expansion. Similarly as for the spectrum, we see that keeping more terms drastically improves the accuracy of the asymptotic approximation. In Figure \ref{fig:ope_dominant}, we show exact OPE coefficients for the dominant tensor structure $\overline{\lambda^0_{\phi^2\mathcal{OO}}}(\Delta,\ell)$ in different spin sectors.

\begin{figure}[h!]
    \centering
    \begin{minipage}[b]{0.95\textwidth}
        \includegraphics[width=\textwidth]{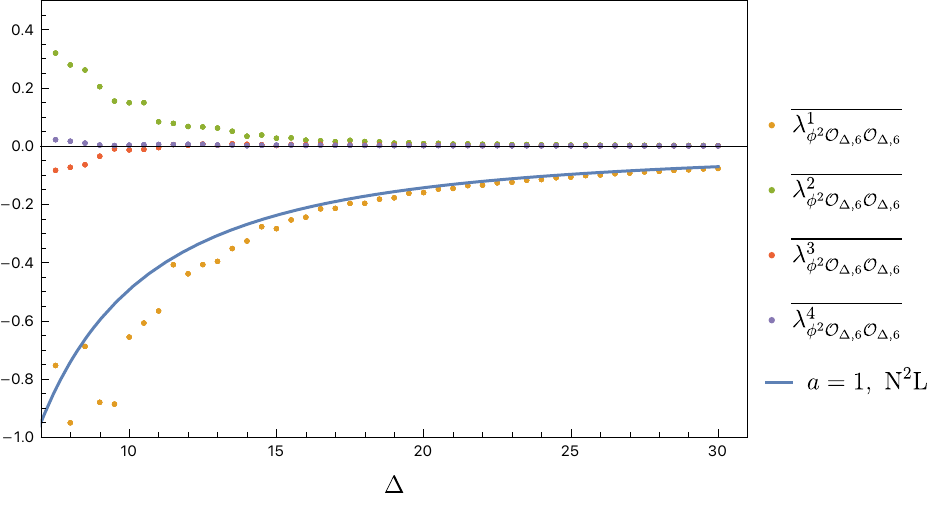}
    \end{minipage}
    \hfill
    \caption{Averaged OPE coefficients for some ($a\leq 4$) of the subdominant tensor structures, for internal spin six. The other subdominant OPE coefficients with $a=5,6,7$ are too small to be visualised properly, and have been therefore excluded from the plot. The curve is the asymptotic formula for $\overline{\lambda_{\phi^2 \mathcal{O}_{\Delta,6}\mathcal{O}_{\Delta,6}}^1}$ with three terms in the expansion, and has an error of 8.6\% at $\Delta=30$.}
    \label{fig:ope_spin6SubdominantWAsymp}
\end{figure}

Different sectors are indicated by different colours, with blue corresponding to spin zero, orange to spin one and green to spin two. The curve represents the fourth order asymptotic formula for these coefficients - the asymptotic behaviour to this order is spin-independent, so we only require a single curve. Finally, let us also discuss the asymptotics of non-dominant OPE coefficients. In Figure \ref{fig:ope_spin6SubdominantWAsymp}, we show several families of subdominant OPE coefficients, for spin $\ell=6$ exchange. Different values of $a$ are distinguished by colours and start with the first subdominant family, $a=1$. The curve represents the asymptotic formula for $\ell=6$ and $a=1$. Again, we observe a good agreement with the exact data. We leave the computation of higher subleading corrections for $a=1$, as well as expansions of coefficients with higher values of $a$, for future work.

\section{Conclusions}
\label{S:Conclusions}

In this work, we have extended the theory of one-point conformal blocks on the geometry $S^1 \times S^2$, initiated in \cite{Buric:2024kxo}, and used it to obtain new asymptotic formulas for HHL OPE coefficients in any conformal field theory. Our new results for conformal blocks include analysis of orthogonality and the inversion formula for one-point functions; recursion relations that give significantly improved expansions of blocks at small $q$; and a systematic asymptotic expansion of blocks at large $\Delta$. An important ingredient for some of these developments is a new basis of three-point tensor structures in which thermal blocks enjoy various remarkable properties.
\smallskip

Using these results, we have obtained the asymptotic formula \eqref{OPE_coeff_asymptotics_gapped} for averaged OPE coefficients $\overline{\lambda^a_{\phi\mathcal{OO}}}$ in the regime $\Delta\to\infty$. With the special choice of tensor structures that we use, the OPE coefficients exhibit hierarchy and become more and more suppressed as the tensor structure label $a$ is increased. The behaviour of dominant ($a=0$) OPE coefficients has often appeared in the literature in relation with the ETH. Our more general formula \eqref{OPE_coeff_asymptotics_gapped} ties together these predictions and some of their refinements, made in \cite{Lashkari:2016vgj,Gobeil:2018fzy,Delacretaz:2020nit,Buric:2024kxo} from a variety of approaches. In addition to the leading asymptotics, our method systematically computes subleading corrections to OPE coefficients as a power series in $\Delta^{-1/3}$. As the powers increase, the coefficients in the expansion depend on more and more dynamical data. While our focus was on OPE coefficients, we have also obtained the analogous expansion of the density of primaries \eqref{rho-gapped}, extending the existing results on the asymptotics of the latter.
\vskip0.15cm

We tested asymptotic formulas for the density of primaries and the OPE coefficients on the example of the free scalar field. To this end, we developed all-order asymptotic expansions of the partition function and the one-point function of $\phi^2$ as $\beta\to0$ at finite chemical potential $\Omega$, thus leading to asymptotic CFT data to an arbitrarily high order in $\Delta^{-1/3}$. On the other hand, we performed the character decomposition of $\mathcal{Z}$ and the block decomposition of $\langle\phi^2\rangle$ to high orders in $q$, the later expansion being possible due to the recursion relations for thermal blocks developed in Section \ref{S:Expansions of Thermal One-point Blocks}. This allowed us to compare exact CFT data to asymptotic formulas and we observed excellent agreement between the two. The comparison also showed that subleading corrections in the asymptotic expansions play an important role and significantly improve the accuracy of the asymptotic approximation for reasonable values of $\Delta$.
\medskip

The present work admits various refinements, extensions and applications. For one, let us note that the expansions that appeared in this paper are asymptotic and the question of their convergence was not discussed. The nature of the expansions and potentially their optimal truncations or resummations remain interesting open questions. Secondly, while we have focused on scalar external operators, extending the results to external spinning fields should be straightforward. In particular, this would allow one to study OPE coefficients $\lambda_{T\mathcal{OO}}^a$ involving the stress tensor (which, for a spinning field $\mathcal{O}$ are not fixed by Ward identities). Low temperature expansions of blocks for external scalars have already been obtained in \cite{Buric:2024kxo}. As already pointed out in \cite{Buric:2024kxo}, a naive application of the asymptotic expansion in the case of conserved $U(1)$ currents reproduces the semi-classical result, according to which the dimension of large-charge $U(1)$ operators scales as $\Delta \sim Q^{2/3}$, \cite{Hellerman:2015nra,Monin:2016jmo}. It would be interesting to compute this relation more precisely and verify whether the subleading terms, as computed in the literature using effective theories, continue to match our expansion.
\smallskip

Both the low and the high temperature expansions may readily be applied to any CFT in which the thermal partition function/one-point function in the relevant regime is available. As in the case of free theory, the former gives exact low lying CFT data, while the latter determines its large $\Delta$ asymptotics. We shall report on results of this kind for generalised free theories and perturbative QFTs in AdS$_4$ in future work. For these cases, the asymptotic analysis gets slightly modified because the theories are non-local and have a large-$\Delta$ density of primaries different from what was assumed in Section \ref{S:Asymptotic CFT data}. Nevertheless, the arguments that section need only minimal adaptations in order to apply. Other interesting applications that seem within reach are to $O(N)$ models, that were recently studied on the $S^1\times S^2$ geometry in \cite{David:2024pir}, as well as holographic CFTs, for which the one-point functions may be extracted from solutions of the Klein-Gordon equation on the AdS-Kerr black hole. 
\smallskip

The high temperature expansion of the partition function is neatly encoded in the thermal EFT, whose Wilson coefficients control the large-$\Delta$ density of states. It should be possible to develop a similar effective theory for one-point functions as well. In both cases, it remains an open question what further bounds may be placed on the EFT coefficients, if any, for generic gapped theories (e.g. see \cite{Allameh:2024qqp,Banihashemi:2025qqi,David:2024pir} for discussions of the coefficient $c_1$ in various models).
\smallskip

More generally, conformal blocks for one-point functions on $S^1\times S^2$ represent the first step towards developing two- or higher-point partial wave decompositions on this geometry, potentially leading to bootstrap studies. This remains a compelling direction for future work.
\smallskip

\paragraph{Acknowledgements:} We would like to thank Julien Barrat, Deniz Bozkurt, Ivan Gusev, Enrico Marchetto, Alessio Miscioscia, Elli Pomoni, Andrei Parnachev, Junchen Rong and Balt van Rees for discussions. IB and FR wish to thank DESY, where part of this work was done, for hospitality. IB is funded by Taighde Éireann – Research Ireland under Grant number SFI-22/FFP-P/11444. FM and VS are funded by the German Research Foundation DFG – SFB 1624 – “Higher structures, moduli spaces and integrability” – 506632645. This project also received funding from the German Research Foundation DFG under Germany’s Excellence Strategy - EXC 2121 Quantum Universe - 390833306. FR is supported by the ERC project number 101087025 “QFTinAdS”.

\appendix

\section{Harmonic Analysis}
\label{A:Harmonic analysis}

In this appendix, we discuss orthogonality of thermal conformal blocks from the point of view of spherical functions. We show how the measure \eqref{measure-blocks} arises from harmonic analysis. We shall be working in Euclidean signature - therefore the present analysis does not prove eq.~ \eqref{orthogonality_blocks_integrated}, which further needs a careful analytic continuation to the Lorentzian setting. In particular, the complex conjugation encountered below is replaced by the shadow transform in the latter.
\smallskip

We use the same notation as in Section 3 of \cite{Buric:2024kxo}. Thus, let $K = SO(4,1)$ be the Euclidean conformal group and $G = K\times K$. Denote by $V$ the carrier space of a principal series representation $\rho$ of $K$, whose conformal weight is of the form $\Delta\in 3/2 + i\mathbb{R}$. Consider the space of spherical functions
\begin{equation}\label{space-of-spherical-functions}
    \Gamma_\rho = \{f : G \to V\ |\ f(k_l g k_r) = \rho(k_l) f(g)\}\,, \qquad g\in G,\ k_{l,r}\in K_d\ .
\end{equation}
Here, $K_d$ stands for the $SO(4,1)$ subgroup of $G$ that is diagonally embedded, i.e. consisting of the elements $(k,k)$. The space \eqref{space-of-spherical-functions} carries the inner product
\begin{equation}\label{inner-product-spherical-abstract}
    \langle f_1, f_2 \rangle = \int\limits_G dg\ (f_1(g),f_2(g))\,,
\end{equation}
where $(,)$ denotes the inner product of vectors in $V$. The group $G$ admits the Cartan decomposition $G = K_d A_p K_d$, where $A_p$ is the abelian subgroup generated by differences of Cartan generators of the two copies of $K$. Spherical functions satisfy
\begin{equation}\label{f-reduction-to-F}
    f(k_l\, a\, k_r) = \rho(k_l) F(a)\,,
\end{equation}
where we have denoted the restriction of the spherical function $f$ to $A_p$ by $F$. It can be shown that the restriction takes values in the space of invariants $V^M$, where $M$ is the centraliser of $A_p$ in $K$. With the choice of $A_p$ made above, $M$ coincides with the Cartan subgroup of $K_d$. In other words, $V^M$ is the zero-weight subspace $V^{[0]}$ of $V$. The relation \eqref{f-reduction-to-F} allows to write the inner product of spherical functions \eqref{inner-product-spherical-abstract} in terms of their restrictions to $A_p$,
\begin{equation}\label{inner-product-intermediate}
    \langle f_1, f_2 \rangle = \int\limits_G dg\ \big(\rho(k_l) F_1(a), \rho(k_l) F_2(a)\big) = \int\limits_G dg\ \left(F_1(a),F_2(a)\right) \ .
\end{equation}
In the last equality we used the fact that the representation $\rho$ is unitary. The integral \eqref{inner-product-intermediate} is certainly divergent because it involves integration over the non-compact group $K$. However, we see that this infinite factor does not depend on functions $f_{1,2}$. Therefore, we will define the reduced integral
\begin{equation}\label{reduced-inner-product-abstract}
    \langle F_1, F_2 \rangle \equiv \int\limits_{A_p} da\ \left(F_1(a),F_2(a)\right)\ .
\end{equation}
Here, $da$ is the reduced Haar measure, i.e. the same one that appears in the orthogonality relations for characters, \eqref{Haar-measure}. Let us now turn to the inner product $(,)$ on $V$. For any $g\in G$, $f_i(g)$ is a function of $x\in\mathbb{R}^3$. Therefore, we may write $f(g;x)$ and similarly $F(a;x)$. With this notation, the inner product \eqref{reduced-inner-product-abstract} becomes
\begin{equation}
    \langle F_1, F_2 \rangle = \int\limits_{A_p} da\ \int\limits_{\mathbb{R}^3} d^3 x\ F_1(a;x)^\ast F_2(a;x)\,,
\end{equation}
where $\ast$ denotes complex conjugation. For a scalar field representation $\rho$, the zero-weight subspace $V^{[0]}$ can be described as
\begin{equation}
    V^{[0]} = \{ \Phi : \mathbb{R}^3 \to \mathbb{C}\ \ | \ \ \Phi(r,\theta,\varphi) = r^{-\Delta}\phi(\theta) \}\,,
\end{equation}
where, as usual, $(t,\theta,\varphi)$ are the spherical polar coordinates. Putting everything together
\begin{align}
    \langle F_1, F_2 \rangle & = \int \omega(\beta,\mu)d\beta\, d\mu \int r^2\sin\theta\, dr\, d\theta\, d\varphi\ \left(r^{-\Delta}F_1(\beta,\mu,\theta)\right)^\ast \left(r^{-\Delta}F_2(\beta,\mu,\theta)\right)\nonumber\\
    & = 2\pi \int \omega(\beta,\mu)d\beta\, d\mu \int \frac{dr}{r} \sin\theta d\theta\, F_1(\beta,\mu,\theta)^\ast F_2(\beta,\mu,\theta)\ .
\end{align}
To pass to the second line, we used the from of the conformal dimensions $\Delta\in3/2+i\mathbb{R}$. Again, since the integration over $r$ gives a divergent factor that does not depend on $F_{1,2}$, it is natural to drop it. We learn
\begin{equation}\label{Euclidean-inner-product}
    \langle F_1, F_2 \rangle^{\text{reg}} = \int \omega(\beta,\mu)d\beta\, d\mu \int\limits_0^\pi \frac{\sin\theta}{2} d\theta\, F_1(\beta,\mu,\theta)^\ast F_2(\beta,\mu,\theta)\ .
\end{equation}
We fixed the normalisation in the last line by requiring that for $\theta$-independent functions, the integral reduces to that over $\omega(\beta,\mu)$. We have arrived at the measure \eqref{measure-blocks} used in the main text. 

\section{An Identity for Thermal Blocks}
\label{A:Proof-hierarchy}

In this Appendix, we prove the statement \eqref{block_hierarchy}. The latter is a direct consequence of the following property of the blocks in their $1/\Delta$ expansion:
\begin{equation}\label{property_delta_expansion}
\left(F^{\Delta_\phi,a}_{\Delta,\ell}\right)^{(n)}(q,u,s)=\sum_{i=-n}^n \left(h^{\Delta_\phi,a,n}_{\Delta,\ell}\right)^{(i)}(q,u)\,P_{a+i}^{(0,-\frac12)}(1-2s)\,,
\end{equation}
where we are using the convention $P_m^{(\alpha,\beta)}\equiv 0$ if $m<0$. The explicit form of the functions $h^{(i)}(q,u)$ will not be important. Orthogonality of Jacobi polynomials, together with the definition \eqref{measure-blocks} of the measure, implies that the integral in eq.~\eqref{block_hierarchy} is zero if the function $F^{(n)}(q,u,s)$ has an expansion in the basis of Jacobi polynomials $P_{m}^{(0,-\frac{1}{2})}(1-2s)$ starting with a polynomial $P_{m_0}$ with index $m_0>j$. For a given choice of tensor structure label $a$, equation \eqref{property_delta_expansion} tells us then that this condition holds for all orders $n<a-j$. Thus, it remains to show eq.~ \eqref{property_delta_expansion}.
\smallskip

To do this, we focus on the $s$ dependence of the blocks in their $1/\Delta$ expansion. By the equation \eqref{zeroth-order-solution}, the block at order zero is simply
\begin{equation}\label{block_order_delta_zero}
\left(F^{\Delta_\phi,a}_{\Delta,\ell}\right)^{(0)} \propto P_a^{(0,-\frac12)}(1-2s)\ .
\end{equation}
Due to the structure of the recurrence relation \eqref{large-DeltaO-structure} that we use to find $F^{(n+1),a}$ in terms of $F^{(n),a}$, it is clear that the $s$ dependence of $F^{(n+1),a}$ is completely encoded in the action of the differential operator $\mathcal{D}$ on $F^{(n),a}$, together with the boundary condition at $q=0$. Recalling eq.~\eqref{tensor_structures_basis}, the latter fixes the projection of $F^{(n+1),a}$ to the Jacobi polynomial $P_a^{(0,-\frac12)}(1-2s)$. The (non trivial part of the) action of $\mathcal{D}$ in the $s$ space instead, comes through the differential operators $\mathcal{D}^{(1)}_{\Delta_\phi},\mathcal{D}^{(2)}_{\Delta_\phi}$, given explicitly in eqs.~\eqref{D1-p-variable} and \eqref{D2-p-variable}. Using standard properties of Jacobi polynomials, we can write the action of these operators in the $P_m^{(0,-\frac12)}(1-2s)$ basis as
\begin{align}
    \mathcal{D}^{(1)}_{\Delta_\phi} P_m^{(0,-\frac12)}(1-2s)& =  c_{+1}P_{m+1}^{(0,-\frac12)}(1-2s)+c_0 P_m^{(0,-\frac12)}(1-2s)+c_{-1}P_{m-1}^{(0,-\frac12)}(1-2s)
    \, ,\\[2pt]
    \mathcal{D}^{(2)}_{\Delta_\phi}P_m^{(0,-\frac12)}(1-2s) & =4m(1+2m)P_m^{(0,-\frac12)}(1-2s)\,,
\end{align}
where
\begin{gather}
    c_{+1}=\frac{4(1+m)(1+2m)(2m+\Delta_\phi)^2}{(1+4m)(3+4m)}\,,\qquad
    c_{-1}=\frac{4m(1-2m)(1+2m-\Delta_\phi)^2}{1-16m^2}\,,\\[2ex]
    c_0 = \frac{64m^4+64m^3-8m^2(1+2(\Delta_\phi-3)\Delta_\phi)-4m(3+2(\Delta_\phi-3)\Delta_\phi)+2(\Delta_\phi-3)\Delta_\phi}{(1-4m)(3+4m)}\ .
\end{gather}
The precise expressions for the coefficients $c_0$ and $c_\pm$ do not matter - what is important is that $\mathcal{D}P_a$ is a linear combination of $P_{a\pm1}$ and $P_a$. Therefore, remembering eq.~\eqref{block_order_delta_zero}, we see that $F^{(1),a}$ is expressible as a linear combination of $P_a,P_{a\pm1}$, $F^{(2),a}$ as a combination of $P_a,P_{a\pm1},P_{a\pm2}$ and so on. The property \eqref{property_delta_expansion} readily follows.

\section{Details on the Computation of OPE Asymptotics}
\label{A:Details on the computations of OPE asymptotics}

In this appendix, we provide more details supporting the discussion of Section \ref{S:Asymptotic OPE coefficients} that leads to the asymptotic formula \eqref{OPE_coeff_asymptotics_gapped} for OPE coefficients. As described in the main text, the method to obtain the final result can be summarised in the following steps
\begin{enumerate}
    \item start from the inversion formula \eqref{inversion-formula}, and perform the change of variables \[(\beta,\mu,\theta)\rightarrow(x_L,x_R,\theta)\,,\] following eqs.~\eqref{betaLR-variables}, \eqref{chage-xL-xR}
    \item expand the integrand in a power series in $\Delta$, and swap the expansion with the integration in $(x_L,x_R,\theta)$
    \item perform the integration term by term
\end{enumerate}
The non-trivial part of the analysis lies in the last step, as the leading terms of the series can integrate to zero, implying that one has to keep under control also the subleading contributions. Special care needs to be taken in particular with the integration over $\theta$, due to the properties of blocks in our tensor structures basis \eqref{tensor_structures_basis}. Let us elaborate on this last statement. Apart from the simple factor in the integration measure \eqref{measure-blocks}, all the $\theta$ dependence in eq.~\eqref{inversion-formula} comes from the one-point function and the block.  After switching to coordinates $x_L,x_R$ and expanding in $\Delta$, the one-point function \eqref{1pt-expansion-ansatz} can be schematically written as
\begin{equation}\label{one-point-expansion-appendix}
\langle\phi(x)\rangle_{\beta,\Omega}=\Delta^{\frac{\Delta_\phi}{3}}\sum_{j=0}^{\infty}b_{\Delta_\phi}^{(j)}(x_L,x_R,s)\Delta^{-\frac{j}{3}}\,,
\end{equation}
where $b_{\Delta_\phi}^{(j)}(x_L,x_R,s)$ is a polynomial in $s$ of order $\left\lfloor{\frac{j}{2}}\right\rfloor$. For the block we have instead (recall eqs.~\eqref{shadow_block_delta_expansion1} and \eqref{shadow_block_delta_expansion2}, keeping in mind that the following is an ansatz based on observations up to $a=4$)
\begin{equation}\label{block-expansion-appendix}
     q^{\Delta-3} g^{3-\Delta_\phi,a}_{3-\Delta,\ell}(x_L,x_R,s) = \Delta^{1 - \frac43 a}\sum_{n,k=0}^{\infty} f_{a,\ell,\Delta_\phi}^{(n,k)}(x_L,x_R,s)\Delta^{- \frac23 n+\frac{2}{3}\min\left(\left\lfloor{\frac{n}{3}}\right\rfloor,a\right) -\frac{k}{3}}\ .
\end{equation}
The product of eqs.~\eqref{one-point-expansion-appendix} and \eqref{block-expansion-appendix}, which is the quantity entering the inversion formula, therefore expands as
\begin{align}\label{expansion_1pt_and_block}
     & q^{\Delta-3} g^{3-\Delta_\phi,a}_{3-\Delta,\ell}(x_L,x_R,s)\langle\phi(x)\rangle_{\beta,\Omega}=\\[3pt]
     & \hskip3.6cm = \Delta^{1+\frac{\Delta_\phi}{3} - \frac43 a}\sum_{j,n,k=0}^{\infty} f_{a,\ell,\Delta_\phi}^{(n,k)}(s)\,b_{\Delta_\phi}^{(j)}(s)\,\Delta^{- \frac13\left(2n+j+k-2\min\left(\left\lfloor{\frac{n}{3}}\right\rfloor,a\right)\right)}\,, \nonumber
\end{align}
where we left understood the dependence of functions $f_{a,\ell,\Delta_\phi}^{(n,k)},b_{\Delta_\phi}^{(j)}$ on $x_L,x_R$ for ease of notation, as it is not relevant for this discussion.
\smallskip

If the blocks were generic functions of $s$, then the first term of the above series, namely the pair $\{f_{a,\ell,\Delta_\phi}^{(0,0)},b_{\Delta_\phi}^{(0)}\}$ would determine, after integration over $s$, the leading order asymptotics of OPE coefficients, for all values of the tensor structure index $a$. As already anticipated, however, some additional features enter into play. To begin with, the property \eqref{block_hierarchy} of blocks, implies the following “selection rule”
\begin{equation}\label{selection_rule}
    \int_0^\pi d\theta\,\Measure(\beta,\mu,\theta)\, f_{a,\ell,\Delta_\phi}^{(n,k)}(\theta)\,b_{\Delta_\phi}^{(j)}(\theta) = 0\,, \quad \text{if} \quad n+\left\lfloor{\frac{j}{2}}\right\rfloor<a, \;\forall \,k\ .
\end{equation}
This bounds the first possible contribution from the sum in eq.~\eqref{expansion_1pt_and_block} to have a scaling $\Delta^{-\alpha}$ with a power
\begin{equation}\label{scaling1}
    \alpha\geq \frac23\left(a-\left\lfloor{\frac{a}{3}}\right\rfloor\right)\,,
\end{equation}
and also bounds the set of couples $\{f_{a,\ell,\Delta_\phi}^{(n,k)},b_{\Delta_\phi}^{(j)}\}$ that can contribute to each order. For example, it reduces the number of terms with $\alpha$ saturating the bound to $2^{\,a\,\mathrm{mod}\, 3}$. Inspecting then explicitly, for $a$ up to 4, these couples that enter into the leading orders, we notice further cancellations that are brought by the integration over $\theta$, leading to two major final simplifications:

\begin{enumerate}
    \item The scaling $\Delta^{-\alpha}$ of the dominant contribution is always
    \begin{equation}\label{scaling2}
            \alpha=\frac23 a\,,
    \end{equation}
    and in particular does not always saturate the bound \eqref{scaling1}. This exponent, combined with that coming from the overall $\Delta$ factor in front of the sum in eq.~\eqref{expansion_1pt_and_block}, implies an overall leading asymptotic for eq.~\eqref{expansion_1pt_and_block} of $\sim \Delta^{1+\frac{\Delta_\phi}{3} - 2a}$. Pairing this to the leading behaviour of the other contributions entering into the inversion formula \eqref{inversion-formula}, namely the measure \eqref{omega-xL-xR}, the partition function \eqref{Zgapped-xL-xR} and the density \eqref{rho-gapped}, one obtains the scaling of OPE coefficients written in eq.~\eqref{OPE_coeff_asymptotics_gapped}.
    \item Of all possible couples $\{f_{a,\ell,\Delta_\phi}^{(n,k)},b_{\Delta_\phi}^{(j)}\}$ that could contribute to the term with the above scaling \eqref{scaling2}, only the single couple $\{f_{a,\ell,\Delta_\phi}^{(0,0)},b_{\Delta_\phi}^{(2a)}\}$ survives the integration over $\theta$.\\
    As a by-product, since the $s$ dependence of the term $f_{a,\ell,\Delta_\phi}^{(0,0)}$ is completely encoded in the Jacobi polynomial $P_a^{(1,-\frac12)}(1-2s)$ (recall eq.~\eqref{f_0_0}), each OPE coefficients will depend, to leading order in $\Delta$, only on a single one of the dynamical coefficients parametrising the one-point function high temperature expansion \eqref{1pt-expansion-ansatz}. Namely the coefficient in question is $b_{0,2a,a}$ in the notation \eqref{1pt-coeff-s-expansion}, as this is the only coefficient multiplying the power $s^{a}$ inside the polynomial function $b_{\Delta_\phi}^{(2a)}$.
\end{enumerate}
All the information needed to verify these last two observations is summarised in the Table \ref{tab:blocks_expansion_s_dependence} below. The table shows the $s$ dependence of the coefficients $f_{a,\ell,\Delta_\phi}^{(n,k)}(s)$ entering in eq.~\eqref{expansion_1pt_and_block} whose corresponding powers $\Delta^{-\alpha}$ satisfy
\begin{equation}\label{alpha_bound}
        \frac23\left(a-\left\lfloor{\frac{a}{3}}\right\rfloor\right)\leq \alpha \leq \frac23a\ .
\end{equation}
Furthermore, we do not consider those terms that are already excluded by the selection rule \eqref{selection_rule}. We expand each $f_{a,\ell,\Delta_\phi}^{(n,k)}(s)$ in the basis of Jacobi polynomials $P_m^{(0,-\frac12)}(1-2s)$, or rather, record those Jacobi polynomials with which it has non-zero overlap. In addition, for each $f_{a,\ell,\Delta_\phi}^{(n,k)}$, we include the information about which coefficients $b_{\Delta_\phi}^{(j)}$ it is paired with. Having collected this data, to check the two statements made above one has to recall that $b_{\Delta_\phi}^{(j)}$ is a polynomial in $s$ of order $\left\lfloor{\frac{j}{2}}\right\rfloor$. As the basis $\{P_m^{(0,-\frac12)}(1-2s)\}$ of Jacobi polynomials is orthogonal in $\theta$ space with respect to our measure \eqref{measure-blocks}, we have that a term $f_{a,\ell,\Delta_\phi}^{(n,k)}$ whose expansion in the basis of Jacobi polynomials starts with a certain polynomial $P_{m_0}$, integrates to zero when paired with a term $b_{\Delta_\phi}^{(j)}$ unless $m_0\leq \left\lfloor{\frac{j}{2}}\right\rfloor$.\\

Inspecting the table, one can easily notice that this non-vanishing condition holds only for the terms in the first column. These are the terms $f_{a,\ell,\Delta_\phi}^{(n=0,k=0)}$, which are paired to the coefficients $b_{\Delta_\phi}^{(2a)}$ coming from the one-point function. This motivates our claim that for a fixed tensor structure label $a$, it is only the single couple $\{f_{a,\ell,\Delta_\phi}^{(0,0)},b_{\Delta_\phi}^{(2a)}\}$ that controls the leading large $\Delta$ asymptotics of eq.~\eqref{expansion_1pt_and_block}.

\begin{table}
\centering
\renewcommand{\arraystretch}{2}
\resizebox{1\textwidth}{!}{
\begin{tabular}{|C|C|C|C|C|C|C|C|C|C|C|}
\hline
\multirow{2}{*}{\text{\diagbox[width=3.5em, height=5.4em]{$\quad a$}{$n\;$}}} & 0 & 1 & 2 & \multicolumn{3}{C|}{3}& \multicolumn{3}{C|}{4} & 5\\
\cline{2-11}
&  \scalebox{0.95}{$k=0$} & \scalebox{0.95}{$0$} & \scalebox{0.95}{$0$} & \scalebox{0.95}{$0$} & \scalebox{0.95}{$1$} & \scalebox{0.95}{$2$}  & \scalebox{0.95}{$0$} & \scalebox{0.95}{$1$} & \scalebox{0.95}{$2$} & \scalebox{0.95}{$0$} \\
\hline
 0 & P_0\;_{\{0\}} & & & & & & & & & \\
 1 & P_1\;_{\{2\}} & P_1\;_{\{0\}} & & & & & & & & \\
 2 & P_2\;_{\{4\}} & P_2\;_{\{2\}} & P_{1,2}\;_{\{0\}} & P_1\;_{\{0\}} & & & & & & \\
 3 & P_3\;_{\{6\}} & P_3\;_{\{4\}} & P_{2,3}\;_{\{2\}} &  P_2\;_{\{0,1,2\}} & P_2\;_{\{0,1\}} & P_{2,3}\;_{\{0\}} &  P_2\;_{\{0\}} & & &\\
 4 & P_4\;_{\{8\}} & P_4\;_{\{6\}} & P_{3,4}\;_{\{4\}} &  P_3\;_{\{2,3,4\}} & P_3\;_{\{2,3\}} & P_{3,4}\;_{\{2\}}  & P_3\;_{\{0,1,2\}}  & P_3\;_{\{0,1\}}  & P_{2,3,4}\;_{\{0\}}  & P_{2,3}\;_{\{0\}}  \\
\hline
\end{tabular}
}
\caption{The table describes the $s$ dependence of the functions $f_{a,\ell,\Delta_\phi}^{(n,k)}$ that in eq.~\eqref{expansion_1pt_and_block} contribute to the terms having a scaling $\Delta^{-\alpha}$ with $\alpha$ bounded as in eq.~\eqref{alpha_bound}, not considering the $f_{a,\ell,\Delta_\phi}^{(n,k)}(s)$ that are already excluded by the selection rule \eqref{selection_rule}. The $s$ dependence here is encoded in terms of the basis of Jacobi polynomials $P_{m}^{(0,-\frac12)}(1-2s)$; when writing $P_{1,2}$, we mean that the function can be expanded in terms of the two Jacobi polynomials with index $m=1$ and $m=2$. For completeness, we have also included the information about which coefficients $b_{\Delta_\phi}^{(j)}$ each of the $f_{a,\ell,\Delta_\phi}^{(n,k)}$ is paired with. The numbers in curly brackets represent the corresponding set of indices $j$.}
\label{tab:blocks_expansion_s_dependence}
\end{table}

\section{More on the Free Theory}
\label{A:More on the free theory}

In this appendix, we include several details related to our analysis of the free field theory. 

\subsection{Asymptotic behaviour of certain infinite sums}
\label{AA:Asymptotic behaviour of certain infinite sums}

We review the asymptotic expansions of sums of the form
\begin{equation}\label{g-sum-of-f}
    g(\beta) = \sum_{n=1}^\infty f(n\beta)\,,
\end{equation}
following \cite{Zagier}. The function $f$ is required to be sufficiently rapidly decaying at $\beta\to\infty$ so that the series \eqref{g-sum-of-f} converges. Assume further that $f$ admits an asymptotic expansion at $\beta=0$ of the form
\begin{equation}\label{f-Zagier-asymptotics}
    f(\beta) \sim \sum_{\lambda<-1} a_\lambda \beta^\lambda + \sum_{\lambda\geq-1} b_\lambda \beta^\lambda\,,
\end{equation}
with only finitely many terms in the first sum. Let us denote the first finite sum by $f_-(\beta)$ and put $f_+ = f - f_-$.  Then, the function $g(\beta)$ admits the following asymptotic expansion at $\beta=0$
\begin{equation}\label{Zagier-formula}
    g(\beta) = \sum_{\lambda<-1} a_\lambda\zeta(-\lambda) \beta^\lambda - b_{-1} \frac{\log\beta}{\beta} + \frac{I_f}{\beta} + \sum_{\lambda>-1}^\infty b_\lambda \zeta(-\lambda) (-\beta)^\lambda\,,
\end{equation}
where $I_f$ is given by
\begin{equation}
    I_f = \int\limits_0^\infty d\beta\left(f_+(\beta) - \frac{b_{-1}e^{-\beta}}{\beta}\right)\ .
\end{equation}
In the main text, we apply formula \eqref{Zagier-formula} to sums with the function $f$ given by eqs.~\eqref{f-partition-function} and \eqref{f-one-point}. These functions are exponentially decaying as $\beta\to\infty$, so eq.~\eqref{g-sum-of-f} converges. They also admit asymptotic expansions of the form \eqref{f-Zagier-asymptotics}.

\subsection{Higher order corrections in the free theory's asymptotic formulas}
\label{app:Asymp_freeTheory}
Here we provide explicit expressions for additional subleading corrections in the large-$\Delta$ asymptotic expansions of the density of primaries and OPE coefficients in free theory, beyond those already shown in the main text. These, along with further terms, can be found also in an auxiliary Mathematica file, at \href{https://gitlab.com/russofrancesco1995/thermal_blocks}{gitlab.com/russofrancesco1995/thermalblocks}.

\subsubsection{Density of primaries}
\label{AAA:Free-spectrum-asymptotics}

We refine the equation \eqref{eq::rho_refined_leading} for the asymptotic density of primaries, and show some of the corrections that are plotted in Figures \ref{fig:spectrum0_free} and \ref{fig:spectrum2_free}. The complete list of higher-order corrections used in the plots is provided in the accompanying Mathematica file; we have chosen to omit part of them from this appendix to avoid overly lengthy expressions. We write the expansion as
\begin{equation}
    \rho^{\text{free}}(\Delta,\ell) \equiv \frac{e^{-\zeta^\prime(-1)}(2\ell+1)}{\sqrt{3}}\Delta^{-\frac{131}{36}}\,e^{3\pi^{\frac13}f^{\frac13}\Delta^\frac23}\sum_{n=0}^{\infty} \rho_n(\Delta,\ell) \Delta^{-n/3}\ .
\end{equation}
where the residual $\Delta$ dependence of the terms $\rho_n(\Delta,\ell)$ is at most of power law type in $\log\Delta$, so that each correction is indeed asymptotically subleading with respect to the previous one.
The first few terms read
\begin{align}
    \rho_0(\Delta,\ell) & = 2^{17/6} f^{59/36} \pi^{23/36}  \, , \\[1ex]
    \rho_1(\Delta,\ell) & = - 3 \cdot 2^{17/6} f^{71/36}  \pi^{35/36} \, , \\[1ex]
    \rho_2(\Delta,\ell) & = \frac{f^{47/36}\pi^{11/36}}{1080\cdot 2^{1/6}}  \left( 360 \log \left(\frac{8\Delta}{\pi f} \right)+43179 \pi f +  1080 \gamma -31445 \right)  \, , \\[1ex]
    \rho_3(\Delta,\ell) & = -\frac{f^{59/36}\pi^{23/36}}{360\cdot 2^{1/6}}  \left( 360 \log \left(\frac{8\Delta}{\pi f} \right)+17259 \pi  f + 1080 \gamma -47165\right)  \, , \\[1ex]
   \rho_4(\Delta,\ell) & = \frac{f^{35/36}}{43545600\cdot 2^{1/6}\pi^{1/36}} \bigg( 504000 \log^2\left( \frac{8 \Delta}{\pi f} \right)\nonumber\\
   &+ 1680 \left(-33485 + 1800\gamma + 43179 \pi f \right) \log\left( \frac{8 \Delta}{\pi f} \right)\nonumber\\
   & + 25200\gamma \left(-6697 + 180\gamma \right) - 13308952790 \pi f + 217622160\gamma \pi f + 1993044069 \pi^2 f^2\nonumber\\
    &- 348364800\pi f \ell (1 + \ell)  + 2073684725+ 133358400 \zeta(3) \bigg)\,, \\[1ex]
    \rho_5(\Delta,\ell) & = -\frac{f^{47/36} \pi^{11/36}}{14515200\cdot 2^{1/6}}  \bigg(504000 \log^2\left( \frac{8 \Delta}{\pi f} \right) \nonumber\\
    &+ 1680(-52085 + 1800\gamma + 17259\pi f) \log\left( \frac{8 \Delta}{\pi f} \right)\nonumber\\
    &+ 25200\gamma(-10417 + 180\gamma )- 7026281990\pi f + 86985360\gamma\pi f + 532452069\pi^2 f^2 \nonumber\\
    &- 348364800\pi f \ell(1 + \ell) + 4810368325 + 133358400 \zeta(3) \bigg) \,.
\end{align}

\subsubsection{OPE coefficients}
\label{AAA:Free-OPE-asymptotics}

For selected values of the tensor structure and spin labels, we provide a refinement of the equation \eqref{OPE-coefficients-dominant-asymptotics-free} describing the asymptotic expansion of OPE coefficients, to show the corrections that are plotted in Figures \ref{fig:ope_ScalarAsymp}, \ref{fig:ope_dominant}, and \ref{fig:ope_spin6SubdominantWAsymp}. The functions displayed below can also be found in the accompanying Mathematica file. We will parametrise the expansion as
\begin{equation}    \overline{\lambda^a_{\mathcal{O}\mathcal{O}\phi}}(\Delta,\ell) = \left( \frac{\Delta}{\pi f} \right)^{1/3}\Delta^{-2a}\sum_{n=0}^{\infty} \overline{\lambda^a_n}(\Delta,\ell) \,\Delta^{-n/3}\,,
\end{equation}
where, as for the case of the density of primaries, the residual $\Delta$ dependence of the terms $\overline{\lambda^a_n}(\Delta,\ell)$ is at most of logarithmic type.

\paragraph{Dominant tensor structure (for all spin values)}

\begin{align}
    \overline{\lambda_0^0}(\Delta,\ell) &= \frac13 \log\left(\frac{8\Delta}{\pi f}\right)+\gamma \,, \\[1ex]
    \overline{\lambda_1^0}(\Delta,\ell)&=0\,,\\[1ex]
    \overline{\lambda_2^0}(\Delta,\ell) &= \frac{1}{216 \pi ^{1/3}
   f^{1/3}}\left(119 \log\left(\frac{8\Delta}{\pi f}\right)+3 \pi  f+189 \zeta (3)+357 \gamma +177\right) \,, \\[1ex]
    \overline{\lambda_3^0}(\Delta,\ell) &= -\frac13 \log\left(\frac{8\Delta}{\pi f}\right)+\gamma +\frac12  \,, \\[1ex]
    \overline{\lambda_4^0}(\Delta,\ell) &= -8 \log\left(\frac{8 \Delta}{f \pi}\right)
- 3 \left(2 + 8 \gamma - 7 \zeta(3)\right)\ .\\
\end{align}
\paragraph{First subdominant (for selected spin values)}~\\[2ex]
\textit{Spin one}
\begin{align}
    \overline{\lambda_0^1}(\Delta,1) &= -\frac{1 }{36 \sqrt{2}} \left(4 \log\left(\frac{8\Delta}{\pi f}\right)-21 \zeta (3)+12 \gamma +18\right)\, , \\[1ex]
    \overline{\lambda_1^1}(\Delta,1) &=0\, ,\\[1ex]
    \overline{\lambda_2^1}(\Delta,1) &= -\frac{1}{2592 \sqrt{2} \pi^{1/3} f^{1/3} } \bigg(16 \log^2\left(\frac{8 \Delta}{\pi f}\right)\\
    & +12 \log\left(\frac{8 \Delta}{\pi f}\right)\left(51 + 8 \gamma - 7 \zeta(3)\right)\\
    & +3 \left(1262 + 12 \pi f + 12 \gamma \left(51 + 4 \gamma - 7 \zeta(3)\right) + 5355 \zeta(3) - 7254 \zeta(5)\right)\bigg) \,, \\[1ex]
    \overline{\lambda_3^1}(\Delta,1) &=-\frac{1 }{72 \sqrt{2}}\left( 8 \log\left(\frac{8 \Delta}{\pi f}\right)+3 \left(2 + 8 \gamma - 7 \zeta(3)\right)\right) \, .
\end{align}
\textit{Spin six}
\begin{align}
    \overline{\lambda_0^1}(\Delta,6) &= -\frac{\sqrt{77}}{12 \sqrt{2}} \left(4 \log\left(\frac{8\Delta}{\pi f}\right)-21 \zeta (3)+12 \gamma +18\right)\, , \\[1ex]
    \overline{\lambda_1^1}(\Delta,6) &= 0 \,,\\[1ex]
    \overline{\lambda_2^1}(\Delta,6) &= -\frac{\sqrt{77}}{864 \sqrt{2} \pi^{1/3} f^{1/3} } \bigg( 16 \log^2\left(\frac{8 \Delta}{\pi f}\right)\\
    &+ 12 \log\left(\frac{8 \Delta}{\pi f}\right) \left(51 + 8 \gamma - 7 \zeta(3)\right)\\
    &+ 3 \left(1262 + 12 \pi f + 12 \gamma \left(51 + 4 \gamma - 7 \zeta(3)\right) + 5355 \zeta(3) - 7254 \zeta(5)\right)\bigg) \,, \\[1ex]
    \overline{\lambda_3^1}(\Delta,6) &=-\frac{1}{24\sqrt{2}}\left(8 \log\left(\frac{8\Delta}{\pi f}\right)+3 \left(2 + 8\gamma - 7\zeta(3)\right)\right) \,.\\
\end{align}
\paragraph{Second subdominant (for spin six only)}~\\
\begin{align}
    \overline{\lambda_0^2}(\Delta,6) &= \frac{3\sqrt{187} }{16 \sqrt{7}} \left(16 \log\left(\frac{8\Delta}{\pi f}\right)-840\zeta(3)+837\zeta(5)+48\gamma+100\right)\, , \\[1ex]
    \overline{\lambda_1^2}(\Delta,6) &=0\, ,\\[1ex]
    \overline{\lambda_2^2}(\Delta,6) &= \frac{\sqrt{187}}{384\sqrt{7} \pi^{1/3}  f^{1/3}} \bigg( 128 \log^2\left(\frac{8\Delta}{\pi f}\right)\\
    & + 24 \log\left(\frac{8\Delta}{\pi f}\right) 
\left(278 + 32\, \gamma - 280\, \zeta(3) + 279\, \zeta(5)\right)\\
    & + 240 f \pi+ 1152\gamma^2- 72\gamma \left(-278 + 280\zeta(3) - 279\zeta(5)\right)\\
    &+160440\zeta(3) - 4882221\zeta(5) + 4783455\zeta(7) + 39532\bigg) \,.
\end{align}

\subsection{Wick's theorem at finite temperature}
\label{AA:Wick's theorem at finite temperature}

In this appendix, we demonstrate the validity of Wick's theorem for finite temperature four-point functions in the free field theory. By the plane-cylinder map, we may consider the theory on $\mathbb{R}\times S^2$. Let us denote the fundamental field by on this background by $\Phi$. The action for the conformally coupled scalar reads
\begin{equation}\label{classical-action-free}
    S = \int\sin\theta\, d\tau\, d\theta\, d\varphi\  \left(\frac12\dot\Phi^2 + \Phi\Delta_{S^2}\Phi - \frac18\Phi^2\right)\ .
\end{equation}
Using the canonical quantisation, the field is expanded in modes
\begin{equation}
    \Phi = \sum_{l,m} \frac{1}{\sqrt{2\omega_l}}\left(a_{l,m} e^{-i\omega_l \tau} Y_{l,m} + a_{l,m}^\dagger e^{i\omega_l \tau} Y_{l,m}^\ast\right)\,, \qquad \omega_l = l+\frac12\,,
\end{equation}
where $Y_{l,m}$ are the usual spherical harmonics. The modes are promoted to operators acting on the Fock space and satisfying the canonical commutation relations
\begin{equation}
    [a_{l,m},a^\dagger_{l',m'}] = \delta_{ll'}\delta_{mm'}, \quad [a,a] = [a^\dagger,a^\dagger] = 0\ .
\end{equation}
Let us consider the four-point function at finite temperature and chemical potential, $\langle\Phi\Phi\Phi\Phi\rangle_{q,y}$. This correlator may be expanded in four-point functions of creation and annihilation operators. By orthogonality of states with different particle numbers, the only non-vanishing such four-point functions contain two $a$-s and two $a^\dagger$-s. Therefore, let us consider
\begin{equation}\label{4pt-creation-annihilation}
    \langle a_{l_1,m_1}a_{l_2,m_2} a_{l_3,m_3}^\dagger a_{l_4,m_4}^\dagger\rangle_{q,y} \equiv \langle123^\dagger4^\dagger\rangle_{q,y}\ .
\end{equation}
We shall also write
\begin{equation}
     \langle a_{l_i,m_i}a_{l_j,m_j}^\dagger\rangle_{q,y} \equiv \langle ij^\dagger\rangle\ .
\end{equation}
With this notation, it was shown in \cite{Buric:2024kxo} that finite temperature two-point functions read
\begin{equation}\label{2pt-creation-annihilation}
    \langle ij^\dagger\rangle_{q,y} = \frac{\delta_{ij}}{1 - q^{\omega_{l_i}}y^{m_i}}\,, \qquad \langle i^\dagger j\rangle_{q,y} = \frac{q^{\omega_{l_i}} y^{m_i}}{1 - q^{\omega_{l_i}}y^{m_i}}\delta_{ij}\,,
\end{equation}
where we have written $\delta_{ij}$ as a shorthand for $\delta_{l_i l_j}\delta_{m_i m_j}$. In order to compute eq.~\eqref{4pt-creation-annihilation}, we also look at correlators $\langle13^\dagger24^\dagger\rangle_{q,y}$ and $\langle3^\dagger124^\dagger\rangle_{q,y}$. These satisfy the equations
\begin{align}
    \langle 1 2 3^\dagger 4^\dagger \rangle_{q,y} - \langle 1 3^\dagger 2 4^\dagger \rangle_{q,y} & = \delta_{23} \langle 14^\dagger\rangle_{q,y}\,,\\
    \langle 1 3^\dagger 2 4^\dagger \rangle_{q,y} - \langle 3^\dagger 1 2 4^\dagger \rangle_{q,y} & = \delta_{13} \langle 24^\dagger\rangle_{q,y}\,,\\
    \langle 3^\dagger 1 2 4^\dagger \rangle_{q,y} & = q y^{m_3} \langle 1 2 3^\dagger 4^\dagger\rangle_{q,y}\ .
\end{align}
The first two equations follow by canonical commutation relations and the third one is derived using cyclicity of the trace and the Baker-Campbell-Hausdorff formula. We get
\begin{equation}
    \langle123^\dagger4^\dagger\rangle_{q,y} = \frac{\delta_{23}\langle14^\dagger\rangle_{q,y} + \delta_{13}\langle24^\dagger\rangle_{q,y}}{1- q^{\omega_{l_3}} y^{m_3}} = \langle23^\dagger\rangle_{q,y}\langle14^\dagger\rangle_{q,y} + \langle13^\dagger\rangle_{q,y} \langle24^\dagger\rangle_{q,y}\ .
\end{equation}
In the second step, we have used the result \eqref{2pt-creation-annihilation}. By similar manipulations, one shows
\begin{align}
    & \langle1^\dagger234^\dagger\rangle_{q,y} = \langle1^\dagger3\rangle_{q,y}\langle24^\dagger\rangle_{q,y} +  \langle1^\dagger2\rangle_{q,y}\langle34^\dagger\rangle_{q,y}\,,\\
    & \langle12^\dagger34^\dagger\rangle_{q,y} = \langle2^\dagger3\rangle_{q,y}\langle14^\dagger\rangle_{q,y} + \langle12^\dagger\rangle_{q,y}\langle34^\dagger\rangle_{q,y}\,,\\
    & \langle12^\dagger3^\dagger4\rangle_{q,y} = \langle2^\dagger4\rangle_{q,y}\langle13^\dagger\rangle_{q,y} + \langle12^\dagger\rangle_{q,y}\langle3^\dagger4\rangle_{q,y}\,,\\
    & \langle1^\dagger23^\dagger4\rangle_{q,y} = \langle1^\dagger4\rangle_{q,y} \langle23^\dagger\rangle_{q,y} + \langle1^\dagger2\rangle_{q,y} \langle3^\dagger4\rangle_{q,y} \,,\\
    & \langle1^\dagger2^\dagger34\rangle_{q,y} = \langle1^\dagger4\rangle_{q,y} \langle2^\dagger3\rangle_{q,y} + \langle1^\dagger 3\rangle_{q,y}\langle2^\dagger4\rangle_{q,y}  \ .
\end{align}
These equations are at the basis of Wick's theorem for four-point functions Let us put
\begin{align}
    \llb 12^\dagger\rrb & = \frac{e^{-i\omega_{l_1}\tau_1 + i \omega_{l_2}\tau_2}}{2\sqrt{\omega_{l_1}\omega_{l_2}}} Y(\vartheta_1)Y^\ast(\vartheta_2)\langle12^\dagger\rangle_{q,y}\,,\\
    \llb 123^\dagger4^\dagger\rrb & = \frac{e^{-i\omega_{l_1}\tau_1 - i\omega_{l_2}\tau_2 + i\omega_{l_3}\tau_3 + i \omega_{l_4}\tau_4}}{4\sqrt{\omega_{l_1}\dots\omega_{l_4}}} Y(\vartheta_1)Y(\vartheta_2)Y^\ast(\vartheta_3)Y^\ast(\vartheta_4)\langle123^\dagger4^\dagger\rangle_{q,y}\ .
\end{align}
We have denoted the coordinates on the sphere $S^2$ collectively by $\vartheta$. The two-point function from \cite{Buric:2024kxo} can be re-written as
\begin{equation}
    \langle\Phi(\tau_1,\vartheta_1)\Phi(\tau_2,\vartheta_2)\rangle_{q,y} = \sum_{l_i,m_j} \left( \llb12^\dagger\rrb + \llb1^\dagger2\rrb \right)\ .
\end{equation}
On the other hand, the four-point function reads
\begin{align}
    \langle \Phi(\tau_1,\vartheta_1)\dots\Phi(\tau_4,\vartheta_4)\rangle_{q,y} & = \sum_{l_i,m_j}  \Big( \llb123^\dagger4^\dagger\rrb + \llb12^\dagger34^\dagger\rrb\\
    & + \llb12^\dagger3^\dagger4\rrb + \llb1^\dagger234^\dagger\rrb + \llb1^\dagger23^\dagger4\rrb +\llb1^\dagger2^\dagger34\rrb\Big)\ .
\end{align}
Therefore, using the above properties of four-point functions of modes, we obtain that, as at zero temperature, the four-point function can be written as a sum over three Wick contractions
\begin{equation}
    \langle \Phi(\tau_1,\vartheta_1)\dots\Phi(\tau_4,\vartheta_4)\rangle_{q,y} = \langle\Phi(\tau_1,\vartheta_1)\Phi(\tau_2,\vartheta_2)\rangle_{q,y} \langle\Phi(\tau_3,\vartheta_3)\Phi(\tau_4,\vartheta_4)\rangle_{q,y} + \text{perms}\ .
\end{equation}

\subsection{One-point functions of more operators}
\label{SS:Higher-point functions and more operators}

In the main text, we have found the one-point function of $\phi^2$ by computing the two-point function of $\phi$ and using the OPE. Similarly, the same two-point function \eqref{two-point-function} can be used to obtain the thermal one-point functions $\langle\mathcal{O}\rangle_{q,y}$ of all operators that appear in the $\phi\times\phi$ OPE. This straightforward procedure was illustrated on the example of the stress tensor in \cite{Buric:2024kxo}. In order to compute one-point functions of operators not present in the $\phi\times\phi$ OPE and extract associated CFT data, one may start from higher-point functions instead of eq.~\eqref{two-point-function}. Let us illustrate this on the example of $\langle\phi^4\rangle_{q,y}$, which is extracted from the four-point function of $\phi$. The latter is obtained by an application of Wick's theorem, as we have just reviewed (see also \cite{Evans:1996bha}). Starting with this four-point functions, we can fuse the fields at points 1 and 2 to get
\begin{align}
    & \langle \phi(x_1)\phi(x_2)\phi(x_3)\phi(x_4) \rangle_{q,y} = \sum_{\mathcal{O}} \lambda_{\phi\phi\mathcal{O}} \frac{C_{\mathcal{O}}^{\mu_1\dots\mu_\ell}(x_{12},\partial_2)}{|x_{12}|^{2\Delta_\phi-\Delta+\ell}} \langle\mathcal{O}_{\mu_1\dots\mu_\ell}(x_2)\phi(x_3)\phi(x_4)\rangle_{q,y}\nonumber\\
    & = \frac{\langle\phi(x_3)\phi(x_4)\rangle_{q,y}}{|x_{12}|^{2\Delta_\phi}}  + \lambda_{\phi\phi\phi^2} \frac{C_{\phi^2}(x_{12},\partial_2)}{|x_{12}|^{2\Delta_\phi-\Delta_{\phi^2}}} \langle\phi^2(x_2)\phi(x_3)\phi(x_4)\rangle_{q,y} + \dots\\
    & = \frac{\langle\phi(x_3)\phi(x_4)\rangle_{q,y}}{|x_{12}|^{2\Delta_\phi}}  + \lambda_{\phi\phi\phi^2}\langle\phi^2(x_2)\phi(x_3)\phi(x_4)\rangle_{q,y} + \dots\ . \nonumber
\end{align}
In the first step, we have expanded the OPE of fields at points 1 and 2 to the second order. To get to the last line, the leading term of the OPE differential operator is taken. We obtain the three-point function
\begin{equation}\label{3-point-function-exact}
    \langle\phi^2(x_2)\phi(x_3)\phi(x_4)\rangle_{q,y} = \frac{\left(\langle \phi(x_1)\phi(x_2)\phi(x_2)\phi(x_4) \rangle_{q,y} - \frac{\langle\phi(x_3)\phi(x_4)\rangle_{q,y}}{|x_{12}|^{2\Delta_\phi}}\right)_{x_1=x_2}}{\lambda_{\phi\phi\phi^2}}\,,
\end{equation}
in terms of the already found two- and the four-point functions of the fundamental field. The procedure can readily be iterated. By performing the OPE of fields at points 3 and 4, we extract the two-point function of $\phi^2$
\begin{equation}\label{2-point-function-exact}
     \langle\phi^2(x_2)\phi^2(x_4)\rangle_{q,y} = \frac{\left(\langle \phi^2(x_2)\phi(x_3)\phi(x_4) \rangle_{q,y} - \frac{\langle\phi^2(x_2)\rangle_{q,y}}{|x_{34}|^{2\Delta_\phi}}\right)_{x_3 = x_4}}{\lambda_{\phi\phi\phi^2}}\,,
\end{equation}
in terms of the three-point function \eqref{3-point-function-exact} and the one-point function \eqref{phi2-one-pt-function}. Finally, we perform the $\phi^2\times\phi^2$ OPE and extract the one-point function of $\phi^4$
\begin{equation}\label{1-point-function-exact}
     \langle\phi^4(x_4)\rangle_{q,y} = \frac{\left(\langle \phi^2(x_2)\phi^2(x_4)\rangle_{q,y} - \frac{1}{|x_{24}|^{4\Delta_{\phi}}} - \lambda_{\phi^2\phi^2\phi^2} \frac{\left(1+\frac12 x_{24}\cdot\partial_4\right)}{|x_{24}|^{2\Delta_{\phi}}} \langle\phi^2(x_4)\rangle_{q,y}\right)_{x_2=x_4}}{\lambda_{\phi^2\phi^2\phi^4}}\,,
\end{equation}
in terms of previously determined data \eqref{2-point-function-exact} and \eqref{phi2-one-pt-function}.

\bibliographystyle{JHEP}
\bibliography{bibliography}
\end{document}